%% file: main.tex
\documentclass[notitlepage,12pt]{article}
\usepackage{fancyvrb, verbatim, listings}                                  % Various for inserting/hosting code
\usepackage{setspace}                                                      % Space between lines (customisable)
\usepackage[margin=2.54cm]{geometry}                                       % Set margins (standard is 2.54cm)
\usepackage[UKenglish,cleanlook]{isodate}                                  % Set default date and date display
\usepackage{fancyhdr} 
\usepackage[justify]{ragged2e}                                             % control over text alignment
\usepackage[activate={true,nocompatibility},final,tracking=true,kerning=true,spacing=true,factor=1100,stretch=10,shrink=10]{microtype}
\microtypecontext{spacing=nonfrench}                                       % Change font and minor spacings to look nicer
\usepackage{multirow}
\usepackage{graphicx}                                                      % control over the import of graphics
\usepackage[table,dvipsnames]{xcolor}                                      % Define tables (with colour control)
\usepackage{float}                                                         % Control over graphics + tables float
\usepackage{booktabs}                                                      % caption graphics + tables (with name in bold)
\usepackage{csvsimple}                                                     % Host a table in LaTeX.
\usepackage[labelfont=bf]{caption}                                         % Caption for objects
\usepackage{subcaption}                                                    % Caption for sub objects
\usepackage{mathtools}                                                     % Various maths functions
\usepackage{amssymb}                                                       % Various maths functions
\usepackage{amsmath}                                                       % Various maths functions
\usepackage{dsfont}                                                        % Various maths functions
\usepackage{centernot}                                                     % center \not usage
\usepackage{siunitx} 
\usepackage[normalem]{ulem} % Strike-through package
\renewcommand{\vec}[1]{\boldsymbol{\mathit{#1}}}                           % vector notation shortcut
\newcommand{\Prob}[1]{\Pr\left( #1 \right)}                         % SHortcut for probability notation
\newcommand{\Probgiven}[2]{\Pr\left( #1 \, \middle\vert \, #2 \right)} % SHortcut for probability notation, given
\newcommand{\E}[2][]{\mathbb{E}_{#1} \left[ #2 \right]}                    % Expectation (with optional subscript) shortcut
\newcommand{\Egiven}[3][]{\mathbb{E}_{#1} \left[ #2 \, \middle\vert \, #3 \right]} % Expectation given (with optional subscript) shortcut
\newcommand{\indicator}[1]{\mathds{1}\left\{ #1 \right\}}                  % SHortcut for indicator function
\renewcommand{\hat}[1]{\widehat{#1}}                                       % Default estimator notation is widehat
\renewcommand{\bar}[1]{\overline{#1}}                                      % Make over bar look nicer
\usepackage{natbib}%[longnamesfirst]                                        % Citation package, see https://en.wikibooks.org/wiki/LaTeX/Bibliography_Management#Natbib
\usepackage{hyperref}                                        % Allow for links across the text, with colour options
\hypersetup{colorlinks=true, linkcolor=blue, citecolor=blue, filecolor=magenta, urlcolor=blue}
\def\sectionautorefname~#1\null{Section~#1\null}                           % Fix autoref for sections
\def\subsectionautorefname~#1\null{Subsection~#1\null}                     % Fix autoref for sections
\def\equationautorefname~#1\null{Equation~(#1)\null}                       % Fix autoref for equations
\newcommand{\appendixref}[1]{\hyperref[#1]{Appendix~\ref*{#1}}}            % Appendix references.
\author{
    Senan Hogan-Hennessy\thanks{Economics Department, Cornell University.
        Corresponding author, \href{mailto:seh325@cornell.edu}{\nolinkurl{seh325@cornell.edu}.}
        Authors are listed in descending order of contribution.}
    \and
    Seungmin Lee \thanks{Keough School of Global Affairs, University of Notre Dame. \url{slee76@nd.edu}.}
    \and
    Christopher B. Barrett
    \thanks{Charles H. Dyson School of Applied Economics and Management, and Jeb E. Brooks School of Public Policy, Cornell University.  \url{cbb2@cornell.edu}.}
    \and  
    John Hoddinott \thanks{Division of Nutritional Sciences, Charles H. Dyson School of Applied Economics and Management, and Ashley School of Global Development and the Environment, Cornell University.  \url{jfh246@cornell.edu}.}
    \and
    Matthew P. Rabbitt \thanks{Urban Institute and Charles H. Dyson School of Applied Economics and Management, Cornell University.  \url{mrabbitt@urban.org}.}
}
\title{Food Insecurity Among Military Veterans\thanks{
        This work was supported by USDA Economic Research Service Cooperative agreement 58-4000-3-0073-R and the Cornell University Agricultural Experiment Station, using Federal Capacity Funds from the US Department of Agriculture's National Institute of Food and Agriculture, through project award NYC-121946.
        Analysis of restricted Panel Study of Income Dynamics (PSID) data was authorized under a data use agreement with the University of Michigan, for Cornell University IRB protocol 0148833.
        PSID restricted use data are produced and distributed by the \cite{PSID}, Institute for Social Research, University of Michigan, Ann Arbor MI, USA.
        We thank Doug Miller and Zhuan Pei for helpful comments, as well as Tom Crossley, Noura Insolera and their team at PSID for help creating the new restricted PSID variable.
        The findings and conclusions in this publication are those of the authors and should not be construed to represent any official USDA or US Government determination or policy. 
    }
}
\date{\today}
\begin{document}
\maketitle
\thispagestyle{empty}
% Abstract
\begin{abstract}
    \noindent
    \input{sections/00-abstract.tex}
\end{abstract}

%%%%%%%%%%%%%%%%%%%%%%%%%%%%%%%%%%%%%%%%%
%% Starting the paper
\newpage
\setcounter{page}{1}
\onehalfspacing
% Introduction section
\noindent
\input{sections/01-introduction.tex}
% Data section + Conceptual Framework
\input{sections/02-data.tex}
% Empirical methods
\input{sections/03-empirical.tex}
% Empirical results
\input{sections/04-results.tex}
% Sensitivity and Robustness
\input{sections/05-sensitivity}
% Conclusion
\input{sections/06-conclusion.tex}

% Bibliography
\singlespacing
\bibliographystyle{agsm}
\bibliography{sections/07-bibliography.bib}
% Appendix
\newpage
\input{sections/08-appendix.tex}
\end{document}

%% file: sections/00-abstract.tex
Veterans face multiple hardships after they leave the military.
Is food insecurity one of these hardships?
This paper examines the long-term effects of military service on food insecurity using new data from the PSID.
Military veteran-headed households experienced lower food insecurity than non-veteran-headed households, and participated less in federal nutrition assistance programs.
We use the Vietnam-era lottery draft to identify causal effects, and find no evidence that military service adversely affects food insecurity.
Our results suggest policies that assist currently enlisted service members, and veterans unlikely to head a traditional household, may most effectively tackle food insecurity among military veterans.

\vspace{0.25cm}

\vspace{0.05cm}

%% file: sections/01-introduction.tex
%%%%%%%%%%%%%%%%%%%%%%%%%%%%%%%%%%%%%%%%%
%% Introduction section
%\section{Introduction}
%\label{sec:intro}
%Writing plan follows the introduction formula \url{https://blogs.ubc.ca/khead/research/research-advice/formula}.
%\subsection{Hook:}
Military service can have lasting consequences for the economic well-being of veterans and their families in the United States (US).
Physical and mental health challenges, disruptions to labor market participation, housing instability, and difficulties returning to civilian life may leave some veterans especially vulnerable to food insecurity \citep{cohen2022risk,widome2015food,betancourt2021exploring,fargo2012prevalence}.
This possibility has drawn increasing policy attention, particularly given evidence that active duty service members and their families experience higher rates of food insecurity than the civilian population \citep{rabbitt2024comparing,jennings2025scoping,hales2026simulating}.

This paper asks whether military service has a causal effect on food insecurity after veterans return to civilian life.
Food insecurity is both a consequence and a potential cause of broader socioeconomic disadvantage, with well-documented links to adverse health, educational, and labor market outcomes \citep{gundersen2011economics,gundersen2015food}.
We answer this question using longitudinal data from the Panel Study of Income Dynamics, and random variation in military service generated by the Vietnam-era draft lottery.

The effect of military service on later food insecurity is theoretically ambiguous. Military service may compound early life disadvantage, particularly because men raised in poverty are more likely to serve in the US military \citep{bareis2016relationship}. 
Conversely, service can increase educational attainment and civilian earnings, particularly among minorities, while providing access to benefits such as the GI Bill, Veterans Administration health care, and military pensions \citep{routon2014effect}. Observed differences in food insecurity may also reflect selection into service rather than the effects of service itself.
Veterans may also select into service on unobserved characteristics associated with better (or worse) socioeconomic outcomes, including resilience, physical fitness, and stronger labor market attachment.
The causal effect of military service on food insecurity therefore cannot be inferred from either economic theory or simple correlational comparisons between veterans and non-veterans.

Our empirical analysis has two parts. First, we use longitudinal data from the Panel Study of Income Dynamics (PSID) covering 1979--2019 to document differences in food insecurity between veteran- and non-veteran-headed households over four decades, a longer period of coverage than prior studies.
We show that veteran-headed households report lower levels of food insecurity on average, including lower probabilities of worrying that food will run out or cutting back on meals.
Using the extended PSID series of household-specific food insecurity measures developed by \cite{lee2024food} and \cite{lee2025probability}, we show that veteran-headed households consistently exhibit lower probabilities of food insecurity and lower participation in food assistance programs than non-veteran-headed households.

Second, we exploit random variation generated by the Vietnam-era draft lottery to estimate the causal effect of military service on later-life food insecurity.
Previous studies of food insecurity among veterans have primarily documented associations because they lack a credible source of exogenous variation in military service.
We find no evidence that Vietnam-era military service increased food insecurity among men who later became household heads.
The draft-lottery estimates are less adverse than the near-zero OLS estimates, consistent with men from lower socioeconomic backgrounds having been more likely to enter military service during this period.
We show that this conclusion is robust to using a more precisely estimated external first stage, weak-instrument-robust inference, and sensitivity analyses addressing the omission of veterans who do not become PSID household heads.

Our findings contrast with correlational evidence that active duty service members and their families experience higher rates of food insecurity than the civilian population \citep{rabbitt2024comparing,jennings2025scoping}.
They are consistent with evidence from the Current Population Survey Food Security Supplement (CPS-FSS) that unadjusted food insecurity prevalence is similar among veterans and non-veterans and, more specifically, that prevalence among Vietnam-era veterans is statistically similar to that among non-veterans \citep{rabbitt2021food}.\footnote{For more recent cohorts than those we study, however, \cite{rabbitt2021food} find that working-age veterans are more likely to experience food insecurity than observably similar non-veterans. Evidence that self-reported SNAP participation rises after veterans leave active duty also suggests that food-related hardship may be concentrated around the transition out of military service \citep{london2015supplemental}.
}
Our findings are also consistent with earlier evidence that Vietnam-era veteran-headed households do not face systematically higher rates of food insecurity when measured using standard categorical outcomes \citep{miller2016food}.
%Evidence of elevated food insecurity among active duty personnel, recently separated veterans, and more recent veteran cohorts therefore does not establish that military service causes a persistent increase in food insecurity among Vietnam-era veterans later in life.

Our results highlight an important distinction in the policy debate.
Military service may have lasting adverse effects for individuals who are not represented among the PSID household heads.
Among the veteran-headed households represented in the PSID, however, we find no evidence of elevated food insecurity and no evidence that Vietnam-era military service causally increased later-life food insecurity.
Policies addressing food insecurity in the military population may therefore need to distinguish between currently serving personnel, veterans at the point of separation, and veterans outside the traditional household structures captured by surveys of household heads, rather than presuming uniform disadvantage among veterans.

%\subsection{Value-added:}
This paper makes two main contributions to the study of veterans' food insecurity.
First, we extend the literature on the relationship between military service and food insecurity after separation by providing novel longitudinal evidence on food insecurity among veteran-headed households.
Previous research has relied primarily on static, cross-sectional measures that provide limited evidence on the persistence of food insecurity over time.
We move beyond this limitation using the estimated probability of food insecurity and the dynamic measures it enables, following \citet{lee2024food} and \citet{lee2025probability}.
We apply these measures to a longitudinal sample of US households covering four decades, allowing us to track food insecurity over time and estimate panel models that capture persistent differences between veteran- and non-veteran-headed households.
We supplement these measures with evidence on household food spending, providing an additional indication of whether veteran-headed households reliably meet their basic nutritional needs.
Together, these analyses show that the lower food insecurity observed among veteran-headed households is not limited to a particular cross-sectional sample or a single categorical measure of food insecurity.

Our second main contribution is to extend the Vietnam draft-lottery literature that uses draft eligibility as a source of random variation in Vietnam-era military service.
The canonical application is \cite{angrist1990lifetime}, who uses the lottery to estimate the effect of military service on income.
Subsequent work has used the same research design to study schooling, long-run earnings \citep{angrist2011schooling}, disability outcomes \citep{angrist2011long}, health and disability outcomes \citep{angrist2010sicker,wang2025effects}, crime \citep{lindo2014drawn}, demographics \citep{conley2012long}, intergenerational labor market effects \citep{goodman2020unfortunate}, and racial integration \citep{bleemer2026vietnam}.

We use restricted PSID variables that classify men exposed to the Vietnam-era draft into draft number categories, allowing us to construct interval measures of draft risk within the PSID.
This construction is not possible using the public PSID files because exact birth dates are not otherwise available.
The newly created variable allows us to use draft-induced variation in military service within a longitudinal household panel containing repeated measures of economic resources, family structure, and food insecurity.
These outcomes are generally not observed in the administrative and Census-based datasets used in much of the previous draft-lottery literature.
We extend the outcomes studied using the Vietnam-era draft lottery to include food insecurity later in life.

%\subsection{Road-map:}
This paper proceeds as follows.
\autoref{sec:data} describes the PSID data, the probability of food insecurity measure and the dynamic measures we study.
\autoref{sec:empirics} lays out the empirical framework for using the Vietnam-era draft lottery to infer the causal effects of military service on food insecurity.
\autoref{sec:results} reports trends in different measures of food insecurity for veterans and non-veterans and presents the estimates of causal effects.
\autoref{sec:sensitivity} examines the robustness of the causal results to uncertainty in the PSID first-stage, weak identification, and the non-random observation of military veterans in data on PSID households.
\autoref{sec:conclusion} concludes.

%% file: sections/02-data.tex
%%%%%%%%%%%%%%%%%%%%%%%%%%%%%%%%%%%%%%%%%
%% Data section
\section{Data and Methods}
\label{sec:data}

\subsection{Panel Study of Income Dynamics}
The PSID is a longitudinal survey of individuals, their families, and their descendants (annual 1968--1997, biennial since), tracking 85,536 individuals over 55 years, from 1968 to 2023 \citep{PSID}.
The longest-running longitudinal household survey in the world, PSID collects detailed information on household composition and socioeconomic status, including income, expenditures, and employment.
We use the 26 survey waves of PSID data from 1979--2019, limiting our sample to males born between 1930--1970 because we later instrument for military service using the Vietnam-era draft, which only conscripted males.
The PSID includes a question on past military service to the individuals described as a ``reference person,'' a term we use interchangeably with ``household head.''
An important implication of this for our analysis is that our definition of a military veteran is someone who became a household head after returning from active-duty service.
We return to the potential for selection associated with this definition later in the paper.
Part of our analysis focuses on the 1950--1952 cohort; these are veterans who returned from Vietnam War service and eventually became reference persons.
For males born between 1930 and 1970, we have data on 8,392 individuals, comprising 2,466 veterans and 5,926 non-veterans.
For the 1950--1952 cohort, we have 816 observations, of which 29\% were draft eligible and 31\% served as veterans.

\subsection{Outcome Measure: Probability of Food Insecurity}
\label{sec:pfi}
Our measure of food insecurity needs to satisfy several criteria: (a) it needs to be available for the time period covered in this study; (b) it tracks the prevalence of food insecurity in the US; and (c) it allows us to assess both the incidence of food insecurity and the extent to which it is chronic versus transitory.
One measure that satisfies these criteria is a simple transform of the Probability of Food Security (PFS), recently introduced by \citet{lee2024food,lee2025probability}.
The PFS estimates the conditional probability that a household's food expenditures equal or exceed the cost of the USDA Thrifty Food Plan (TFP), the US government's estimate of the minimum food spending required to sustain a healthy diet for a household of its size and location in a given year.
Benefits under the Supplemental Nutrition Assistance Program (SNAP), the federal government's main program for reducing food insecurity, are tied directly to TFP estimates.
Food-expenditure data are collected in all waves of the PSID, and PFS is calibrated to match the official, CPS-based food insecurity prevalence estimates.

We use a simple transform of PFS, the Probability of Food Insecurity, $\text{PFI}=1-\text{PFS}$, the probability that a household's food expenditures are less than the corresponding TFP value in a given year.
We classify individual $i$ in household $h(i)$ and year $t$ as living in a food insecure or food secure household by comparing the household's food spending $FS_{h(i),t}$ with the relevant TFP cost, $\underline{P}_{h(i),t}$.
\begin{equation}
    \label{eqn:pfi-definition}
    \text{PFI}_{i,t}
    =
    \Probgiven{
        FS_{h(i),t}<\underline{P}_{h(i),t}
    }{
        \vec W_{h(i),t}, FS_{h(i),t-1}
    },
\end{equation}
where $\vec W_{h(i),t}$ is a vector of characteristics: the household head's age, gender, and educational status, household size, and household income.
This measure compares observed household food spending against the USDA measure for the minimum amount of food spending required for a household of its size and location to sustain an healthy diet.
We use estimates of this conditional probability in the PSID data; more detailed discussion of PFI calculation can be found in \appendixref{appendix:hfsm}, and \cite{lee2025probability}.

We use $\text{PFI}_{i,t}$ to classify an household as food insecure by comparing the PFI against USDA annual population estimates of food insecurity, $P_t$.
Specifically, we calculate $\text{FI}_{i,t} = \indicator{P_t < \text{PFI}_{i,t}}$, where $\indicator{.}$ is the indicator function and $P_t$ is the probability that calibrates the share of PFI food-insecure households to the annual USDA nationwide food insecurity prevalence \citep{rabbitt2025household}.\footnote{
    Before USDA began reporting food insecurity prevalence estimates in 1995, we use the thresholds that \citet{lee2025probability} estimated from a regression model of post-1995 PFS thresholds on national SNAP participation rates (for $P_t$ with $t < 1995$).
}

The continuous PFI and binary FI classification measures are year-specific estimates.
\cite{lee2024food} also introduce measures that summarize the persistence and transience of food insecurity within households over time.
Using these measures, we categorize men according to their long-run food insecurity dynamics: (1) chronic and persistently food insecure, (2) chronic and not persistent, (3) transiently food insecure, or (4) persistently food secure.
\appendixref{appendix:chronic} provides further details on the construction and interpretation of these dynamic food insecurity measures.

\subsection{Descriptive Statistics}

\begin{table}[h!]
    \singlespacing
    \centering
    \caption{Descriptive Statistics, PSID Households Headed by Veterans and Non-Veterans.}
    \small
    \makebox[\textwidth][c]{
        \begin{tabular}{l c c c c c c c c}
            \\[-1.8ex]\hline \hline \\[-1.8ex] 
            & \multicolumn{3}{c}{Men born 1930--1970} & \multicolumn{3}{c}{Men born 1950--1952} \\
            \cmidrule(lr){2-4} \cmidrule(lr){5-7}
            & All & Veteran & Non-veteran & All & Veteran & Non-veteran \\
            \\[-1.8ex]\hline \\[-1.8ex] 
            \multicolumn{2}{l}{\textbf{Panel A. Military Service}} \\
            \input{sections/tables/psid-summary-panelA.tex}
            \\[-1.8ex]\hline \\[-1.8ex] 
            \multicolumn{2}{l}{\textbf{Panel B. Food Insecurity}} \\
            \input{sections/tables/psid-summary-panelB.tex}
            %\\[-1.8ex]\hline \\[-1.8ex] 
            %\textbf{Panel C. Welfare Participation} \\
            \input{sections/tables/psid-summary-panelC.tex}

            \\[-1.8ex]\hline \\[-1.8ex] 
            \multicolumn{2}{l}{\textbf{Panel C. Demographic and Labor}} \\
            \input{sections/tables/psid-summary-panelD.tex}
            \\[-1.8ex]\hline \\[-1.8ex] 
        \end{tabular}
    }
    \vspace{-0.25cm}
    \label{tab:psid-summary}
    \justify
    \footnotesize
    \textbf{Note:}
    Observation counts are the number of men with non-missing data for each variable category;
    outcomes that vary year-on-year (in the PSID panel, such as income) are averaged across individual--years, and we report the mean of this value. 
    All \$ variables are adjusted for 2023 CPI-U, mean across non-missing survey years conditional on non-zero values; household food spending is measured in survey reported monthly spending per household member, income figures are annual.
    RSN refers to Random Sequence Number for the Vietnam-era draft lottery, further explained in \autoref{sec:draft-iv}.
\end{table}

\noindent
We first show descriptive statistics for PSID household heads, for men subject to the Vietnam draft as well as men born over a longer time period; \autoref{tab:psid-summary} shows these summary statistics.
Among men born between 1930 and 1970, the average age of a veteran-headed household is 44.4 years, five years higher than that of a non-veteran-headed household.
But that is the only noteworthy difference.
Veteran- and non-veteran-headed households are similar in years of schooling (13.3), the likelihood of being employed (84\%), household size (three persons), and household real income per capita (\$44,015 versus \$45,741 in 2023 US dollars).
However, marked differences exist between veteran and non-veteran households in the subsample of men born between 1950 and 1952.
Non-veterans are more likely to be employed (87\% versus 81\%), have higher levels of individual income (\$89,832 versus \$65,873 for veterans), and live in households with higher levels of per-capita income (\$51,938 versus \$40,264 for veterans).

Panel A reports descriptive information on the timing and duration of military service.
Service length was 4.6 years among veterans born between 1930 and 1970 and 4.8 years among veterans in the subsample born between 1950 and 1952.
In the latter sample, the share of men in Category 1, indicating a high risk of being drafted, was 0.26 among veterans, compared with 0.20 among non-veterans.

Panel B reports the sample means for our food insecurity outcomes, including SNAP participation.
Among men born between 1930 and 1970, there is no indication that veteran-headed households experience worse food insecurity, as measured by mean PFI, the percentage of years food secure, and the number of years food insecure.
If anything, the mean values of these outcomes among veteran-headed households---0.15, 95\%, and 0.8 years, respectively---are slightly better than the corresponding values among non-veteran-headed households---0.17, 91\%, and 1.2 years, respectively.
Veteran-headed households also have slightly lower rates of SNAP participation than non-veteran-headed households.

Panel C reports demographic and labor-market characteristics.
Among men born between 1930 and 1970, veteran- and non-veteran-headed households are similar in years of schooling, employment, household size, and real household income per capita, although veterans are older on average.
Larger differences appear within the 1950--1952 birth cohorts, where veterans have lower employment, individual income, and household income per capita than non-veterans.

%% file: sections/tables/psid-summary-panelA.tex
% latex table generated in R 4.4.1 by xtable 1.8-8 package
% Wed Jul 22 16:56:19 2026
 Veteran status        &     0.3 &     1 &   0 &     0.31 &     1 & 0 \\ 
  Service start, year   & 1963.0 & 1963 &   & 1971 & 1971 &    \\ 
  Service end, year     & 1968.0 & 1968 &   & 1975 & 1975 &    \\ 
  Service length, years &     0.9 &     4.57 &   0 &     0.94 &     4.80 & 0 \\ 
  Draft eligibility     &       &        &   &     0.29 &     0.35 & 0.27 \\ 
  RSN 1--95             &       &        &   &     0.22 &     0.26 & 0.20 \\ 
  RSN 96--195           &       &        &   &     0.22 &     0.23 & 0.22 \\ 
  RSN 196--366          &       &        &   &     0.36 &     0.29 & 0.39 \\ 
  

%% file: sections/tables/psid-summary-panelB.tex
% latex table generated in R 4.4.1 by xtable 1.8-8 package
% Wed Jul 22 16:56:21 2026
 PFI                                         &   0.16 &   0.15 &   0.17 &   0.15 &   0.16 &   0.15 \\ 
  Years observed in PSID                      &  16.41 &  16.01 &  16.58 &  17.89 &  17.17 &  18.19 \\ 
  Years food secure                           &  15.34 &  15.25 &  15.37 &  17.02 &  16.36 &  17.29 \\ 
  Percent of years food secure                &   0.92 &   0.95 &   0.91 &   0.94 &   0.94 &   0.94 \\ 
  Years food insecure                         &   1.08 &   0.76 &   1.22 &   0.88 &   0.82 &   0.90 \\ 
  Percent of years food insecure              &   0.08 &   0.05 &   0.09 &   0.06 &   0.06 &   0.06 \\ 
  Household food spending, per person & 390 & 391 & 389 & 385 & 364 & 394 \\ 
  SNAP, yearly participation & 0.05 & 0.04 & 0.06 & 0.05 & 0.06 & 0.04 \\ 
SNAP, ever participated    & 0.22 & 0.21 & 0.22 & 0.20 & 0.28 & 0.17 \\ 
  

%% file: sections/tables/psid-summary-panelC.tex
% latex table generated in R 4.4.1 by xtable 1.8-8 package
% Wed Jul 22 16:56:25 2026

%% file: sections/tables/psid-summary-panelD.tex
% latex table generated in R 4.4.1 by xtable 1.8-8 package
% Wed Jul 22 16:56:30 2026
 Age                          &      41.01 &      44.49 &      39.49 &      42.67 &      42.01 &      42.95 \\ 
  Race $=$ white               &       0.86 &       0.88 &       0.84 &       0.86 &       0.84 &       0.87 \\ 
  Education years              &      13.31 &      13.33 &      13.29 &      13.65 &      13.18 &      13.85 \\ 
  Employed                     &       0.84 &       0.84 &       0.84 &       0.85 &       0.81 &       0.87 \\ 
  Individual income            &  73,690 &  77,793 &  71,888 &  82,833 &  65,873 &  89,832 \\ 
  Household income             & 121,187 & 118,317 & 122,433 & 127,407 & 107,532 & 135,608 \\ 
  Household income, per person  &  45,217 &  44,015 &  45,741 &  48,478 &  40,264 &  51938 \\ 
  Household size               &       3.06 &       3.03 &       3.06 &       3.03 &       2.95 &       3.06 \\ 
  \hline \\ Observations    &   8,392 &   2,466 &   5,926 &     816 &     274 &     542 \\ 
  

%% file: sections/03-empirical.tex
%%%%%%%%%%%%%%%%%%%%%%%%%%%%%%%%%%%%%%%%%
%% Empirical section
\section{Causal Effects of Military Service}
\label{sec:empirics}
Correlational food insecurity differences between veteran and non-veteran headed households cannot be interpreted as causal effects of military service, as selection into military service is not random.

Men may enlist voluntarily, intentionally avoid service, or qualify for military-service exemptions for reasons correlated with later-life food insecurity.
For example, men from lower-socioeconomic-status backgrounds may be more likely to enter the military as a path to stable employment, training, or escape from constrained local opportunities.
Those same backgrounds may also increase subsequent food insecurity risk.
Other forms of selection may work in the opposite direction, since military service also requires good physical health, screening, and attachment to formal institutions.
Veteran-headed households appearing less food insecure than non-veteran-headed households is therefore not, by itself, evidence that military service reduces food insecurity.
The observed difference reflects the unknown causal effect of military service combined with pre-existing differences between veterans and non-veterans, compounded by selection into the set of veterans who later survive and head households.

To separate the causal effect of military service from these competing selection channels, we go beyond correlational analysis and exploit random variation in military service generated by the Vietnam-era draft lottery.
This design allows us to estimate the causal effect of military service for the relevant cohort of US military veterans.

\subsection{The Vietnam-Era Draft Lottery}
\label{sec:draft-iv}
The US military operated a randomized draft to conscript young men for the Vietnam War effort.
The draft lotteries were conducted annually from 1970 through 1972 for American men born between 1944 and 1952.\footnote{
    Draft lotteries were also conducted from 1973 through 1975 for men born between 1953 and 1956, although none were conscripted because the draft and the Vietnam War were ending.
    We therefore do not include these men in the IV analysis.
}

Based on their exact birth dates, men were assigned Random Sequence Numbers (RSNs) according to the order in which lottery balls were drawn.
For example, the first lottery ball drawn in the 1970 draft corresponded to 14 September, so men born on 14 September in the relevant birth cohorts were assigned RSN 1.
These men were the first birthdate cohort subject to conscription as the Department of Defense enlisted additional service members.
In the 1971 draft, the second lottery ball drawn corresponded to 24 December, so men born on 24 December 1951 were assigned RSN 2.
In the 1972 draft, the third lottery ball drawn corresponded to 15 December, so men born on 15 December 1952 were assigned RSN 3.
Corresponding lottery numbers were assigned to every birth date in the relevant birth cohorts, and men were conscripted into the US military in the order determined by the randomly drawn RSNs.

Men born between 1944 and 1950 with an RSN $\leq 195$ were subject to conscription unless they held a military exemption.
Men born in 1951 were subject to conscription if their RSN was $\leq 125$, while men born in 1952 were subject to conscription if their RSN was $\leq 95$.
These draft limits were determined only at the end of their respective years, based on Department of Defense anticipated personnel needs.
It was therefore not apparent at the beginning of a year whether intermediate RSNs---those between 95 and 195---would result in conscription.

It was clear, however, that men with extremely low RSNs faced a high probability of being conscripted.
This often induced voluntary enlistment into preferred military occupational specialties and service branches to preempt conscription into specialties and units likely to face combat, such as service as a US Army infantry rifleman.
This volunteering response, together with direct conscription based on the random lottery draw, forms part of the causal effect of the Vietnam-era draft lottery.
The randomly assigned RSNs induced military service through both involuntary conscription and voluntary enlistment.

To instrument for military service, we use a newly created restricted-access PSID variable.
At our request, the PSID constructed a variable reporting the RSN category of men born between 1944 and 1952 using their exact birth dates; this variable is available only through the restricted-access PSID enclave.\footnote{
    Some of the data used in this analysis are derived from restricted PSID data files obtained under special contractual arrangements designed to protect the anonymity of respondents.
    These data are not available from the authors.
    Persons interested in obtaining restricted PSID data files should contact \href{mailto:PSIDHelp@umich.edu}{PSIDHelp@umich.edu}.
}
The restricted variable divides RSNs into four categories.
Men with RSNs from 1 through 95 are assigned to Category 1 and faced a high risk of being drafted.
Men with RSNs from 96 through 195 are assigned to Category 2 and faced greater uncertainty about whether they would be drafted.
Men with RSNs of 196 or above are assigned to Category 3 and were unlikely to be drafted.
Men whose exact birth dates are unavailable are assigned to Category 4, and have unknown RSNs.

From these RSN categories and the cohort-specific conscription limits, we construct a binary variable indicating whether a male household head born between 1950 and 1952 was subject to conscription.\footnote{
    Men born in 1951 faced an RSN conscription cutoff of 125, but the restricted PSID variable identifies only the broader RSN category from 96 through 195.
    This category therefore includes men on both sides of the 1951 cutoff.
    We code men born in 1951 with RSNs from 96 through 195 as not subject to conscription, $Z_i=0$, because 70\% of the birth dates in this category had RSNs above the conscription cutoff.
    The results are substantively unchanged if we instead code these men as conscripted, or treat their conscription status as missing and exclude them from the analysis.
}

\begin{figure}[htbp!]
    \caption{Vietnam-Era RSNs, Conscription, Military Service.}
    \begin{subfigure}[c]{0.475\textwidth}
        \centering
        %\caption{Conscription definition}
        %\textbf{SUMMARY STAT OUTPUT GOES HERE.}
        \includegraphics[width=\textwidth]{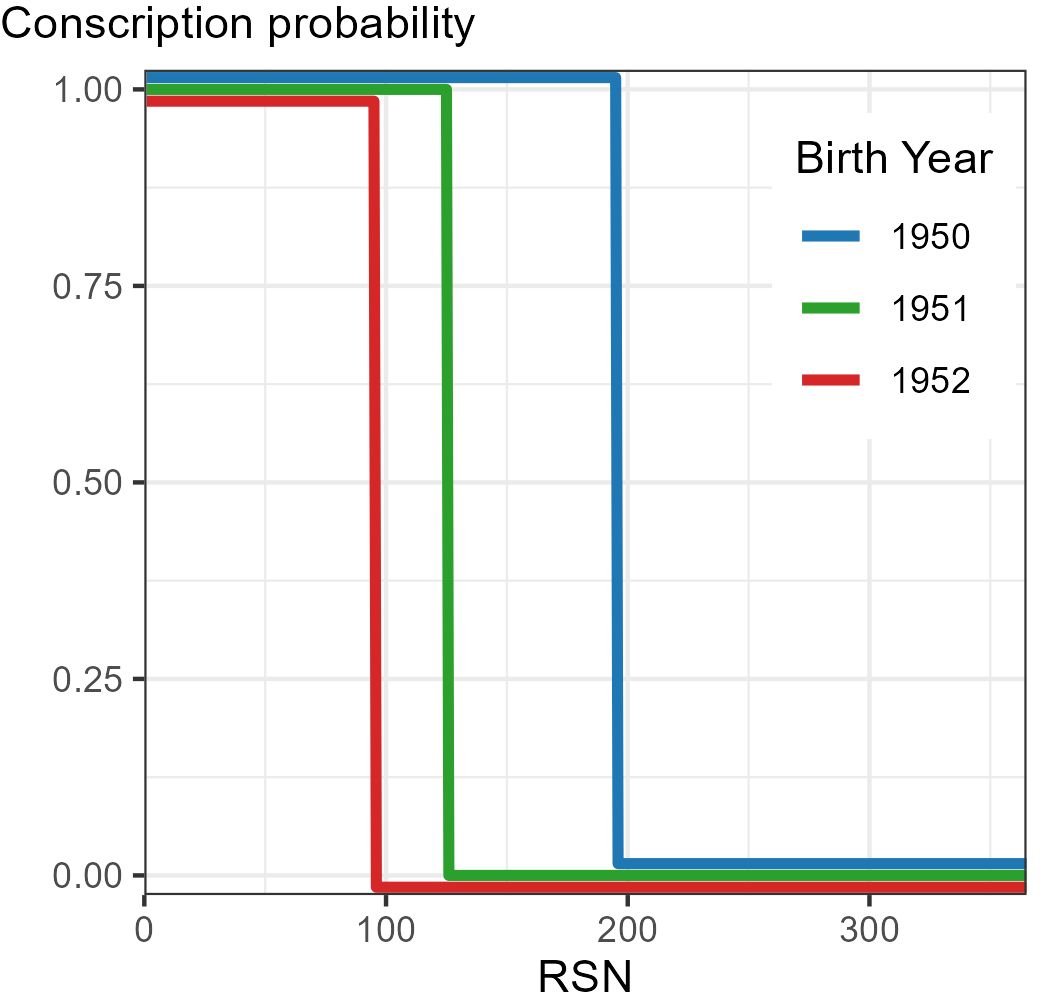}
    \end{subfigure}
    \begin{subfigure}[c]{0.475\textwidth}
        \centering
        %\caption{Military service relative to conscription}
        %\textbf{SUMMARY STAT OUTPUT GOES HERE.}
        \includegraphics[width=\textwidth]{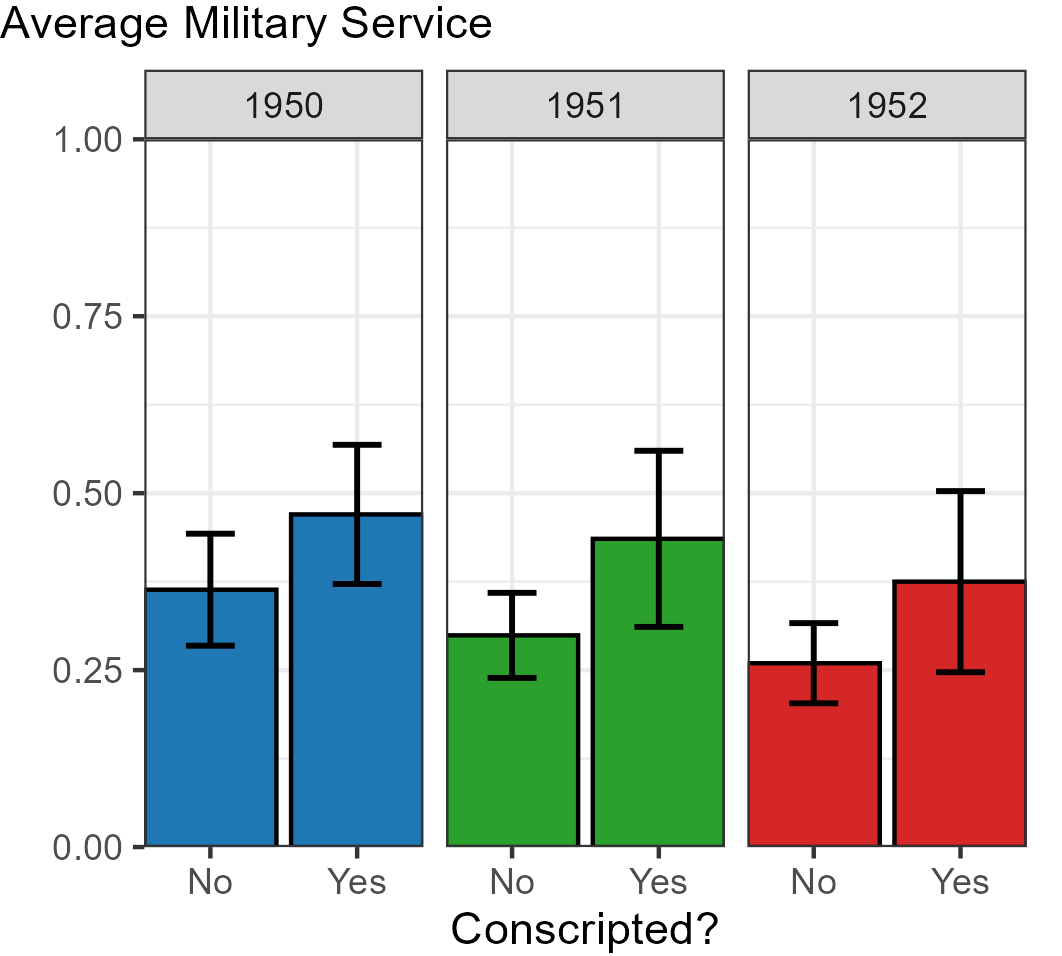}
    \end{subfigure}
    \label{fig:draft-conscripted}
    \justify
    \footnotesize    
    \textbf{Note:}
    These figures show the definition of conscription relative to RSN, and then actual rate of military service among the men born 1950--1952, who go on to head a PSID household.
    Panel A shows the definition of $\Probgiven{\text{Conscripted}_i}{\text{RSN$_i$, birth year$_i$}}$, and B shows estimates of $\Probgiven{\text{Military veteran}_i}{\text{Conscripted$_i$, birth year$_i$}}$ calculated among PSID household head men.
\end{figure}

Let $\text{RSN}_i$ denote man $i$'s Random Sequence Number, $\text{birth year}_i$ denote his year of birth, and $\indicator{.}$ refer to the indicator function.
We define the binary conscription instrument as
\[ Z_i \coloneq \begin{cases}
    \indicator{\text{RSN}_i \leq 195},
        & \text{ if birth year}_i = 1950 \\
    \indicator{\text{RSN}_i \leq 125},
        & \text{ if birth year}_i = 1951 \\
    \indicator{\text{RSN}_i \leq 95},
    & \text{ if birth year}_i = 1952 \\
\end{cases} \]
\autoref{fig:draft-conscripted} illustrates the relationship between RSNs, conscription, and military service among PSID male household heads born between 1950 and 1952.
Panel A shows how conscription status varies with RSN and birth year.
Panel B shows rates of military service by conscription status and birth year in the PSID data.

\autoref{eqn:firststage-draft} gives the first-stage specification, using the binary conscription variable, $Z_i$, as an instrument for whether the male household head served in the military, $D_i$.
\begin{equation}
    \label{eqn:firststage-draft}
    D_i = \pi Z_i + \vec\varphi' \vec X_i + \eta_i,
\end{equation}
where $\vec X_i$ contains controls for birth year.\footnote{
    The physical randomization procedure used in the 1970 lottery may have produced non-random assignment across different lotteries \citep{fienberg1971randomization}.
    We therefore control for birth year, following \cite{berinsky2015empirical}.
}
The coefficient $\pi$ is the average effect of conscription on military service.

\begin{table}[h!]
    \small
    \singlespacing
    \centering
    \caption{Draft-Lottery First-Stage Estimates \\ PSID Household Head Men by Birth Year.}
    \makebox[\textwidth][c]{
        \begin{tabular}{l c c c c c}
            \\[-1.8ex]\hline \hline \\[-1.8ex] 
            & \multicolumn{2}{c}{Pooled cohorts} & \multicolumn{3}{c}{By birth year} \\
            \cmidrule(lr){2-3} \cmidrule(lr){4-6}
            & 1944--1949 & 1950--1952 & 1950 & 1951 & 1952 \\
            \\[-1.8ex]\hline \\[-1.8ex] 
            \textbf{Panel A: Conscription effects:} \\
            \input{sections/tables/firststage-binary-main.tex} \\
            \midrule
            \textbf{Panel B: RSN group effects:} \\
            \input{sections/tables/firststage-categorical-main.tex}
            \\[-1.8ex]\hline \\[-1.8ex]
        \end{tabular}
    }
    \label{tab:firststage}
    \vspace{-0.25cm}
    \justify
    \footnotesize
    \textbf{Note}: This table reports binary draft-eligibility effects and categorical RSN effects on veteran status, estimated in separate regressions for each column.
    Panel A refers to binary draft effect specified in \autoref{eqn:firststage-draft}, and Panel B to the categorical IV specification.
    Each cross-sectional observation is a separate man in the PSID, as the outcome does not vary in different years of the PSID panel.
    The pooled cohort models control for birth year as fixed effects, so estimates are relative to an omitted group of fixed effects.
    The birth year categorical rows show all the fixed effects estimates, so are raw military service rate (not relative to an omitted group); RSN 1--95 is omitted for the 1944--49 cohort because of collinearity with birth year controls.
    Standard Errors (SEs) are reported in parentheses, which are clustered on birth year.
\end{table}

We begin by estimating \autoref{eqn:firststage-draft} to assess whether draft-induced conscription affected military service among men born between 1950 and 1952.
\autoref{tab:firststage} reports the first-stage estimates.
A man born between 1950 and 1952 who was subject to conscription because of a low RSN was, on average, 12.3 percentage points more likely to serve in the military than a man who was not subject to conscription.
This estimate is statistically significant at the one-percent level.
Despite the relatively small PSID sample, the binary instrument produces a first-stage F statistic of 10.

The three birth-year-specific estimates show that the pooled first-stage result is not driven by a single birth cohort.
The estimated effects are necessarily less precise within the smaller birth-year-specific samples, but draft-induced conscription increases military service in each of the three cohorts.
These results are consistent with previous research using the Vietnam-era draft lottery \citep{angrist1990lifetime,angrist2011schooling}.\footnote{
    \cite{angrist1990lifetime} and \cite{angrist2011long} estimate first-stage effects separately for white and non-white men to account for racial differences in compliance with the draft.
    We find little evidence of differences in the first-stage relationship by race among PSID men born between 1950 and 1952, as shown in Appendix~\autoref{tab:firststage-race}.
    We therefore do not conduct the main analysis separately by race.
}
Military service among men born between 1944 and 1949 did not respond strongly to conscription by RSN in the 1970 draft.
This is unsurprising because these older cohorts had already served in the military at higher rates than men born between 1950 and 1952.
We therefore restrict the IV analysis to men born between 1950 and 1952.

The binary conscription instrument is our preferred specification.
It applies the known cohort-specific draft rules and produces a first-stage F statistic of 10 among men born between 1950 and 1952.
By contrast, the more flexible categorical specification estimates separate relationships between RSN categories and military service within each birth year.
Although the categorical specification uses the same underlying draft-lottery variation, its joint first-stage F statistic is substantially smaller, indicating that the additional instruments weakly predict military service.

Using PFI as the outcome, the Hansen test does not reject the over-identifying restrictions imposed by the categorical specification.
This result provides no evidence that the categorical instruments are mutually inconsistent with the model, but it does not overcome the weak first-stage or establish that the additional instruments improve identification.
We therefore use the binary conscription instrument as the sole instrument in the causal analysis.
We retain the categorical first-stage estimates in \autoref{tab:firststage} to document the relationship between RSN categories and military service, but do not use the categorical specification to estimate the effects of military service.

\subsection{Effects on Food Insecurity}
We estimate the effects of military service on food insecurity outcomes using the following model
\begin{equation}
    \label{eqn:secondstage}
    Y_{i,t}
    =
    \beta D_i
    +
    \vec\gamma' \vec X_{i,t}
    +
    \varepsilon_{i,t}.
\end{equation}
The variable $D_i$ is a binary indicator equal to one if man $i$ served in the military.
The outcome $Y_{i,t}$ denotes one of the food insecurity or related outcomes defined above for man $i$ in year $t$.
The vector $\vec X_{i,t}$ includes fixed effects for outcome year, and birth year.
The coefficient $\beta$ is the causal effect of military service identified by the Vietnam-era draft-lottery design.

In the correlational analysis, we estimate \autoref{eqn:secondstage} by OLS using the full sample of men born between 1930 and 1970.
The OLS estimates describe conditional differences in outcomes between veteran- and non-veteran-headed households, but they do not identify the causal effect of military service.
In the causal analysis, we restrict the sample to men born between 1950 and 1952 and instrument for military service, $D_i$, using the binary conscription instrument, $Z_i$.
Thus, \autoref{eqn:firststage-draft} is the first-stage specification and \autoref{eqn:secondstage} is the second-stage specification.

The Instrumental Variable (IV) design requires three identifying assumptions.
First, the randomly assigned RSNs must be independent of men's potential food insecurity outcomes, conditional on birth year.
Second, conscription must affect the outcomes only through its effect on military service.
Third, draft eligibility must not make any man less likely to serve in the military.
Under these independence, exclusion, and monotonicity assumptions, $\beta$ identifies the local average treatment effect of military service among men whose military service was induced by the Vietnam-era draft lottery \citep{angrist1996identification}.

The independence assumption is credible because RSNs were assigned by lottery, with controls for birth year addressing concerns about the physical randomization of the 1970 lottery.
The first-stage estimates in \autoref{tab:firststage} show that the binary conscription instrument is relevant for men born between 1950 and 1952.
The exclusion restriction requires conscription to affect later-life food insecurity only through military service.\footnote{
    \cite{angrist2011schooling} identify education-related responses to the draft as a potentially important concern for the exclusion restriction because men could pursue education to defer or avoid conscription.
    They find that the observed educational differences instead reflect subsidized education following military service, including benefits provided through the GI Bill.
    Under that interpretation, education is a consequence of military service rather than a separate pathway through which draft eligibility affects later outcomes.
}
Monotonicity permits both direct conscription and voluntary enlistment in anticipation of conscription, but rules out men who would serve when not subject to conscription and would not serve when subject to conscription.
We do not use PSID survey weights in the IV analysis.\footnote{
    The draft-lottery design identifies a local average treatment effect for draft compliers among men born between 1950 and 1952 who are observed as PSID household heads.
    Neither this birth cohort nor the draft compliers within it are intended to represent the full US population.
    Reweighting the sample on observed characteristics would not transform this complier-specific estimand into a population average treatment effect.
}
The causal estimates should therefore be interpreted as effects for draft compliers in the 1950--1952 birth cohorts who later become PSID household heads, rather than as population average effects for all veterans.

We conduct three sensitivity exercises to assess the robustness of the core IV results.
First, we assess whether the relatively weak first stage in the PSID affects either the IV point estimates or statistical inference.
We implement a two-sample IV estimator that combines reduced-form estimates from the PSID with more precise external estimates of the effect of draft eligibility on military service from the 2000 Census.
This exercise assesses whether sampling variation in the first-stage estimate from the relatively small PSID draft cohort materially affects the IV point estimates. We complement this analysis with Anderson-Rubin weak-instrument-robust confidence intervals \citep{anderson1949estimation}, which remain valid when the first-stage is weak.
Together, these exercises assess whether the core conclusions are sensitive to either first-stage estimation noise or the potentially poor finite-sample performance of conventional IV inference in the PSID.

Second, we conduct a sensitivity analysis for veterans who are not observed as PSID household heads. This exercise asks how large the unobserved effects among non-head veterans would need to be to produce a substantively adverse population average effect, given the estimated effect among PSID household heads and external estimates of the share of veterans who become household heads.
Third, we conduct a model-informed sensitivity analysis that separately accounts for non-random survival and household headship.
This exercise combines external information on survival and household headship with observed food security differences between PSID household heads and men who never become household heads to assess how selection into the observed sample could change the estimated population effect.

Together, these exercises assess whether the main conclusion is robust to uncertainty in the PSID first-stage, the absence of veterans outside traditional household structures from the analytic sample, and explicit assumptions about the relationships among military service, survival, household headship, and food insecurity.

%% file: sections/tables/firststage-binary-main.tex
% latex table generated in R 4.4.1 by xtable 1.8-8 package
% Wed Jul 22 17:41:00 2026
 Draft eligibility effect & 0.036 & 0.123 & 0.122 & 0.155 & 0.094 \\ 
    & (0.031) & (0.039) & (0.072) & (0.069) & (0.071) \\ 
  F statistic & 1.4 & 10 & 2.9 & 5 & 1.8 \\ 
  Observations & 1,099 & 816 & 243 & 286 & 287 \\ 
  

%% file: sections/tables/firststage-categorical-main.tex
% latex table generated in R 4.4.1 by xtable 1.8-8 package
% Wed Jul 22 16:57:44 2026
 RSN 1--95 &   & 0.416 & 0.436 & 0.435 & 0.375 \\ 
    &   & (0.036) & (0.066) & (0.059) & (0.06) \\ 
  RSN 96--195 & -0.03 & 0.335 & 0.511 & 0.288 & 0.241 \\ 
    & (0.046) & (0.037) & (0.073) & (0.06) & (0.061) \\ 
  RSN 196--366 & -0.02 & 0.279 & 0.289 & 0.223 & 0.321 \\ 
    & (0.043) & (0.028) & (0.053) & (0.048) & (0.043) \\ 
  RSN Unknown & -0.109 & 0.347 & 0.467 & 0.408 & 0.183 \\ 
    & (0.049) & (0.033) & (0.063) & (0.055) & (0.053) \\ 
  F statistic & 1.931 & 2.854 & 2.435 & 3.61 & 1.787 \\ 
  Sargan--Hansen p. value & 0.704 & 0.299 & 0.102 & 0.64 & 0.733 \\ 
  Observations & 1,099 & 816 & 243 & 286 & 287 \\ 
  

%% file: sections/04-results.tex
%%%%%%%%%%%%%%%%%%%%%%%%%%%%%%%%%%%%%%%%%
%% Results section
\section{Empirical Results}
\label{sec:results}

\subsection{Trends in Food Insecurity}
We begin by considering descriptive trends over time.
The static food insecurity measures show that veteran-headed households are not more food insecure than non-veteran-headed households.

\begin{figure}[h!]
    \centering
    \singlespacing
    \caption{Trends in Food Insecurity Outcomes among Veterans and Non-veteran Households.}
    \begin{subfigure}[b]{0.495\textwidth}
        \centering
        %\caption{Probability of food insecurity.}
        \includegraphics[width=\textwidth]{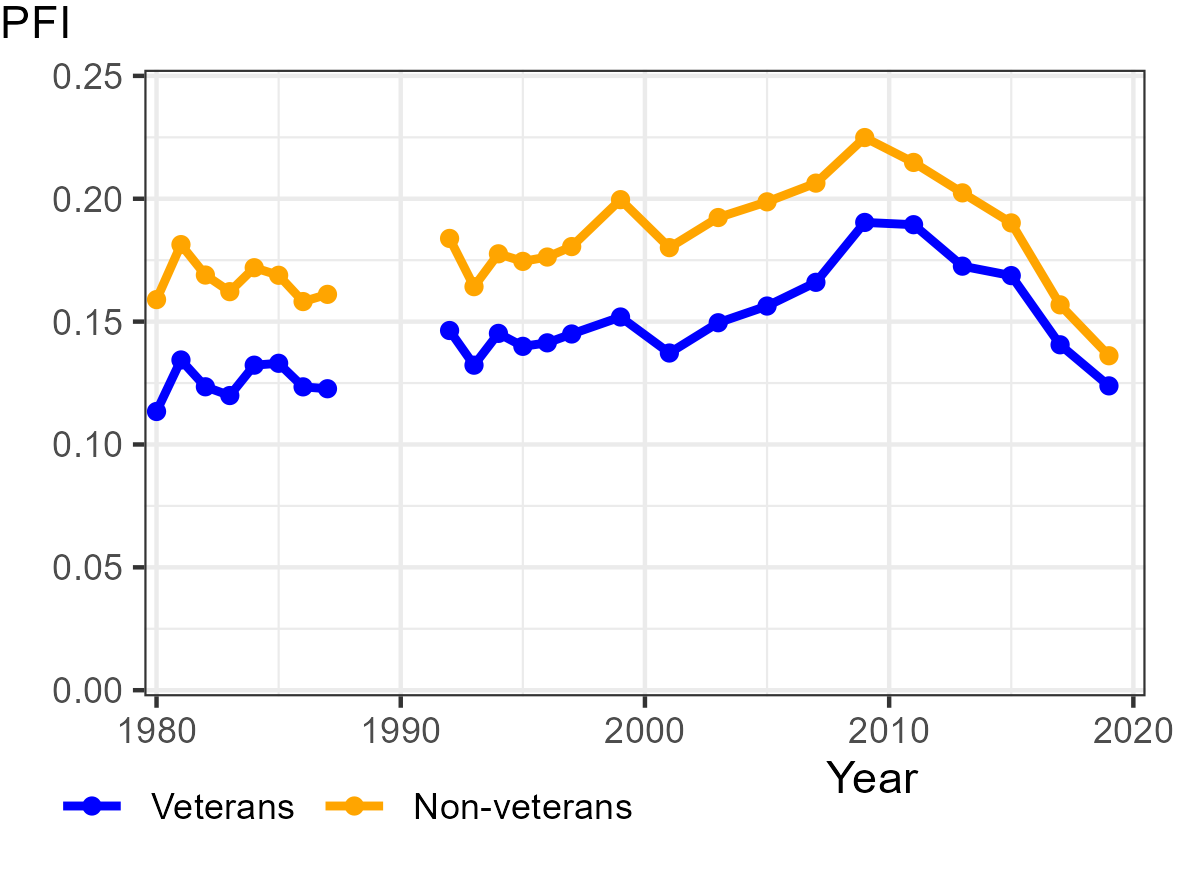}
        \label{fig:fam-pfs}
    \end{subfigure}
    \begin{subfigure}[b]{0.495\textwidth}
        \centering
        %\caption{Food insecurity rate, by PFS.}
        \includegraphics[width=\textwidth]{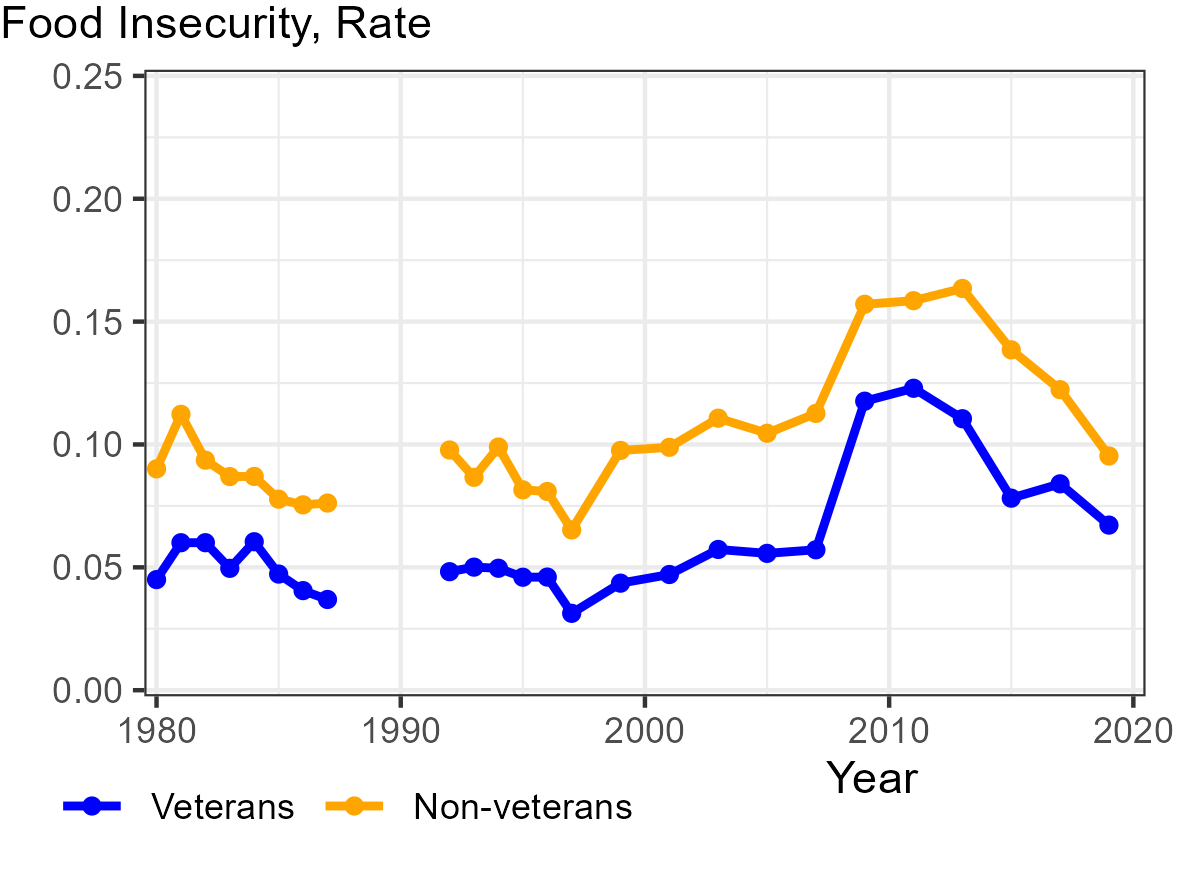}
        \label{fig:fam-pfs-insecure}
    \end{subfigure}
    \begin{subfigure}[b]{0.495\textwidth}
        \centering
        %\caption{Food spending, \$ 2023 CPI-U.}
        \includegraphics[width=\textwidth]{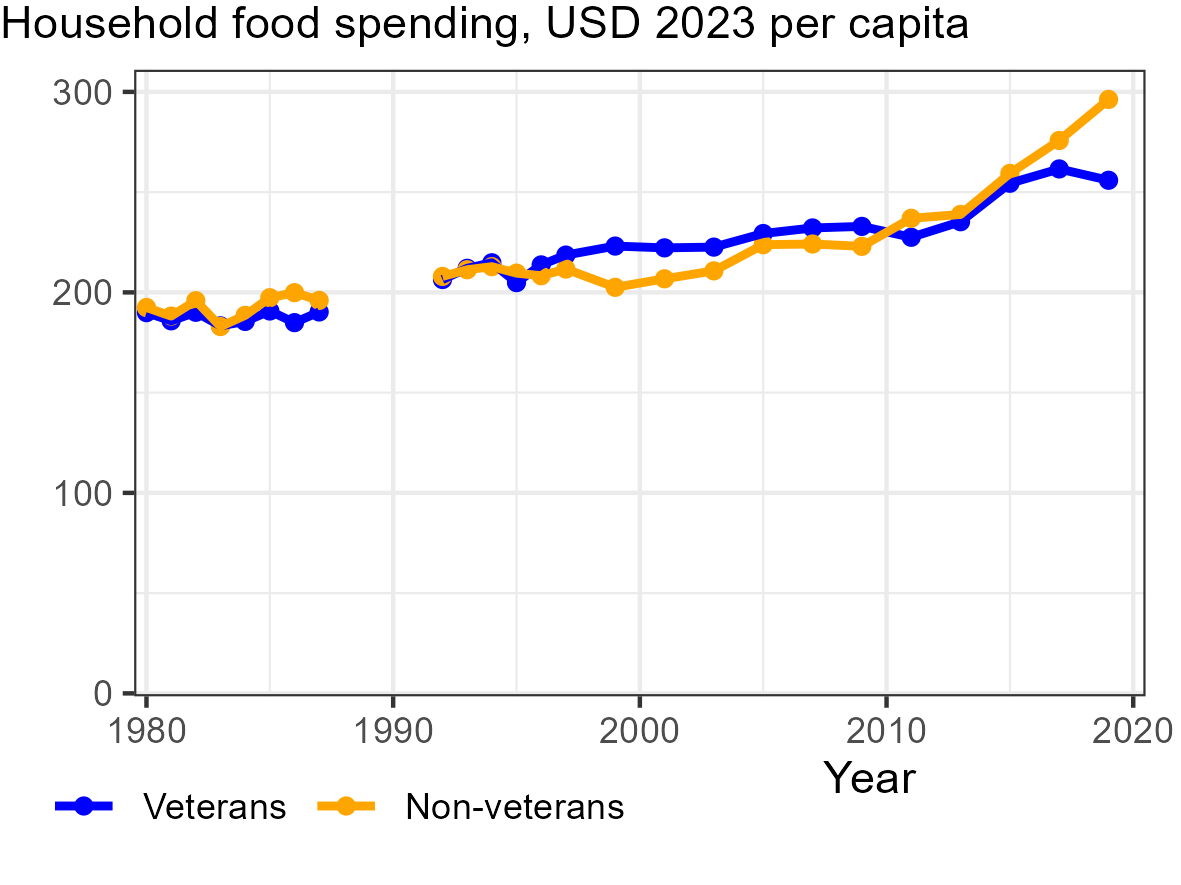}
        \label{fig:fam-foodspending-real}
    \end{subfigure}
    \begin{subfigure}[b]{0.495\textwidth}
        \centering
        %\caption{Participation in SNAP or TANF, rate.}
        \includegraphics[width=\textwidth]{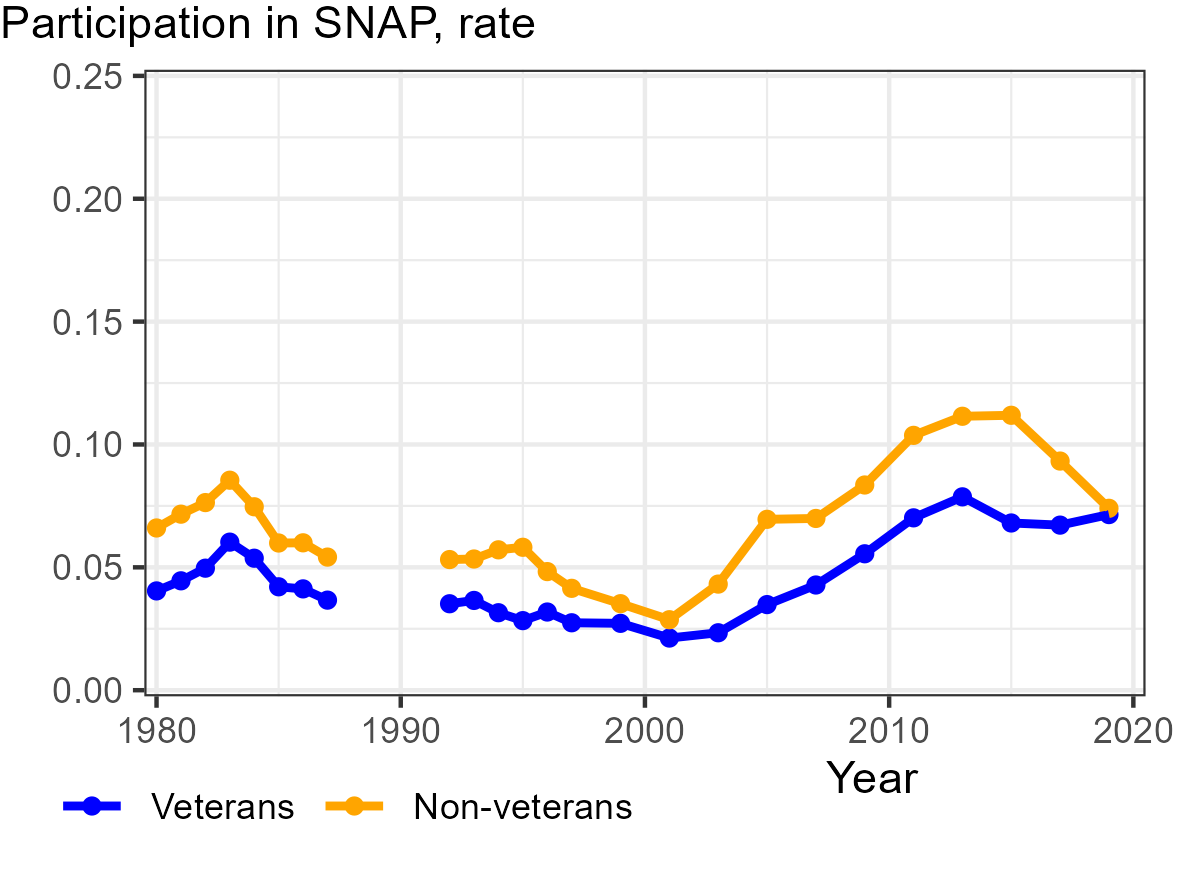}
        \label{fig:fam-welfare-particiaption}
    \end{subfigure}
    \label{fig:fam-food-trends}
    \vspace{-1cm}
    \justify
    \footnotesize
    \textbf{Note:}
    These figures show the average of food insecurity outcomes, for years available in the PSID data separately by veteran status of the household head.
\end{figure}

\autoref{fig:fam-food-trends} shows mean food security outcomes over the study period, separately for veteran- and non-veteran-headed households.
The first two panels show that veteran-headed households consistently have lower rates of food insecurity than non-veteran-headed households.
This difference is visible both in PFI, for which veteran-headed households tend to have lower values, and in the PFI-based food insecurity indicator, FI, for which veteran-headed households tend to have a lower prevalence of food insecurity in each year.
These differences are not driven by a single year or episode.
Both groups follow similar aggregate time-series patterns, including increased food insecurity during periods of macroeconomic stress, such as the 2008--2009 recession, but the difference between veteran- and non-veteran-headed households persists throughout the period.
The bottom panels show that household food spending per capita is broadly similar across the two groups, while SNAP participation is slightly higher among non-veteran-headed households.
Taken together, these associational trends indicate that veteran-headed households do not exhibit higher food insecurity, lower food spending, or greater reliance on SNAP than non-veteran-headed households in the PSID.

We also classify households into long-run food security groups using the measures developed by \cite{lee2024food}.
Veteran-headed households are less likely to be classified as chronically or persistently food insecure than non-veteran-headed households.
\appendixref{appendix:chronic} provides further details on these measures and reports the resulting classifications after averaging across all observed years for each PSID household.

\subsection{Estimates of Causal Effects}
\paragraph{Main Long-Run Effects.}
We next turn to estimates of the effects of military service on long-run food insecurity outcomes.
\autoref{tab:psid-iv-cross} reports the main long-run estimates, where each observation is one man and time-varying outcomes are averaged across all years in which that man is observed.
Columns (1) and (3) report means for non-veterans in the full sample and the Vietnam draft cohorts, respectively.
Columns (2) and (4) report OLS associations for the full sample of men born between 1930 and 1970 and for the Vietnam draft cohorts born between 1950 and 1952, respectively.
Column (5) reports IV estimates for the draft cohorts using the binary conscription instrument described in \autoref{sec:draft-iv}.
The OLS estimates describe differences between veteran- and non-veteran-headed households conditional on observable characteristics, while the IV estimates use draft-induced variation in military service to identify causal effects among draft compliers born between 1950 and 1952.

\begin{table}[!htbp]
    \singlespacing
    \centering
    \vskip-0.75cm
    \caption{Long-Term Effects of Military Service Among PSID Households.}
    \footnotesize
    \makebox[\textwidth][c]{
        \begin{tabular}{l c c c c c c c c}
            \\[-1.8ex]\hline \hline \\[-1.8ex] 
            & \multicolumn{2}{c}{Men Born 1930--1970}
                & \multicolumn{3}{c}{Men Born 1950--1952} \\
            \cmidrule(lr){2-3} \cmidrule(lr){4-6}
            & Non-veteran & OLS & Non-veteran & OLS & Binary \\
            & Mean        &     & Mean        &     & IV   \\
            & [Obs no.]   &     & [Obs no.]   &     & \\
            \\[-1.8ex]\hline \\[-1.8ex]
            \multicolumn{2}{l}{\textbf{Panel A. Food Insecurity}} \\
            \input{sections/tables/psid-iv-crossA.tex}
            \\[-1.8ex]\hline \\[-1.8ex]
            \multicolumn{2}{l}{\textbf{Panel B. Persistence of Food Insecurity}}\\
            \input{sections/tables/psid-iv-crossB.tex}
            %\\[-1.8ex]\hline \\[-1.8ex]
            %\textbf{Panel C. Welfare Participation} \\
            %\input{sections/tables/psid-iv-crossC.tex}
            \\[-1.8ex]\hline \\[-1.8ex]
            \multicolumn{2}{l}{\textbf{Panel C. Demographic and Labor}} \\
            \input{sections/tables/psid-iv-crossD.tex}
            \\[-1.8ex]\hline \\[-1.8ex]
            Birth year controls?
                & & Yes &  & Yes & Yes  \\
            \hline
        \end{tabular}
    }
    \label{tab:psid-iv-cross}
    \justify
    \footnotesize
    \textbf{Note}:
    This table reports associations and IV estimates of the effects of military service on the outcomes listed in the rows.
    Each observation is one man, and time-varying outcomes are averaged across all years in which that man is observed.
    Columns (1) and (3) report non-veteran means, while columns (2) and (4) report OLS estimates for men born between 1930 and 1970 and men born between 1950 and 1952, respectively.
    Column (5) reports IV estimates for men born between 1950 and 1952 using the binary Vietnam-era conscription instrument.
    Household food spending is measured monthly in 2023 US dollars.
    Income outcomes are log transformed, so their coefficients can be interpreted approximately as proportional changes.
    Numbers in square brackets report the number of non-missing observations.
    SEs appear in parentheses and are clustered by birth year.
\end{table}

\autoref{tab:psid-iv-cross} provides no evidence that military service adversely affects long-run food insecurity outcomes.
In the full sample of men born between 1930 and 1970, the OLS associations are small and provide no consistent indication of greater food insecurity among veterans.
Among men born between 1950 and 1952, veteran status is associated with a 0.01 higher mean PFI (SE 0.02), a 0.01 higher share of years food secure (SE 0.03), and a 0.01 lower share of years food insecure (SE 0.03).

The IV estimates move away from adverse effects.
Draft-induced military service is associated with a 0.13 lower mean PFI (SE 0.08), a 0.27 higher share of years food secure (SE 0.23), and a 0.27 lower share of years food insecure (SE 0.23).
These estimates are large relative to the non-veteran means in the draft cohorts, where mean PFI is 0.17 and the average share of years food insecure is 0.09.
The difference between the OLS and IV estimates is consistent with non-random selection into military service, including the greater propensity of men from lower socioeconomic status households to enlist during this period.
The central result is not simply that the IV point estimates are negative.
Rather, the estimates provide no support for the hypothesis that Vietnam-era military service increased later-life food insecurity among men who later became PSID household heads.

The food spending estimates are also inconsistent with persistent material hardship caused by military service.
In the draft cohorts, the OLS estimates suggest that veteran-headed households spent somewhat less on food than non-veteran-headed households.
Monthly household food spending is 18 dollars lower (SE 15.40), while monthly household food spending per capita is 14 dollars lower (SE 9.09).
The corresponding IV estimates are positive, equal to 129 dollars for household food spending (SE 114.14) and 89 dollars for household food spending per capita (SE 84.02).
These estimates are considerably less precise than the PFI and food insecurity classification estimates.
They therefore do not establish that military service increased food spending.
They also do not show the combination of higher PFI, more years food insecure, and lower food spending that one would expect if military service generated persistent material hardship.

The persistence measures likewise provide no evidence that military service shifted men into repeated or durable food insecurity.
Panel B of \autoref{tab:psid-iv-cross} reports effects on the long-run food insecurity classifications derived from the PFI series.
The estimated effect on chronic and persistent food insecurity is $-0.10$ (SE 0.27).
The estimated effect on being chronically but not persistently food insecure is $-0.15$ (SE 0.15).
The estimated effect on transient food insecurity is $-0.06$ (SE 0.02).
The estimated effect on being persistently food secure is positive, at 0.31 (SE 0.42).
These estimates again do not support the hypothesis that military service increased long-run exposure to food insecurity among men observed as household heads.
The point estimates move away from chronic food insecurity, although the persistence classifications should be interpreted cautiously given the modest size of the draft-cohort sample and the imprecision of most estimates.

The SNAP estimates also provide no evidence of greater reliance on federal food assistance.
In the long-run specification, the binary IV estimates are $-0.16$ for ever participating in SNAP (SE 0.18) and $-0.11$ for the share of years participating in SNAP (SE 0.14).\footnote{
    Comparisons with administrative records establish that self-reported SNAP participation is considerably under-reported in surveys \citep{meyer2022errors,gregory2026supplemental}.
    %There is no reason, however, to suspect this measurement error is correlated with veteran status.
    Whether this measurement error is substantially different between veterans and non-veterans is unclear.
}
These estimates should not be interpreted as precise evidence that military service reduced SNAP participation.
Rather, they show that the absence of adverse effects on food insecurity is not masking increased reliance on SNAP among veterans who become household heads in the PSID.

\paragraph{Dynamic Panel Estimates.}
Annual panel estimates lead to the same conclusion as the long-run averages.
\autoref{tab:psid-iv-panel} repeats the analysis in an individual-year panel, where each observation is a man-year rather than a man-specific average.
This specification uses annual variation in household food insecurity and food spending across the PSID panel.
The panel estimates are useful because the long-run averages in \autoref{tab:psid-iv-cross} could obscure effects concentrated in particular periods of adulthood.

\begin{table}[h!]
    \singlespacing
    \centering
    \caption{Dynamic Effects of Military Service among PSID Men.}
    \footnotesize
    \makebox[\textwidth][c]{
        \begin{tabular}{l c c c c c c c c}
            \\[-1.8ex]\hline \hline \\[-1.8ex]
            & \multicolumn{2}{c}{Men Born 1930--1970}
                & \multicolumn{3}{c}{Men Born 1950--1952} \\
            \cmidrule(lr){2-3} \cmidrule(lr){4-6}
            & Non-veteran & OLS & Non-veteran & OLS & Binary \\
            & Mean        &     & Mean        &     & IV \\
            & [Obs no.]   &     & [Obs no.]   &     & \\
            \\[-1.8ex]\hline \\[-1.8ex]
            \multicolumn{2}{l}{\textbf{Panel A. Food Insecurity}} \\
            \input{sections/tables/psid-iv-panelA.tex}
            \\[-1.8ex]\hline \\[-1.8ex]
            %\textbf{Panel B. Welfare Participation} \\
            %\input{sections/tables/psid-iv-panelB.tex}
            %\\[-1.8ex]\hline \\[-1.8ex] 
            \multicolumn{2}{l}{\textbf{Panel B. Demographic and Labor}} \\
            \input{sections/tables/psid-iv-panelC.tex}
            \\[-1.8ex]\hline \\[-1.8ex] 
            Birth year controls?
                & & Yes &  & Yes & Yes \\
            \hline
        \end{tabular}
    }
    \label{tab:psid-iv-panel}
    \justify
    \footnotesize
    \textbf{Note}:
    This table reports associations and IV estimates of the effects of military service on the annual outcomes listed in the rows.
    Each observation is an individual-year, and each man is observed for approximately 18 years on average.
    Columns (1) and (3) report non-veteran means, while columns (2) and (4) report OLS estimates for men born between 1930 and 1970 and men born between 1950 and 1952, respectively.
    Column (5) reports IV estimates for men born between 1950 and 1952 using the binary Vietnam-era conscription instrument.
    Household food spending is measured monthly in 2023 US dollars.
    Income outcomes are log transformed, so their coefficients can be interpreted approximately as proportional changes.
    Numbers in square brackets report the number of non-missing individual-year observations.
    Standard errors appear in parentheses and are clustered by PSID survey year and by birth year.
\end{table}

The annual panel estimates are less precise for some food insecurity outcomes.
The annual specification retains year-to-year variation in PFI, which contains substantial transitory household-level noise.
By contrast, the long-run specification averages outcomes within men before estimation, filtering out much of this transitory variation and estimating effects on persistent differences in food insecurity histories.
This explains why the long-run IV estimates in \autoref{tab:psid-iv-cross} can be more precise than the corresponding annual estimates in \autoref{tab:psid-iv-panel}, despite using fewer observations.

The annual food insecurity estimates do not show that military service increased food insecurity during the years in which men are observed in the PSID.
In the full sample of men born between 1930 and 1970, the OLS estimates show that veteran-headed households have lower PFI and lower annual food insecurity.
Veteran status is associated with a 0.03 lower PFI, a 0.05 higher probability of being food secure, and a 0.05 lower probability of being food insecure.
Among the 1950--1952 draft cohorts, the OLS estimates are smaller.
Veteran status is associated with a 0.02 higher PFI (SE 0.01) and essentially no difference in annual food insecurity classification.

The IV estimates again move away from adverse effects.
The estimated effect on PFI is $-0.09$ (SE 0.09), the estimated effect on annual food security is 0.12 (SE 0.13), and the estimated effect on annual food insecurity is $-0.12$ (SE 0.13).
The annual PFI estimate is somewhat less precise than its long-run counterpart because the panel specification retains annual fluctuations rather than averaging them into a person-level food insecurity history. 
The annual estimates for the binary food security classifications are smaller in magnitude but more precisely estimated than the corresponding long-run estimates.

The annual food spending estimates also do not suggest a persistent resource shortfall among draft-induced veterans.
In the draft cohorts, the OLS estimates imply 26 dollars less in household food spending (SE 11.34) and 13 dollars less in household food spending per capita (SE 10.30) among veterans.
The IV estimates are positive, equal to 97 dollars for household food spending (SE 95.24) and 114 dollars for household food spending per capita (SE 95.42).
These coefficients are imprecisely estimated and should not be interpreted as evidence that military service increased food spending.
Using annual data, the draft-induced estimates do not show the combination of lower food spending and higher food insecurity that would indicate later-life material hardship.

The annual SNAP estimate tells the same broad story.
The estimated effect of military service on annual SNAP participation is $-0.03$ (SE 0.09).
This estimate provides no evidence that the absence of adverse food insecurity effects is explained by draft-induced veterans relying more heavily on SNAP.

\begin{figure}[h!]
    \centering
    \singlespacing
    \caption{Effect of Military Service on PFI by Age.}
    \begin{subfigure}[b]{0.495\textwidth}
        \centering
        %\caption{OLS.}
        \includegraphics[width=\textwidth]{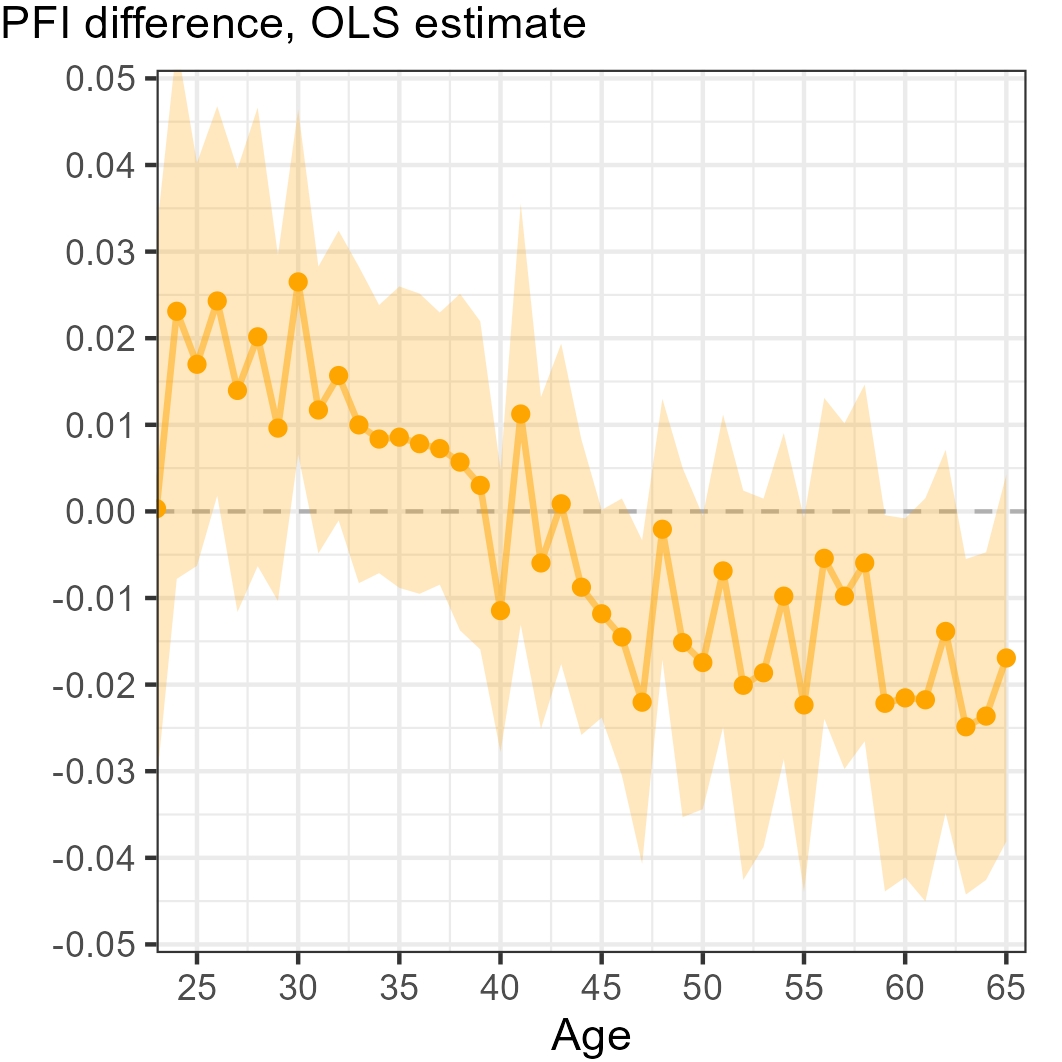}
        \label{fig:pfs-age-ols}
    \end{subfigure}
    \begin{subfigure}[b]{0.495\textwidth}
        \centering
        %\caption{IV.}
        \includegraphics[width=\textwidth]{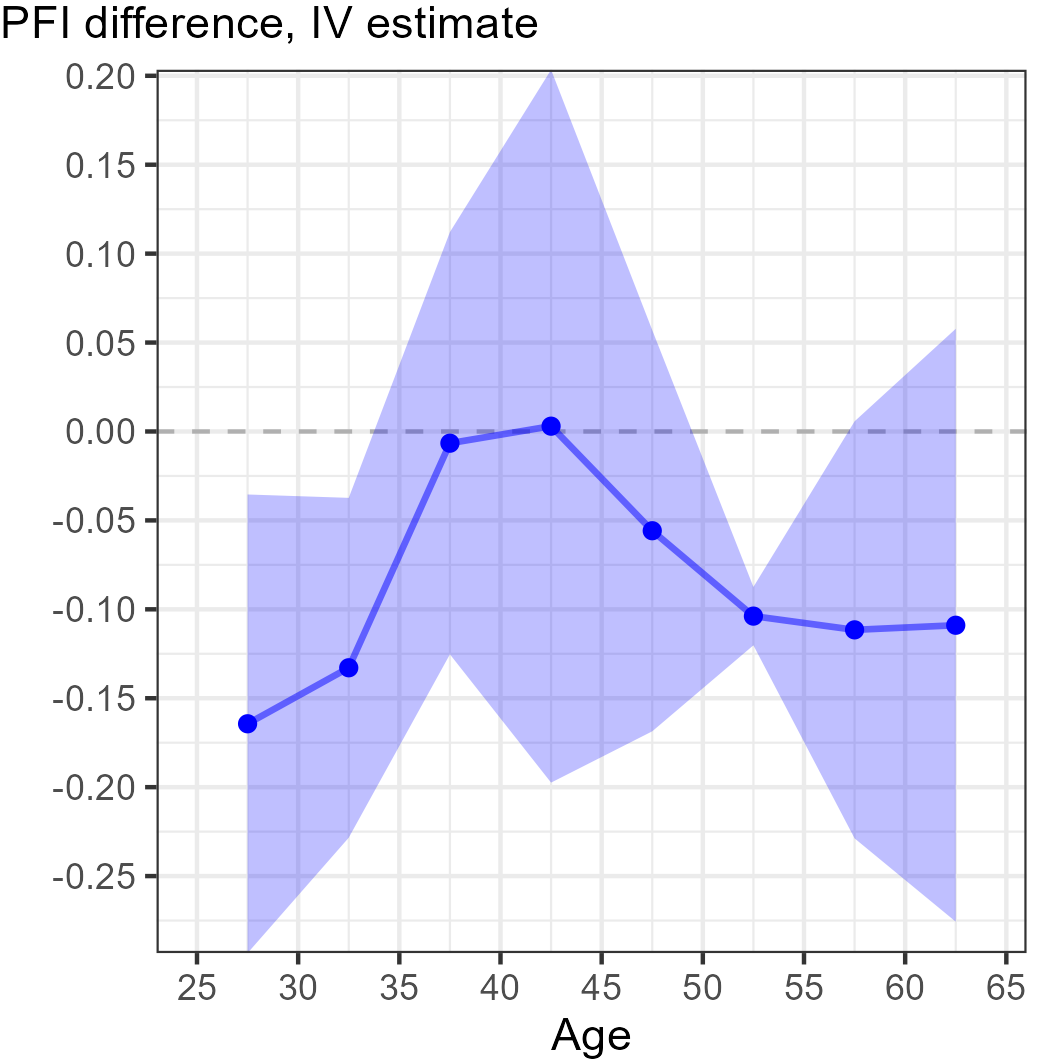}
        \label{fig:pfs-age-iv}
    \end{subfigure}
    \label{fig:pfs-age}
    \vspace{-1cm}
    \justify
    \footnotesize
    \textbf{Note:}
    These panels report estimates of the association or causal effect of military service on PFI at different ages.
    Panel A reports OLS estimates for PSID men born between 1930 and 1970.
    Panel B reports IV estimates for men born between 1950 and 1952, using the binary Vietnam-era conscription instrument for veteran status.
    Each estimate is calculated within a five-year age bin.
    For example, the estimate plotted at age 27.5 is calculated among men aged 25--30.
    The shaded regions represent 95\% confidence intervals based on standard errors clustered by birth year.
\end{figure}

\paragraph{Food Insecurity Over the Life-cycle.}
The average effects in \autoref{tab:psid-iv-cross} and \autoref{tab:psid-iv-panel} could mask periods of heightened food insecurity at particular ages.
Military service could plausibly increase hardship immediately after separation, with effects fading later, or generate health and disability shocks that become more consequential only at older ages.
\autoref{fig:pfs-age} therefore reports differences in PFI between veterans and non-veterans over the life cycle.
Panel A reports OLS associations between veteran status and PFI for men born between 1930 and 1970.
Panel B reports IV estimates for men born between 1950 and 1952 using draft-induced variation in military service.

The OLS estimates suggest that military service is associated with higher PFI into a man's late thirties, after which the association becomes negative.
This pattern mirrors the finding in \cite{angrist2011schooling} that the initial negative income effects of military service among Vietnam veterans disappear once veterans reach their late forties.

The IV estimates are necessarily noisier than the pooled estimates in \autoref{tab:psid-iv-cross} because each point is estimated within a five-year age bin.
Nevertheless, the figure does not reveal an age range in which military service clearly raises PFI.
The estimates provide no clear evidence of an adverse effect at any observed age.
Among men who survive and become PSID household heads, Vietnam-era military service does not appear to produce a detectable food insecurity penalty at any particular stage of observed adulthood.

The life-cycle evidence is consistent with the chronic food insecurity measures.
If military service caused recurrent or durable food hardship, we would expect either higher PFI over sustained age ranges or higher rates of chronic and persistent food insecurity.
We find neither.
In \autoref{tab:psid-iv-cross}, the estimated effect on chronic and persistent food insecurity is $-0.10$ (SE 0.27).
The estimated effect on being chronically but not persistently food insecure is $-0.15$ (SE 0.15).
The estimated effect on transient food insecurity is $-0.06$ (SE 0.02).
The estimated effect on being persistently food secure is positive, at 0.31 (SE 0.42), although it is imprecisely estimated.
These estimates should be interpreted cautiously given the modest size of the draft-cohort sample and the fact that the persistence categories are discrete summaries of long-run food insecurity histories.
Nevertheless, the pattern is directionally inconsistent with the claim that military service causes veterans who later become household heads to face elevated food insecurity risk throughout adulthood.

\paragraph{Labor Market and Household Outcomes.}
The labor market and household-structure outcomes do not show a consistent pattern of later-life economic deterioration.
In the long-run estimates, military service is associated with 2.98 more years of education (SE 1.33), 0.10 higher employment (SE 0.27), $-0.13$ log points of individual income (SE 0.80), 0.50 higher log household income (SE 0.74), and 0.86 higher log household income per capita (SE 0.88).
The estimated effect on household size is 0.27 (SE 0.68).

The annual panel estimates are similarly mixed and imprecise.
The estimated effect is 1.77 years for education (SE 1.51), 0.08 for employment (SE 0.13), $-0.21$ log points for individual income (SE 0.39), $-0.05$ log points for household income (SE 0.45), 0.27 log points for household income per capita (SE 0.45), and $-1.07$ for household size (SE 0.71).
These estimates do not support strong claims about labor market mechanisms.
They are nevertheless useful for interpreting the food insecurity results.
The absence of an adverse food insecurity effect does not appear alongside systematic evidence of lower employment, lower income, larger household burdens, or greater SNAP reliance.
This pattern is consistent with \cite{angrist2011long}, who find that the negative earnings effects of Vietnam-era military service fade by middle age.
It suggests that any shocks associated with military service among this observed group of household heads did not translate into persistent food insecurity disadvantage.

%% file: sections/tables/psid-iv-crossA.tex
% latex table generated in R 4.4.1 by xtable 1.8-8 package
% Wed Jul 22 16:59:33 2026
Mean PFI & 0.19 & 0.009 & 0.17 & 0.01 & -0.13 \\
    & [8,392] & (0.01) & [816] & (0.02) & (0.08) \\
Food secure, percent years & 0.89 & 0.023 & 0.91 & 0.01 & 0.27 \\
    & [8,392] & (0.01) & [816] & (0.03) & (0.23) \\
Food insecure, percent years & 0.11 & -0.026 & 0.09 & -0.01 & -0.27 \\
    & [8,392] & (0.01) & [816] & (0.03) & (0.23) \\
Household food spending & 378.16 & 9.90 & 375.93 & -17.6 & 128.83 \\
    & [8,392] & (5.05) & [816] & (15.4) & (114.14) \\
Household food spending, per person & 186.9 & -4.25 & 183.9 & -14.05 & 88.75 \\
    & [8,392] & (4.94) & [816] & (9.09) & (84.02) \\
SNAP, ever participated & 0.23 & 0.025 & 0.2 & 0.08 & -0.16 \\
    & [8,392] & (0.02) & [816] & (0.04) & (0.18) \\
SNAP, percent years participated & 0.08 & 0.005 & 0.07 & 0.02 & -0.11 \\
    & [8,392] & (0.01) & [816] & (0.02) & (0.14) \\
  

%% file: sections/tables/psid-iv-crossB.tex
% latex table generated in R 4.4.1 by xtable 1.8-8 package
% Wed Jul 22 16:59:35 2026
Chronic and persistent & 0.08 & -0.030 & 0.06 & -0.02 & -0.1 \\
    & [8,392] & (0.01) & [816] & (0.03) & (0.27) \\
Chronic and not persistent & 0.03 & -0.002 & 0.02 & -0.02 & -0.15 \\
    & [8,392] & (0.00) & [816] & (0.01) & (0.15) \\
Transiently insecure & 0 & 0.002 & 0 & 0 & -0.06 \\
    & [8,392] & (0.00) & [816] & (0.01) & (0.02) \\
Persistently secure & 0.89 & 0.030 & 0.92 & 0.04 & 0.31 \\
    & [8,392] & (0.01) & [816] & (0.04) & (0.42) \\

%% file: sections/tables/psid-iv-crossD.tex
% latex table generated in R 4.4.1 by xtable 1.8-8 package
% Wed Jul 22 16:59:40 2026
Education years & 13.33 & 0.196 & 13.74 & 0.05 & 2.98 \\
    & [8,074] & (0.10) & [786] & (0.22) & (1.33) \\
Employed, percent years & 0.83 & -0.003 & 0.84 & -0.04 & 0.1 \\
    & [8,273] & (0.01) & [806] & (0.03) & (0.27) \\
Individual income & 66.43 & -0.010 & 82.78 & -0.16 & -0.13 \\
    & [7,374] & (0.04) & [734] & (0.12) & (0.8) \\
Household income & 113.64 & -0.036 & 122.92 & -0.09 & 0.5 \\
    & [8,391] & (0.03) & [816] & (0.08) & (0.74) \\
Household income, per capita & 43.64 & -0.017 & 48.26 & -0.1 & 0.86 \\
    & [8,391] & (0.03) & [816] & (0.08) & (0.88) \\
Household size & 4.15 & 0.031 & 4.05 & 0.07 & 0.27 \\
    & [8,392] & (0.05) & [816] & (0.08) & (0.68) \\

%% file: sections/tables/psid-iv-panelA.tex
% latex table generated in R 4.4.1 by xtable 1.8-8 package
% Wed Jul 22 17:00:40 2026
 PFI                                 & 0.17     & -0.03  & 0.14   & 0.02    & -0.09   \\ 
                                      & [87,840] & (0)    & [9059] & (0.01)  & (0.09)  \\ 
  Food secure                         & 0.91     & 0.05   & 0.94   & 0       & 0.12    \\ 
                                      & [87,840] & (0.01) & [9059] & (0.02)  & (0.13)  \\ 
  Food insecure                       & 0.09     & -0.05  & 0.06   & 0       & -0.12   \\ 
                                      & [87,840] & (0.01) & [9059] & (0.02)  & (0.13)  \\ 
  Household food spending             & 401.4    & 7.27   & 408.58 & -25.5   & 97.31   \\ 
                                      & [87,840] & (3.8)  & [9059] & (11.34) & (95.24) \\ 
  Household food spending, per person & 213.34   & -4.75  & 219.2  & -12.78  & 114.43  \\ 
                                      & [87,840] & (4.06) & [9059] & (10.3)  & (95.42) \\ 
  SNAP, participation                 & 0.06     & -0.02  & 0.04   & 0.02    & -0.03   \\ 
                                      & [87,840] & (0)    & [9059] & (0.01)  & (0.09)  \\ 
  

%% file: sections/tables/psid-iv-panelC.tex
% latex table generated in R 4.4.1 by xtable 1.8-8 package
% Wed Jul 22 17:00:54 2026
 Education years              & 13.31    & 0.39   & 13.88  & -0.29  & 1.77   \\ 
                               & [71,467] & (0.06) & [7621] & (0.17) & (1.51) \\ 
  Employed                     & 0.83     & 0      & 0.85   & -0.06  & 0.08   \\ 
                               & [85,162] & (0.01) & [8905] & (0.02) & (0.13) \\ 
  Individual income            & 73.21    & 0.24   & 91.7   & -0.18  & -0.21  \\ 
                               & [46,883] & (0.04) & [5314] & (0.06) & (0.39) \\ 
  Household income             & 122.63   & 0.14   & 136.92 & -0.15  & -0.05  \\ 
                               & [85,100] & (0.02) & [8900] & (0.06) & (0.45) \\ 
  Household income, per capita & 41.29     & 0.14   & 46.73   & -0.13  & 0.27   \\ 
                               & [85,100] & (0.02) & [8900] & (0.06) & (0.45) \\ 
  Household size               & 2.97     & -0.03  & 2.93   & -0.07  & -1.07  \\ 
                               & [85,162] & (0.05) & [8905] & (0.09) & (0.71) \\ 
  

%% file: sections/05-sensitivity.tex
%%%%%%%%%%%%%%%%%%%%%%%%%%%%%%%%%%%%%%%%%
%% Sensitivity and Robustness
\section{Sensitivity and Robustness of Results}
\label{sec:sensitivity}

This section examines whether the absence of estimated adverse effects is robust to limitations of the PSID sample and uncertainty in the draft-lottery first-stage.

We consider three sensitivity exercises.
First, we assess whether uncertainty in the first-stage estimated within the PSID affects either the IV point estimates or statistical inference.
We implement a two-sample IV approach that estimates the reduced-form effect of draft eligibility on food insecurity in the PSID, but imports more precise first-stage estimates of the effect of draft eligibility on military service from the 2000 Census.
This exercise assesses whether sampling uncertainty in the first stage estimated from the relatively small PSID draft cohorts materially affects the IV point estimates.
We complement this exercise with Anderson--Rubin confidence intervals that remain valid under weak identification.
Second, we conduct an agnostic sensitivity analysis for veterans who are not observed as household heads.
This exercise asks how large the unobserved effects among non-head veterans would have to be to generate a substantively adverse population average effect, given the estimated effect among PSID household heads and the share of veterans observed as household heads in external data.
Third, we conduct a more structured sensitivity analysis that separately accounts for non-random survival and household headship.
This exercise uses external information on survival and headship, together with observed food security differences between household heads and never-heads in the PSID, to calibrate how much sample selection could change the estimated population effect.

Together, these exercises assess whether the main conclusion is robust to uncertainty in the PSID first-stage and conventional IV inference, the absence of veterans outside traditional household structures from the analytic sample, and explicit assumptions about the relationships among military service, survival, household headship, and food insecurity.

\subsection{More Precise First-Stage Estimates}
The first sensitivity exercise asks whether sampling uncertainty in the first-stage estimated from the relatively small PSID draft cohorts materially affects the main IV estimates.
The main IV specification estimates both the reduced-form and first-stage within the PSID, instrumenting veteran status using the binary Vietnam draft-lottery conscription instrument for a relatively small sample of men.
As a robustness exercise, we instead implement a two-sample indirect least squares (ILS) estimator.

The reduced-form relationship between draft eligibility and later food security outcomes is estimated in the PSID, while the first-stage relationship between draft eligibility and military service is imported from the analysis by \cite{angrist2011schooling}, who use the much larger 2000 Census to estimate the effect of conscription on military service.
Under the assumption that the first-stage relationship estimated in the Census applies to the corresponding PSID draft cohorts, the ratio of the separately estimated reduced-form and first-stage effects identifies the same draft-lottery local average treatment effect.
This approach is also referred to as a two-sample IV estimator and is equivalent, in the just-identified case, to an indirect least squares estimator using separate samples for the numerator and denominator.

The exercise does not provide a new source of identifying variation.
Rather, it asks whether the main results are sensitive to estimating the first-stage within the relatively small PSID draft cohorts.
Although the large Census sample substantially improves the precision of the denominator, the numerator remains the reduced-form effect estimated among PSID household heads.
The overall precision of the estimator therefore continues to depend on the available PSID sample.\footnote{
    \appendixref{sec:iv-binary} presents the two-sample IV estimates and provides details on the estimation procedure.
}

The two-sample IV estimates provide little indication that sampling uncertainty in the PSID first-stage masks adverse effects of military service.
For the main food security outcomes, the two-sample IV estimates do not reveal adverse effects of military service.
If anything, the point estimates move in the opposite direction.
The estimated effect on the probability of food insecurity is negative, the estimated effect on the share of years food secure is positive, and the estimated effect on the share of years food insecure is negative.
These estimates are larger in magnitude than the preferred binary IV estimates in \autoref{tab:psid-iv-cross}, but they do not suggest that the main estimates are masking adverse effects.

The two-sample IV exercise should not be interpreted as overturning or replacing the preferred within-PSID estimates.
Instead, replacing the PSID first-stage with a more precise external first-stage continues to produce no evidence of a long-run food insecurity penalty from military service.
The exercise directly assesses whether sampling variation in the first-stage relationship estimated from the relatively small PSID draft cohorts drives the main findings.
While the precision of the overall estimator remains constrained by the PSID reduced-form, replacing the within-sample first-stage with a more precisely estimated external first-stage provides a direct check that the substantive results are not driven by uncertainty in the PSID first-stage estimate.

Because the first-stage F statistic for the preferred within-PSID specification is 10, \autoref{tab:psid-twosample-cross} also reports Anderson--Rubin 95\% confidence intervals that are robust to weak identification.
For mean PFI, the Anderson--Rubin confidence interval is $[-0.27,-0.02]$.
For the share of years food secure, the confidence interval is $[-0.04,0.77]$, while the corresponding interval for the share of years food insecure is $[-0.76,0.04]$.
These weak-IV-robust confidence intervals provide no evidence that military service adversely affected food insecurity, although the intervals for the binary food security classifications are wide and include zero.
The Anderson--Rubin confidence intervals for household food spending and household food spending per capita are also wide and include zero, consistent with the imprecision of the conventional IV estimates for these outcomes.

The preferred estimates use the binary conscription instrument constructed from the known cohort-specific draft rules.
The two-sample IV estimates use the same binary draft-eligibility design but replace the first-stage estimated in the PSID with corresponding estimates from the substantially larger 2000 Census.
The two approaches therefore differ in the sample used to estimate the first-stage rather than in the underlying source of identifying variation.
Both approaches reach the same substantive conclusion: the estimates do not indicate that Vietnam-era military service increased later-life food insecurity among draft compliers who subsequently became PSID household heads.

\subsection{Unobserved Veterans in the PSID}
The IV estimate is calculated among lottery draft compliers who are observed as PSID household heads, a possibly selected sample.
Its interpretation as an average causal effect (among draft lottery compliers) would require that draft-induced changes in survival or household headship do not create consequential compositional differences between men observed in the PSID and those not.
The following exercises assess how selection into household headship and outcomes among unobserved veterans could affect the interpretation of the estimates.

We use the PSID IV estimates as the estimated effect among observed household heads.
We use March CPS data from 1979--2019 to estimate that 77\% of veteran men born in the Vietnam draft cohorts are household heads.
We then vary the unobserved effect among the remaining 23\% of veterans who are not household heads over a wide range.
The resulting object is a partially identified population effect.
It is partially identified because the PSID does not jointly reveal veteran status and food insecurity outcomes for men who do not become household heads.
\autoref{fig:sens-pfs} reports the resulting range of population effects under alternative assumptions about the effect of military service among these unobserved veterans.\footnote{
    \appendixref{sec:unobserved} describes this sensitivity analysis in full detail.
}

The results show that moderate adverse effects among non-head veterans would not overturn the main conclusion.
Because 77\% of comparable veteran men are household heads in the March CPS, the unobserved non-head effect receives substantially less weight in the population average than the estimated effect among observed heads.
The implied population effect remains close to zero unless military service had very large adverse effects on food security among non-head veterans.
In particular, the population effect on the probability of food security remains statistically indistinguishable from zero when the assumed effect among non-head veterans ranges from a 30\% point reduction to a 30\% point increase in food security.

This exercise does not rule out concentrated harm among the minority of veterans outside traditional household structures.
Instead, it shows that, to overturn our main conclusion for the broader veteran population, such harm would have to be severe, concentrated among the unobserved non-head population, and sufficiently large to offset the estimated effect among the 77\% of veterans who are household heads.

\subsection{Accounting for Survival and Household Headship}
\begin{figure}[h!]
    \centering
    \singlespacing
    \caption{Trends in Food Security Outcomes, Household heads and Never heads.}
    \begin{subfigure}[b]{0.495\textwidth}
        \centering
        %\caption{PFI.}
        \includegraphics[width=\textwidth]{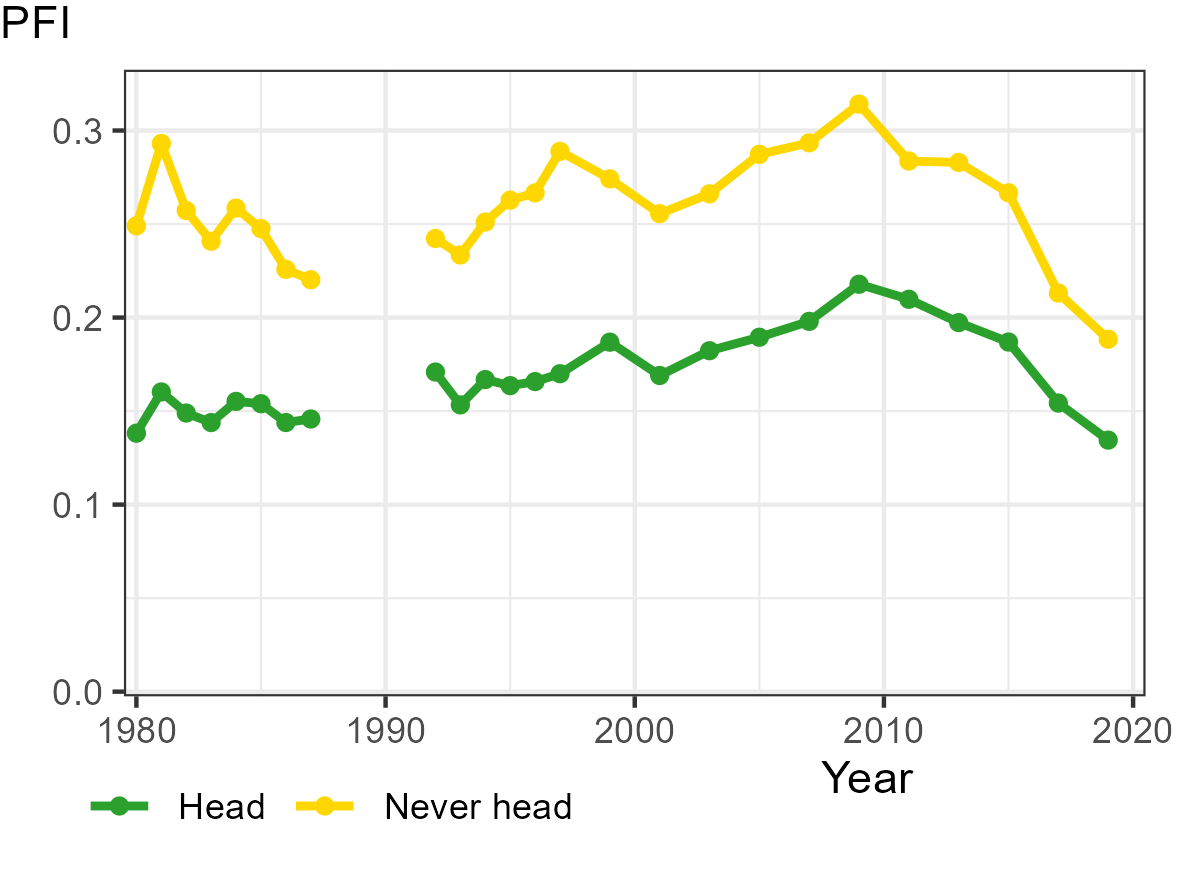}
        \label{fig:head-pfs}
    \end{subfigure}
    \begin{subfigure}[b]{0.495\textwidth}
        \centering
        %\caption{Food insecurity rate.}
        \includegraphics[width=\textwidth]{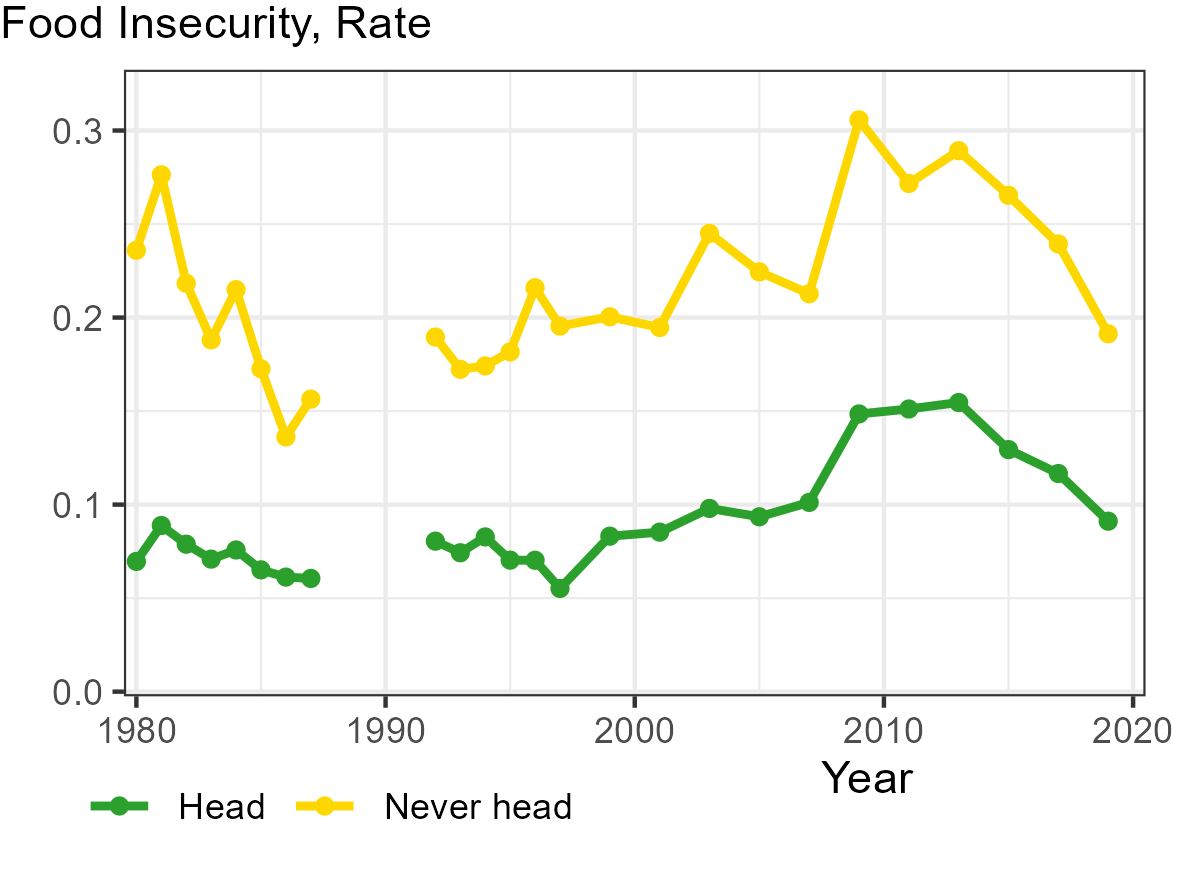}
        \label{fig:head-pfs-insecure}
    \end{subfigure}
    \begin{subfigure}[b]{0.495\textwidth}
        \centering
        %\caption{Food spending, \$ 2023 CPI-U.}
        \includegraphics[width=\textwidth]{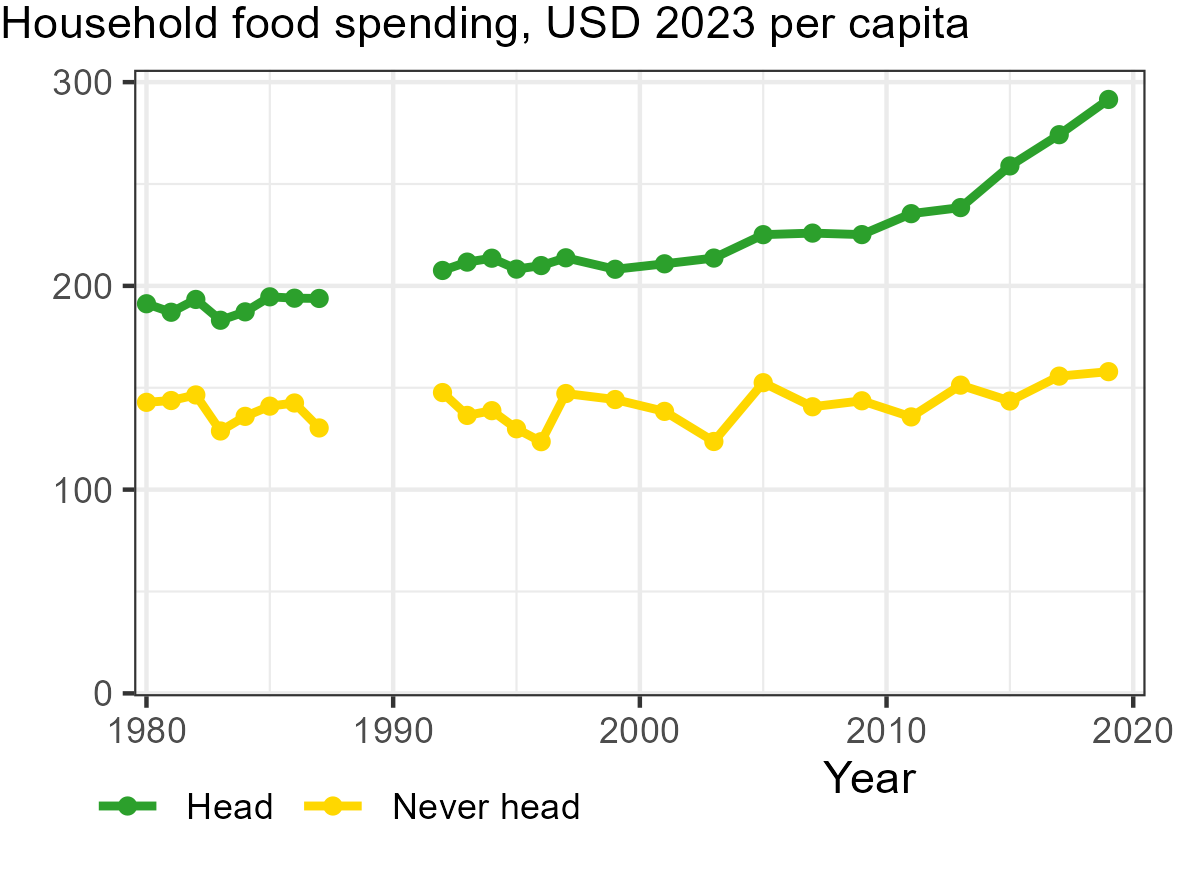}
        \label{fig:head-foodspending-real}
    \end{subfigure}
    \begin{subfigure}[b]{0.495\textwidth}
        \centering
        %\caption{Participation in SNAP, rate.}
        \includegraphics[width=\textwidth]{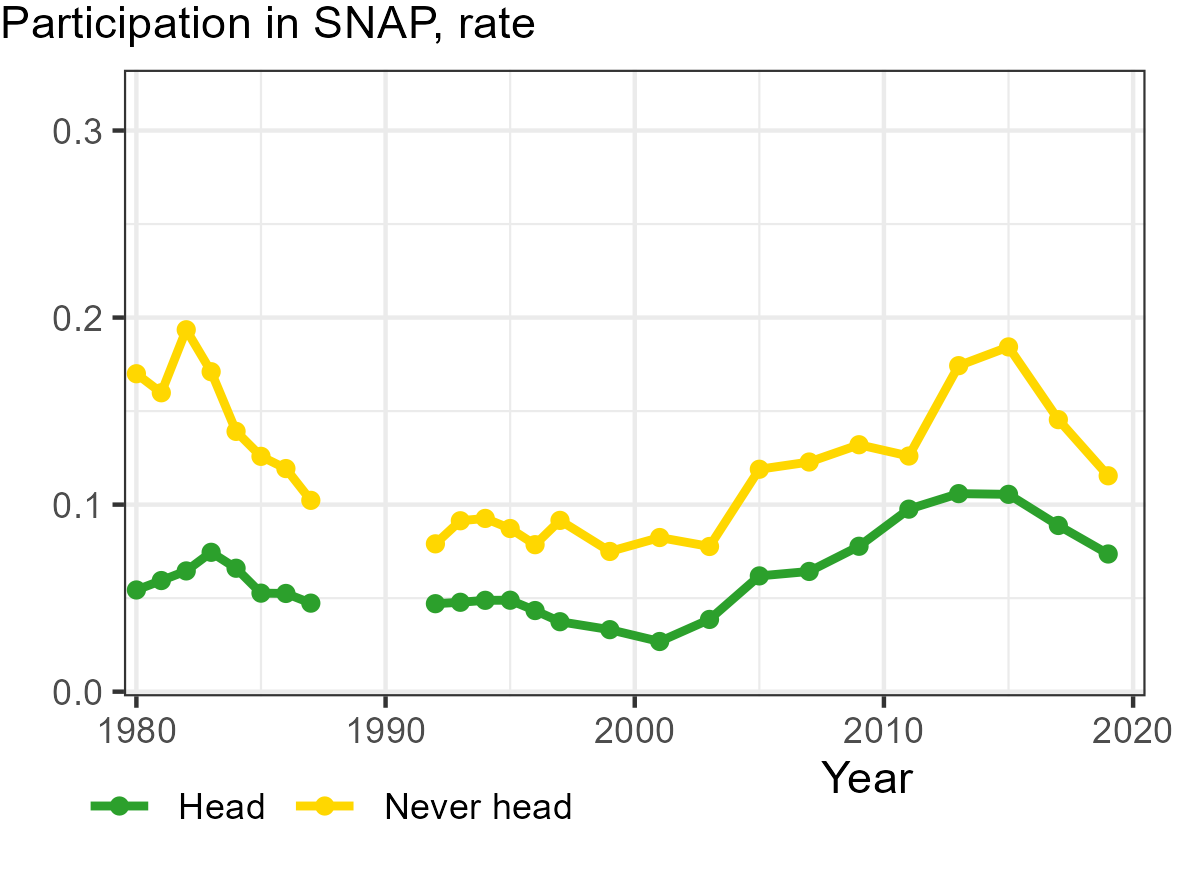}
        \label{fig:head-welfare-particiaption}
    \end{subfigure}
    \label{fig:head-food-trends}
    \vspace{-1cm}
    \justify
    \footnotesize
    \textbf{Note:}
    These figures show the average of food security outcomes, for years available in the PSID data separately among men who ever head a PSID household (Heads) and men in the PSID never observed as a household head (Never heads).
    Veteran status is only observed among PSID men who ever head a household, so is missing among Never heads.
\end{figure}

The main IV estimates identify the effect of military service among men who survive and are observed as household heads in the PSID.
This is the estimand directly available in the main PSID analysis because veteran status is observed only for household heads.
However, this restriction may matter if military service affects whether men later appear in the household-head sample.
\autoref{fig:head-food-trends} illustrates this concern.

Men in the PSID who are never observed as household heads have consistently worse food insecurity outcomes than men who ever become household heads.
They have a higher probability of food insecurity, are classified as food insecure more often, spend less on food, and participate in SNAP at higher rates.
These differences do not identify the causal effect of household headship, nor do they reveal veteran status among never-heads.
They do, however, show that the population omitted from the PSID analysis of veteran household heads could be more food insecure than the population included in the analysis.
If military service reduces survival or the probability of later household headship, the main PSID estimates could understate the population effect of military service on food insecurity.

We assess this concern using a model-informed correction that combines the design-based PSID IV estimate with external information on survival and household headship.
The correction begins with the estimated effect of military service among observed household heads.
It then estimates how military service changes the probability that a man survives and becomes a household head.
The correction translates this selection margin into outcome units using the observed PSID food security gradient between men who ever become household heads and men who never do.
The complete model, assumptions, and variance calculations are presented in \appendixref{appendix:selection-correction}.

The household-headship component is estimated using March CPS data for men born between 1950 and 1952.
Among living men in these cohorts, the estimated household-headship rate is 0.777 among veterans and 0.743 among non-veterans.
Veterans are therefore approximately 3.5 percentage points more likely than non-veterans to be household heads, conditional on survival.
Because the CPS observes only living respondents, it identifies the headship margin conditional on survival but does not identify the survival margin itself.

We calibrate the survival component using mortality evidence from \cite{conley2012long}.
Our preferred specification uses their comparison between the proportions of draft-eligible male and female decedents born between 1950 and 1952.
This comparison provides a reduced-form estimate of the effect of draft eligibility on mortality between ages 39 and 49.
We convert the reported difference in draft-eligible representation among decedents into a mortality-rate difference using the 1990 SSA male period life table \citep{bell2005lifetables}.
We then convert the mortality effect into a survival effect by reversing its sign.
Finally, we divide the resulting reduced-form survival effect by the PSID first-stage effect of draft eligibility on military service.

The estimated survival and household-headship effects are combined into a single selection margin, $\widehat{\Delta}_A$.
This margin measures the effect of military service on the probability of being observed as a surviving household head.
For each outcome, the model-informed estimate is
\begin{equation*}
    \widehat{\beta}^{pop}
        =
        \widehat{\beta}_{PSID}
        -
        \widehat{\Delta}_A
        \widehat{\Delta}_{Y\mid A},
\end{equation*}
where $\widehat{\beta}_{PSID}$ is the binary draft-IV estimate among
observed household heads and $\widehat{\Delta}_{Y\mid A}$ is the
estimated food-security difference between men who are ever an household head in the PSID and those who never are.
%The sign convention implies that the headship gradient is subtracted after being scaled by the effect of military service on sample selection.
SEs are calculated using the delta method, incorporating
uncertainty in the PSID IV estimate, the PSID headship gradient, the CPS
headship rates, the published mortality estimate, and the estimated
first-stage.
This SE calculation treats the external estimates as independent across data
sources and treats the life-table mortality calibration as fixed.

\begin{table}[h!]
    \singlespacing
    \centering
    \caption{Model-Informed Population Effects, Accounting for Survival and Household Headship.}
    \small
    \makebox[\textwidth][c]{
        \begin{tabular}{l c c c c c}
            \\[-1.8ex]\hline \hline \\[-1.8ex]
            & PSID IV  & Headship & Selection & Selection  & Population \\
            & Estimate & Gradient & Margin    & Correction & Estimate \\
            & $\hat{\beta}_{PSID}$ 
            & $\widehat{\Delta}_{Y \mid A}$ 
            & $\widehat{\Delta}_{A}$ 
            & $-\widehat{\Delta}_{A}\widehat{\Delta}_{Y \mid A}$ 
            & $\hat{\beta}^{pop}$ \\
            \\[-1.8ex]\hline \\[-1.8ex] 
            \input{sections/tables/sensitivity-table.tex}
            \\[-1.8ex]\hline \\[-1.8ex] 
        \end{tabular}
    }
    \label{tab:sensitivity-table}
    \justify
    \footnotesize
    \textbf{Note}: This table reports model-informed population effects of military service after accounting for selection into survival and household headship, by a model fully defined in \appendixref{appendix:selection-correction}.
    The PSID IV estimate, $\hat{\beta}_{PSID}$, is the binary draft-IV estimate among men observed as PSID household heads.
    The headship gradient, $\widehat{\Delta}_{Y \mid A}$, is the estimated difference in food-security outcomes (men ever observed as household heads minus men never observed as household heads), controlling for birth year and month.
    The selection margin, $\widehat{\Delta}_{A}$, combines the estimated effect of military service on survival and household headship using CPS and external survival information from Table 2 \cite{conley2012long}.
    The selection correction is $-\widehat{\Delta}_{A}\widehat{\Delta}_{Y \mid A}$, so that positive values indicate corrections toward more food insecure.
    The population estimate is calculated as $\hat{\beta}^{pop} = \hat{\beta}_{PSID} - \widehat{\Delta}_{A}\widehat{\Delta}_{Y \mid A}$, with SEs reported in parentheses.
\end{table}

The selection correction changes the estimated effects very little.
The combined effect of military service on survival and household headship is approximately 0.03 (SE 0.03), so the observed headship gradients receive relatively little weight in the population correction.
For mean PFI, the estimated headship gradient is $-0.07$ (SE 0.01), but the resulting correction rounds to zero (SE 0.00).
The model-informed population estimate is therefore $-0.12$ (SE 0.08), compared with the PSID IV estimate of $-0.13$ (SE 0.08).

The population estimates are similarly close to the IV estimates for the other outcomes.
The population estimate is 0.27 (SE 0.23) for the share of years food secure and $-0.26$ (SE 0.23) for the share of years food insecure.
The corresponding population estimates are 128.98 dollars (SE 114.24) for monthly household food spending and 89.49 dollars (SE 84.07) for monthly household food spending per capita.
Intuitively, the corrections remain small because even meaningful differences between heads and never-heads are multiplied by a small estimated effect of military service on selection into the observed household-head sample.\footnote{
    The results are practically unchanged when we instead use the alternative mortality estimate from \cite{conley2012long}.
    Both published mortality estimates imply only small effects of military service on survival.
    The estimates using the alternative mortality calibration are reported in \appendixref{appendix:selection-correction}, \autoref{tab:sensitivity-table-alt}.
}

This exercise is a model-informed sensitivity analysis rather than a replacement for the draft-lottery research design.
The correction relies on additional assumptions linking the IV estimate among observed household-head compliers to the broader population.
It also uses the observed headship gradient to approximate food security outcomes for men who are marginally omitted from the observed household-head sample.
Nevertheless, the exercise directly addresses whether plausible selection through survival and household headship is quantitatively large enough to overturn the main findings.
Under either mortality calibration, the model-informed population estimates are nearly identical to the design-based IV estimates.
Accounting for survival and household headship therefore does not materially change the paper's conclusions.

%% file: sections/tables/sensitivity-table.tex
% latex table generated in R 4.4.1 by xtable 1.8-8 package
% Wed Jul 22 17:04:04 2026
 Mean PFI & -0.13 & -0.07 & 0.03 & 0 & -0.12 \\ 
    & (0.08) & (0.01) & (0.03) & (0) & (0.08) \\ 
  Food secure, percent years & 0.27 & 0.19 & 0.03 & -0.01 & 0.27 \\ 
    & (0.23) & (0.06) & (0.03) & (0) & (0.23) \\ 
  Food insecure, percent years & -0.27 & -0.19 & 0.03 & 0.01 & -0.26 \\ 
    & (0.23) & (0.06) & (0.03) & (0) & (0.23) \\ 
  Household food spending & 128.83 & -5.01 & 0.03 & 0.16 & 128.98 \\ 
    & (114.14) & (71.16) & (0.03) & (4.88) & (114.24) \\ 
  Household food spending, per capita & 88.75 & -23.89 & 0.03 & 0.74 & 89.49 \\ 
    & (84.02) & (51.21) & (0.03) & (2.91) & (84.07) \\ 
  SNAP, ever participated & -0.16 & 0.12 & 0.03 & 0 & -0.16 \\ 
    & (0.18) & (0.02) & (0.03) & (0) & (0.18) \\ 
  SNAP, percent years participated & -0.11 & -0.06 & 0.03 & 0 & -0.11 \\
    & (0.14) & (0.01) & (0.03) & (0) & (0.14) \\
  

%% file: sections/06-conclusion.tex
%%%%%%%%%%%%%%%%%%%%%%%%%%%%%%%%%%%%%%%%%
%% Conclusion section
\section{Summary and Concluding Remarks}
\label{sec:conclusion}
Care for military veterans is a policy priority in the US, and elsewhere.
Active-duty military personnel experience a significantly higher prevalence of food insecurity than the civilian population does \citep{rabbitt2024comparing}.
No prior causal evidence exists, however, as to whether that relationship persists long-term nor, in particular, as to whether military service causes long-term changes in veterans' food insecurity status.
Using Vietnam-era draft lottery selection to instrument for endogenous military service and an unprecedentedly long time series of individual-level food insecurity measures, we find no evidence that US military service adversely affects food insecurity among draft compliers who later become male household heads in the PSID.

We subject these findings to a battery of sensitivity checks, including correction for non-random survival and selection into household headship.
Although men in the PSID who are never observed as household heads exhibit meaningfully worse food insecurity outcomes than do men who become heads, adjustments to correct for selection into household headship make only very modest differences to our estimates that do not affect the qualitative conclusion.
Military service does not appear to adversely affect food insecurity among Vietnam-era draft compliers who become household heads.
It is, of course, possible that these results are specific to the Vietnam-era cohort of US military veterans.
Indeed, prior correlational studies suggest that post-Vietnam era veterans have meaningfully higher likelihood of food insecurity than both Vietnam-era veterans and non-veterans \citep{miller2016food,rabbitt2021food}, leaving open the question of whether military service in the post-Vietnam, all-volunteer force causes any change in veterans' food security status post-separation.  
The main policy takeaway from our findings is that efforts to support those who serve in the military likely should focus on active-duty personnel and those who prove unable to head a household post-separation from the service.

%% file: sections/08-appendix.tex
%%%%%%%%%%%%%%%%%%%%%%%%%%%%%%%%%%%%%%%%%
%% Appendix section
% Set-up the section.
\appendix
\setcounter{table}{0}
\renewcommand{\thetable}{A\arabic{table}}
\setcounter{figure}{0}
\renewcommand{\thefigure}{A\arabic{figure}}

% Start appendix
\section{Supplementary Appendix}

\subsection{PSID-based Food Insecurity Measures}
\label{appendix:hfsm}
In the 1999--2003 and 2015--2021 survey rounds, the PSID administered the USDA Household Food Security Survey Module (HFSSM), a battery of survey questions used to classify household food insecurity status.
The module consists of 10 questions for households without children and 18 questions for households with children, with food security classifications based on the number of affirmative responses.
The USDA Economic Research Service (ERS) collected HFSSM data in the December Current Population Survey each year to estimate the official prevalence of food insecurity, overall and by subpopulation \citep{rabbitt2025household}.\footnote{
    As of September 2025, USDA discontinued data collection for the HFSSM in the CPS.
}
The limited availability of the HFSSM data, with a long gap between survey periods, does not support the long-run panel analysis central to this study.
This paper therefore relies on the PFI measure, which can be constructed consistently over most of the PSID period, permitting long-run analysis.
\cite{lee2025probability} showed that 86\% of households were classified in the same category, food secure or food insecure, under both the official measure and the PFS-based measure, validating the PFS-based measure for food security analysis.

\begin{figure}[h!]
    \caption{Adult Food Security Survey Answers among PSID Households Headed by Veterans and Non-Veterans, 1999--2003 and 2015--2021.}
    \centering
    \singlespacing    \includegraphics[width=\textwidth]{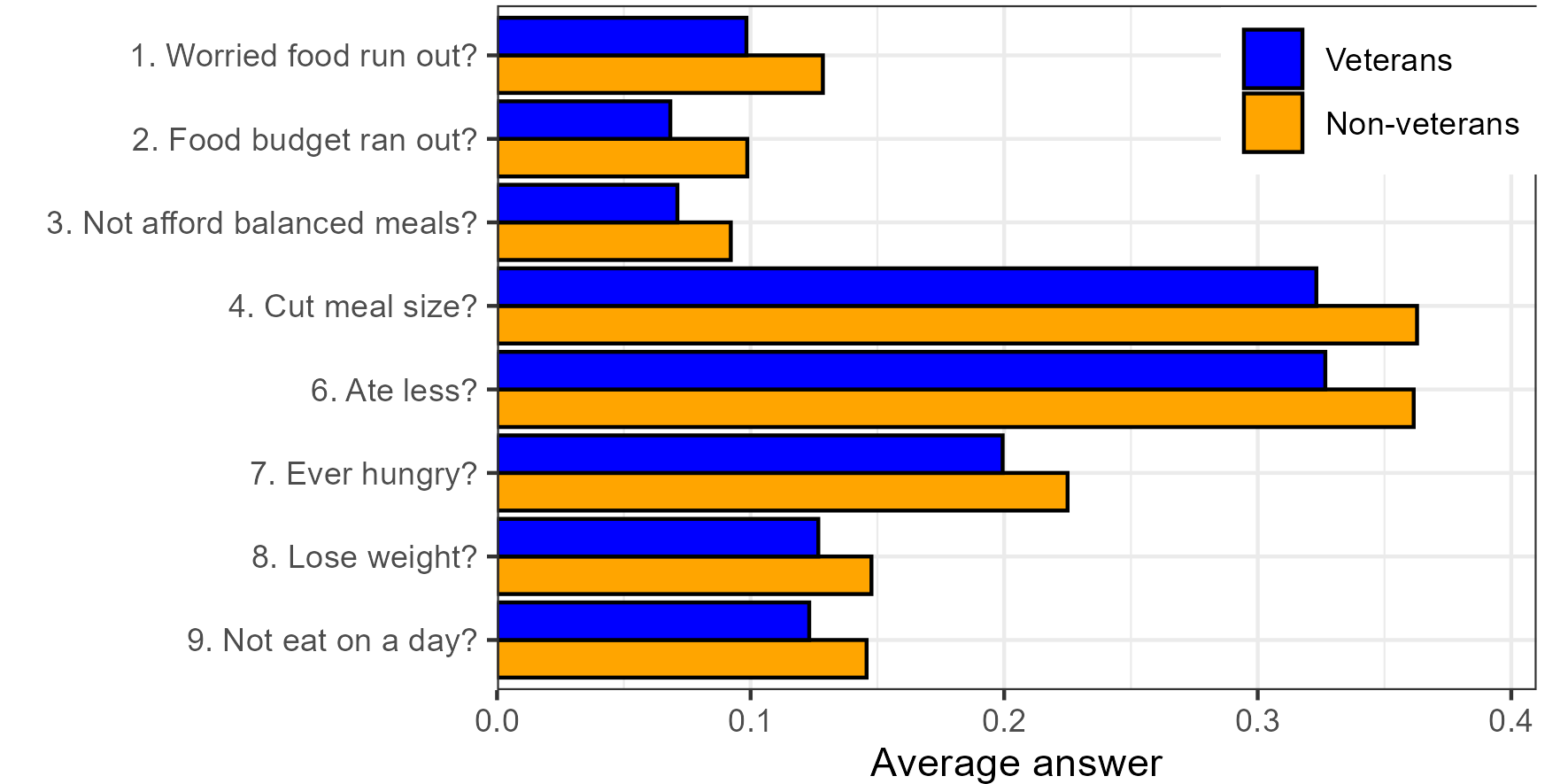}
    \label{fig:psid-surveymean}
    \justify
    \footnotesize
    \textbf{Note:}
    0 counts as answering ``never'' to the questions, 1 as ``sometimes'' or ``often.''
    The food security survey was conducted in 1999--2003 and 2015--2021, and we averaged the responses for those years.
\end{figure}

\autoref{fig:psid-surveymean} presents descriptive evidence from the years in which the HFSSM was available.
Consistent with the PFI results, veteran-headed households were generally less likely than non-veteran-headed households to report food-related hardships in response to the HFSSM questions.
Among the narrower sample of Vietnam draft cohorts, the patterns are less uniform, with veterans exhibiting somewhat lower income and greater SNAP participation.
The limited temporal coverage of the HFSSM and the relatively small number of observations for the draft cohorts make these results primarily descriptive.
More broadly, the HFSSM's restricted availability, together with the announced discontinuation of its collection in the CPS, highlights the value of measures such as PFS and PFI that can be implemented consistently over long time horizons and across future waves of the PSID.

The PSID-based food security measures are constructed as follows.
First, food spending $FS_{h(i),t}$ of person $i$ in household $h(i)$ in year $t$ is modeled as a linear function of lagged food expenditure and its squared term, as well as the household's demographic and socioeconomic variables $\vec W_{h(i),t}$.
This specifically includes the household head's age, gender, and educational status, household size, and household income.
The predicted food spending from this model, $\widehat{FS}_{h(i),t}$, serves as the conditional mean of food spending.

\begin{equation*}
    FS_{h(i),t}
    =
    \alpha_{0}
    +
    \alpha_{1}FS_{h(i),t-1}
    +
    \alpha_{2}FS^{2}_{h(i),t-1}
    +
    \vec \alpha_{3}' \vec W_{h(i),t}
    +
    \epsilon_{h(i),t}.
\end{equation*}

Once food spending is predicted, we model the squared residual from the earlier model, $\widehat{\epsilon}_{h(i),t}^{\,2}$, as a function of the same set of right-hand-side variables.
The predicted outcome serves as the conditional variance of food spending.

\begin{equation*}
    \widehat{\epsilon}_{h(i),t}^{\,2}
    =
    \beta_{0}
    +
    \beta_{1}FS_{h(i),t-1}
    +
    \beta_{2}FS^{2}_{h(i),t-1}
    +
    \vec \beta_{3}' \vec W_{h(i),t}
    +
    \nu_{h(i),t}.
\end{equation*}

Given the conditional means and variances of food spending, we construct a household-year-specific cumulative distribution function (CDF) of food spending, assuming that food spending follows a specific distribution.
We assume that food spending follows a Gamma distribution, $FS_{h(i),t}\sim\operatorname{Gamma}\left(\alpha,\beta\right)$.
We calibrate the parameters using the method of moments such that

\begin{equation*}
    \alpha
    =
    \frac{\widehat{FS}_{h(i),t}^{\,2}}
         {\widehat{\epsilon}_{h(i),t}^{\,2}},
    \qquad
    \beta
    =
    \frac{\widehat{\epsilon}_{h(i),t}^{\,2}}
         {\widehat{FS}_{h(i),t}}.
\end{equation*}

The PFS is defined as one minus the conditional CDF evaluated at the household's TFP cost. Thus, PFI, defined as one minus PFS, is
\begin{equation*}
    \text{PFI}_{i,t} = 1 - PFS_{i,t}
    = F\left( \underline{P}_{h(i),t} \mid FS_{h(i),t-1}, \vec W_{h(i),t} \right) \in [0,1].
\end{equation*}

We classify a household as food insecure if the PFI exceeds the USDA  threshold probability $P_t$ in year $t$.
After 1995, when USDA estimates of official food security prevalence are available, we anchor the threshold probability so that the prevalence of food security under PFS equals the official food security prevalence.
For instance, if 10\% of households were food insecure in year $t$, we set the threshold probability so that 10\% of households were classified as food insecure under the PFS measure.
For pre-1995, before USDA began reporting food insecurity prevalence estimates, we use the $P_t$ thresholds that \citet{lee2025probability} estimated from a regression model of post-1995 PFS thresholds on national SNAP participation rates.

\begin{figure}[h!]
    \caption{Kernel Density Distribution of Probability of Food Insecurity}
    \begin{subfigure}[c]{0.475\textwidth}
        \centering
        \caption{All Panel Data.}
        \includegraphics[width=\textwidth]{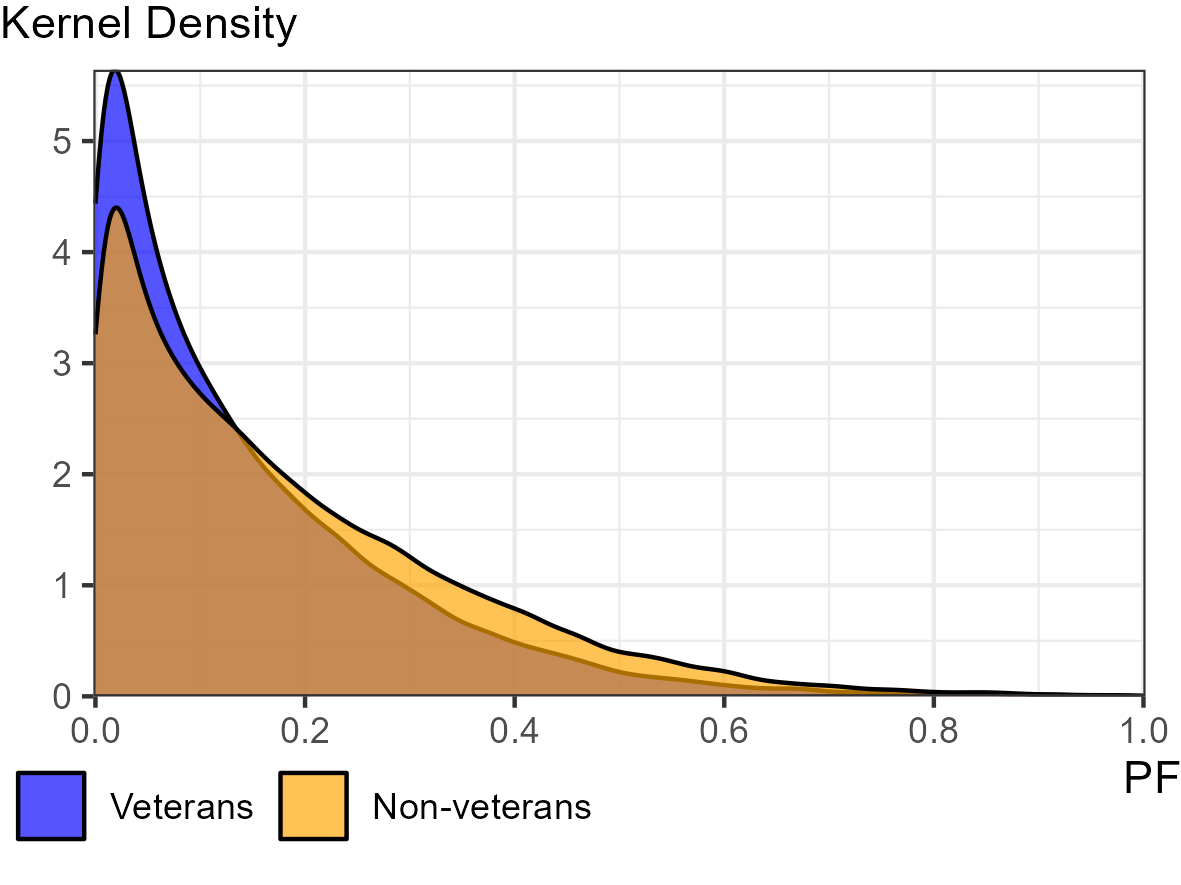}
    \end{subfigure}
    \begin{subfigure}[c]{0.475\textwidth}
        \centering
        \caption{Mean PFI.}
        \includegraphics[width=\textwidth]{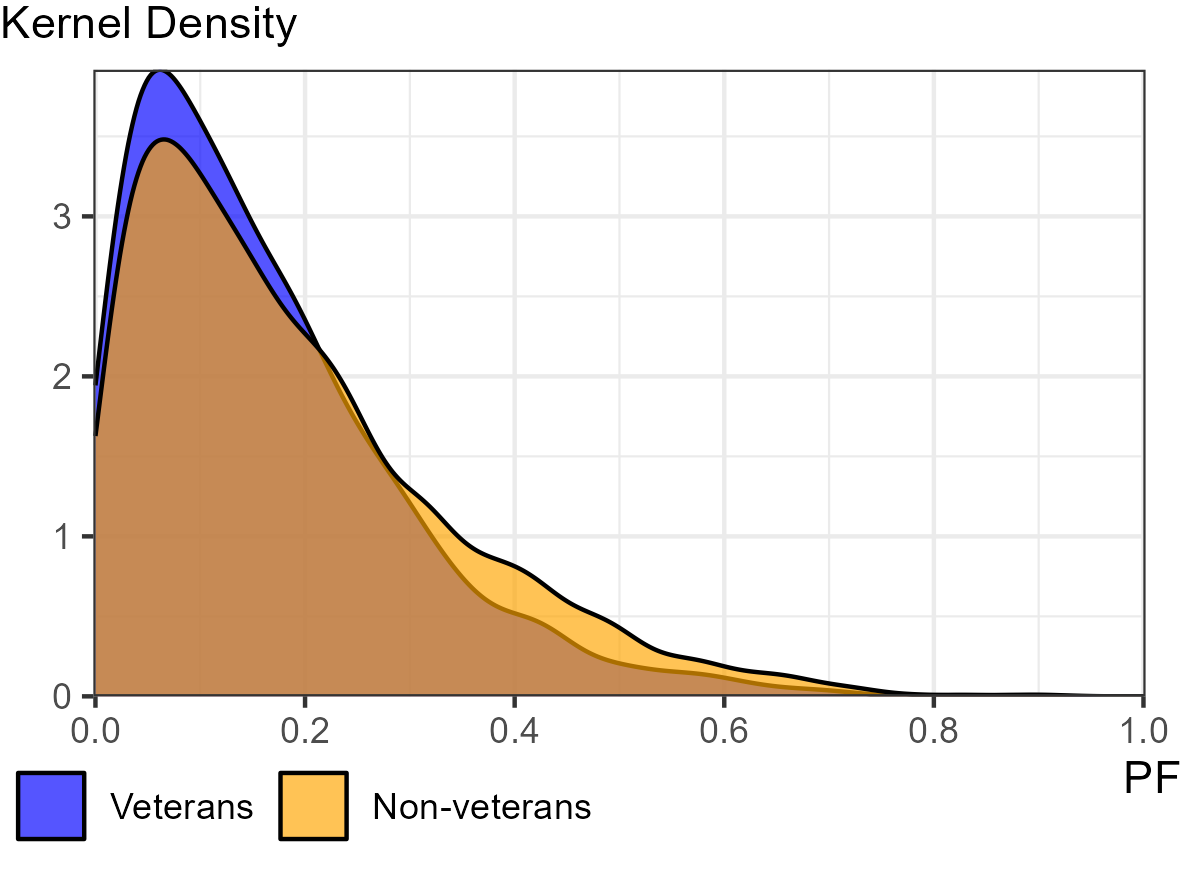}
    \end{subfigure}
    \label{fig:pfs-dist}
    \justify
    \footnotesize    
    \textbf{Note:}
    These figures show the estimated distribution of the PFI among PSID households headed by veterans or non-veterans.
    Panel A uses all observations for all observed years of  households in the PSID panel; Panel B averages the PFI for each man's household across observed years, and then plots the distribution of the man-specific intertemporal mean.
\end{figure}

The resulting distribution of PFI measures from the PSID sample of male household heads is depicted in \autoref{fig:pfs-dist}.

\subsubsection{Chronic Food Insecurity}
\label{appendix:chronic}

Using this PFS measure and the threshold probabilities, we construct two additional measures that capture food security dynamics, Total Food Insecurity (TotalFI) and Chronic Food Insecurity (CFI), as suggested by \cite{lee2024food}, to assess households' long-term food security status.
Both measures capture household-level long-term food security status based on the intertemporal mean PFI and deviations from the threshold probability.
TotalFI is the mean excess of household food insecurity (gap between PFI and the threshold probability) over time, capturing the average food insecurity status, and CFI is the period-mean PFI excess over time of households whose average PFI is above the average threshold probability (CFI is zero if the average PFI is no greater than the average threshold probability);
these definitions have been transformed from the equivalent food security definitions in \cite{lee2024food}.
We use a squared food insecurity aversion parameter, placing greater weight on larger food security shortfalls, so that 2 consecutive years food insecure followed by 1 secure year is worse than 2 separate spells of insecurity broken by a single year of security.

\begin{equation*}
\label{eqn:TotalFI}
    \text{TotalFI}_{i}(\alpha,PFS_{i,1},...,PFS_{i,t})=\frac{1}{T}\sum_{t=1}^{T}\left(1-\frac{\min (PFS_{i,t},\underline P_{h(i),t})}{\underline{P_{t}}}\right)^\alpha
\end{equation*}
\begin{equation*}
\label{eqn:CFI}
    CFI_{i}(\alpha,PFS_{i1},...,PFS_{i,t})=\left(1-\min\left[1,\frac{\sum_{t=1}^{T}PFS_{i,t}}{\sum_{t=1}^{T}\underline P_{h(i),t}}\right]\right)^\alpha
\end{equation*}

We classify a household's long-term food insecurity status into one of four mutually exclusive categories: persistently food secure, if both TotalFI and CFI equal zero; transiently food insecure, if CFI equals zero but PFS falls below the threshold probability in at least one period; chronically but not persistently food insecure, if CFI is non-zero but PFS does not fall below the threshold probability in every period; and chronically and persistently food insecure, if CFI is non-zero and PFS falls below the threshold probability in every period.

\begin{figure}[H]
    \caption{Rates of Chronic Food Insecurity, Veterans and Non-veterans 1980--2021.}
    \centering
    \singlespacing
    \includegraphics[width=\textwidth]{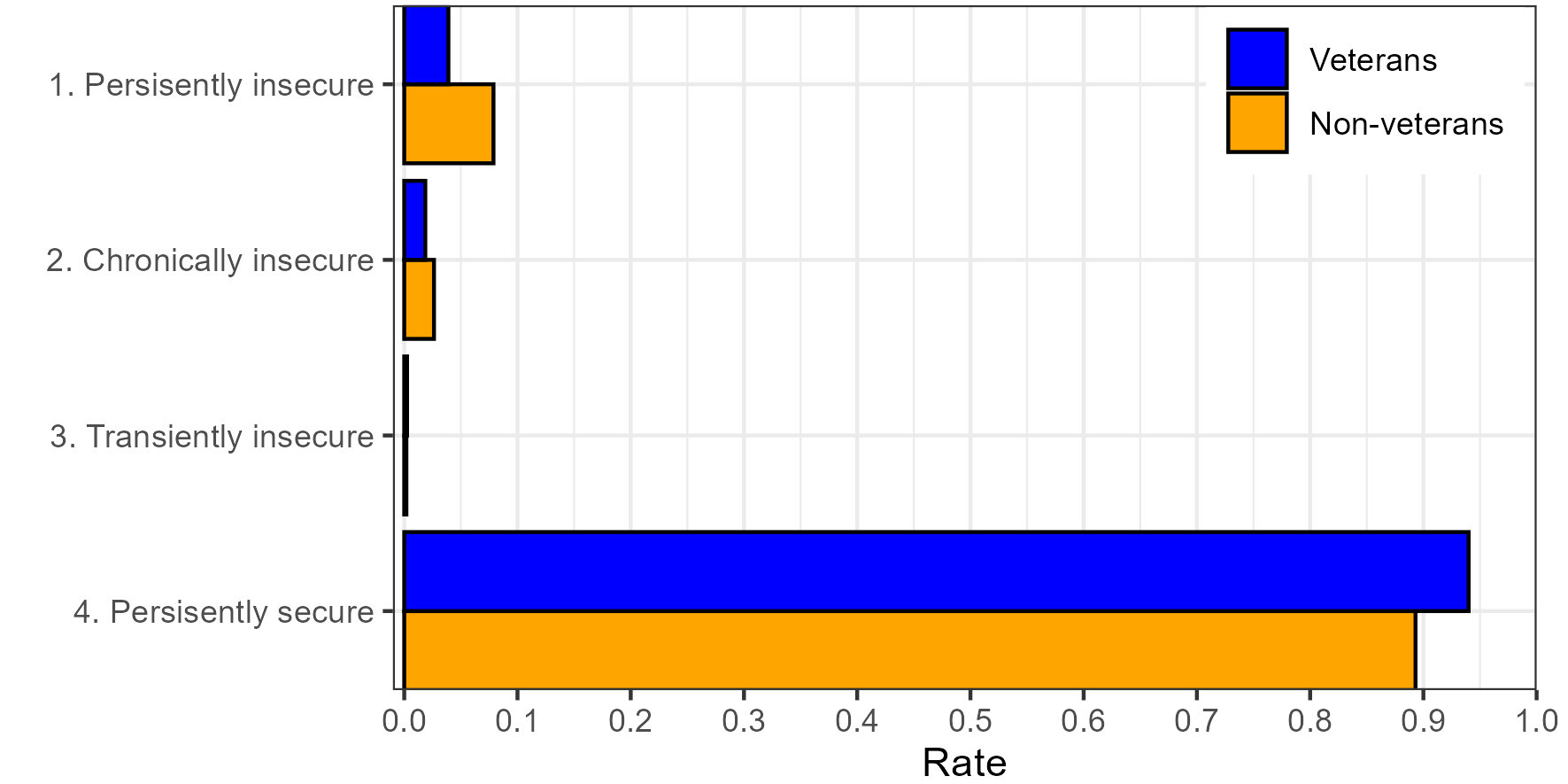}
    \label{fig:psid-chronic}
    \justify
    \footnotesize
    %\textbf{Note:}
\end{figure}

\autoref{fig:psid-chronic} shows households' long-term food security status by veteran status. While the share of persistently food-secure households is similar between groups, the share of persistently food-insecure households among non-veterans is twice that among veterans. These results indicate that veterans have lower rates of food insecurity, in terms of frequency, duration, and severity, over the study period relative to non-veterans.

\subsection{Vietnam-Era Draft Lottery}
\label{sec:appendix-subname}
This subsection provides additional evidence on the relationship between the Vietnam-era draft lottery and military service in the PSID.
The main analysis uses the binary conscription instrument constructed from the cohort-specific draft cutoffs.
As shown in \autoref{tab:firststage}, this instrument produces a substantially stronger first-stage than the more flexible categorical RSN specification.

\begin{table}[h!]
    \small
    \singlespacing
    \centering
    \caption{Draft-Lottery First-Stage, PSID Men Separated by Race.}
    \makebox[\textwidth][c]{
        \begin{tabular}{l c c c c c}
            \\[-1.8ex]\hline \hline \\[-1.8ex] 
            & \multicolumn{2}{c}{Pooled cohorts} & \multicolumn{3}{c}{By birth year} \\
            \cmidrule(lr){2-3} \cmidrule(lr){4-6}
            & 1945--1949 & 1950--1952 & 1950 & 1951 & 1952 \\
            \\[-1.8ex]\hline \\[-1.8ex]
            \textbf{Panel A. Race \(=\) White.} \\
            \input{sections/tables/firststage-binary-panelA.tex}
            & \\
            \textit{RSN categorical effects:} \\
            \input{sections/tables/firststage-categorical-panelA.tex}
            \\[-1.8ex]\hline \\[-1.8ex] 
            \textbf{Panel B. Race $=$ Non-white.} \\
            \input{sections/tables/firststage-binary-panelB.tex}
            & \\
            \textit{RSN categorical effects:} \\
            \input{sections/tables/firststage-categorical-panelB.tex}
            \\[-1.8ex]\hline \\[-1.8ex] 
        \end{tabular}
    }
    \label{tab:firststage-race}
    \justify
    \footnotesize
    \textbf{Note}: This table reports draft-eligibility effects and categorical RSN effects on veteran status, estimated in separate regressions for each column.
    Each observation is a separate man in the PSID, as the outcome does not vary in different years of the PSID panel.
    All models include a full set of dummies for birth year as controls;
    Standard Errors (SEs) are reported in parentheses.
    The empty entries are omitted values, because there are too few race-specific observations in each bin to estimate a coefficient (i.e., collinearity in the OLS regression).
\end{table}

\autoref{tab:firststage-race} reports the binary first-stage estimates separately for white and non-white men.
Previous studies estimate separate first-stages by race because compliance with the Vietnam-era draft differed between white and non-white men \citep{angrist1990lifetime,angrist2011long}.
In the PSID, the estimated effects of conscription on military service are similar across the two groups, although the race-specific estimates are less precise because of the smaller sample sizes.
We therefore use a common first-stage in the main analysis rather than estimating the effects of military service separately by race.

\begin{figure}[h!]
    \caption{Categorical RSN and Years of Military Service, by Birth Year.}
    \begin{subfigure}[c]{0.475\textwidth}
        \centering
        \caption{Military service relative to RSN group.}
        \includegraphics[width=\textwidth]{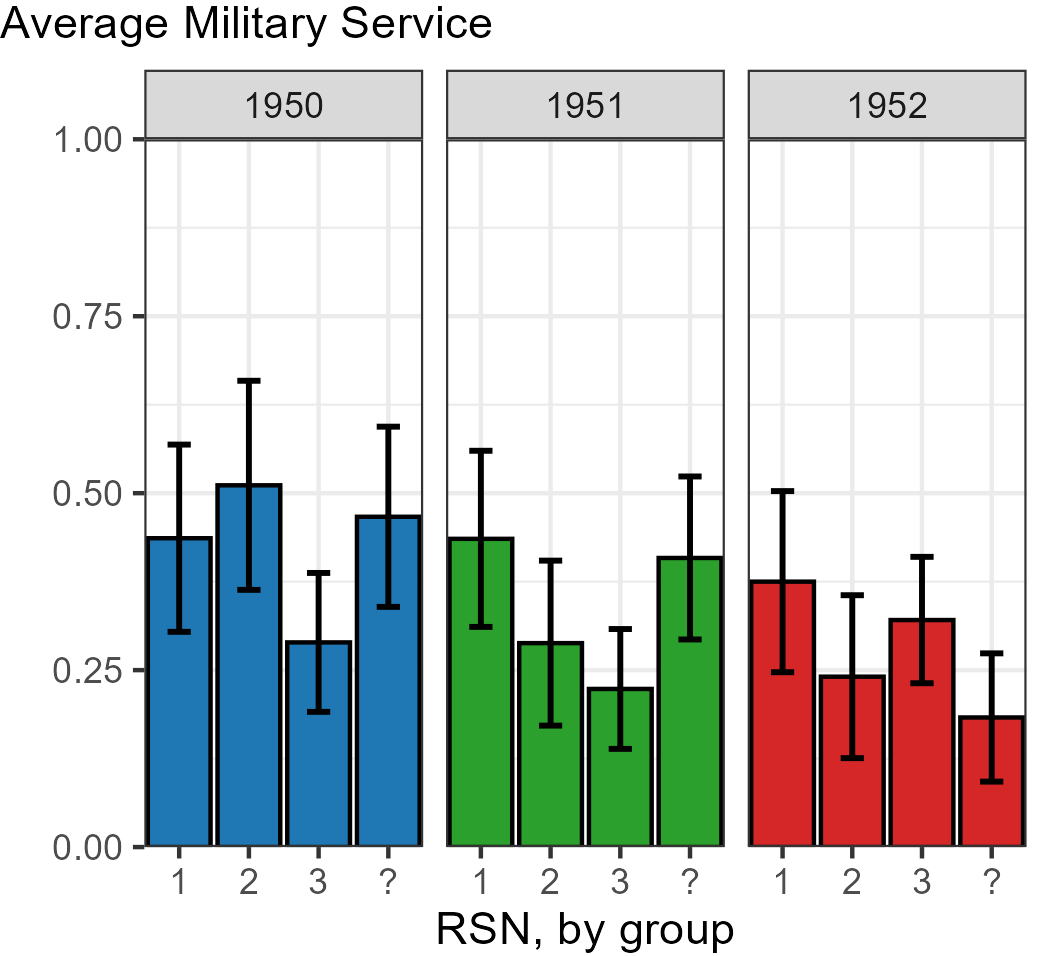}
    \end{subfigure}
    \begin{subfigure}[c]{0.475\textwidth}
        \centering
        \caption{Total years in military relative to Conscription.}
        \includegraphics[width=\textwidth]{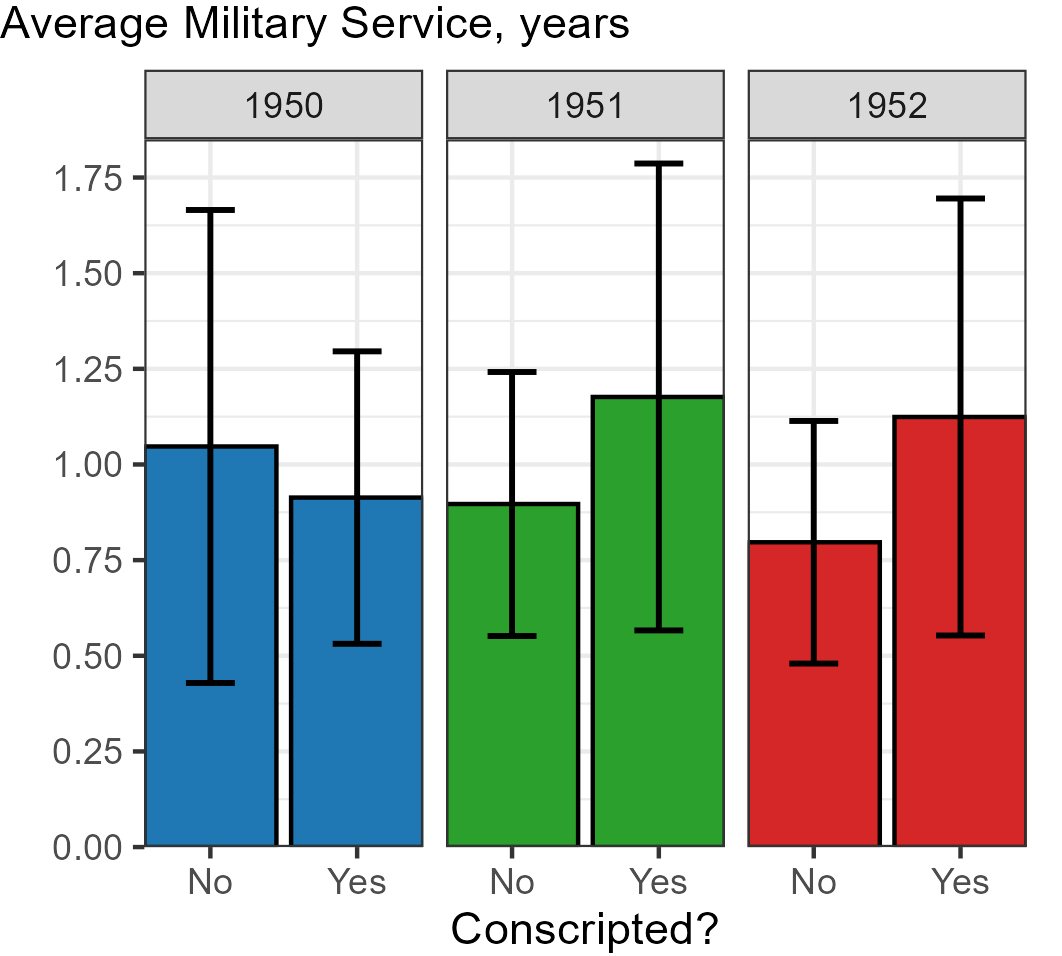}
    \end{subfigure}
    \justify
    \footnotesize    
    \textbf{Note:}
    These figures show military service relative to the categories of RSN values, and the average number of years served in the military whether conscripted or not.
    RSN group 1 means draft number 1--95, group 2 means 96--195, group 3 means 196--366, and ? means an unknown draft number.
\end{figure}

For completeness, \autoref{tab:firststage-race} also reports the relationship between the individual RSN categories and military service.
The categorical specification allows each RSN category to have a separate relationship with military service within each birth cohort.
However, its additional flexibility produces a substantially weaker joint first-stage than the binary specification based on the known conscription rules.
The categorical estimates are therefore useful for describing how military service varies across the underlying RSN categories, but we do not use the categorical specification to estimate the effects of military service on the outcomes studied in the paper.

The figures below provide additional descriptive evidence on the relationship among RSN category, conscription, and military service.
They show how the probability and duration of military service vary across RSN categories and birth cohorts.

\begin{figure}[h!]
    \centering
    \singlespacing
    \caption{Average Military Service Length among Military Veterans, by Birth Year.}
    \includegraphics[width=0.6\textwidth]{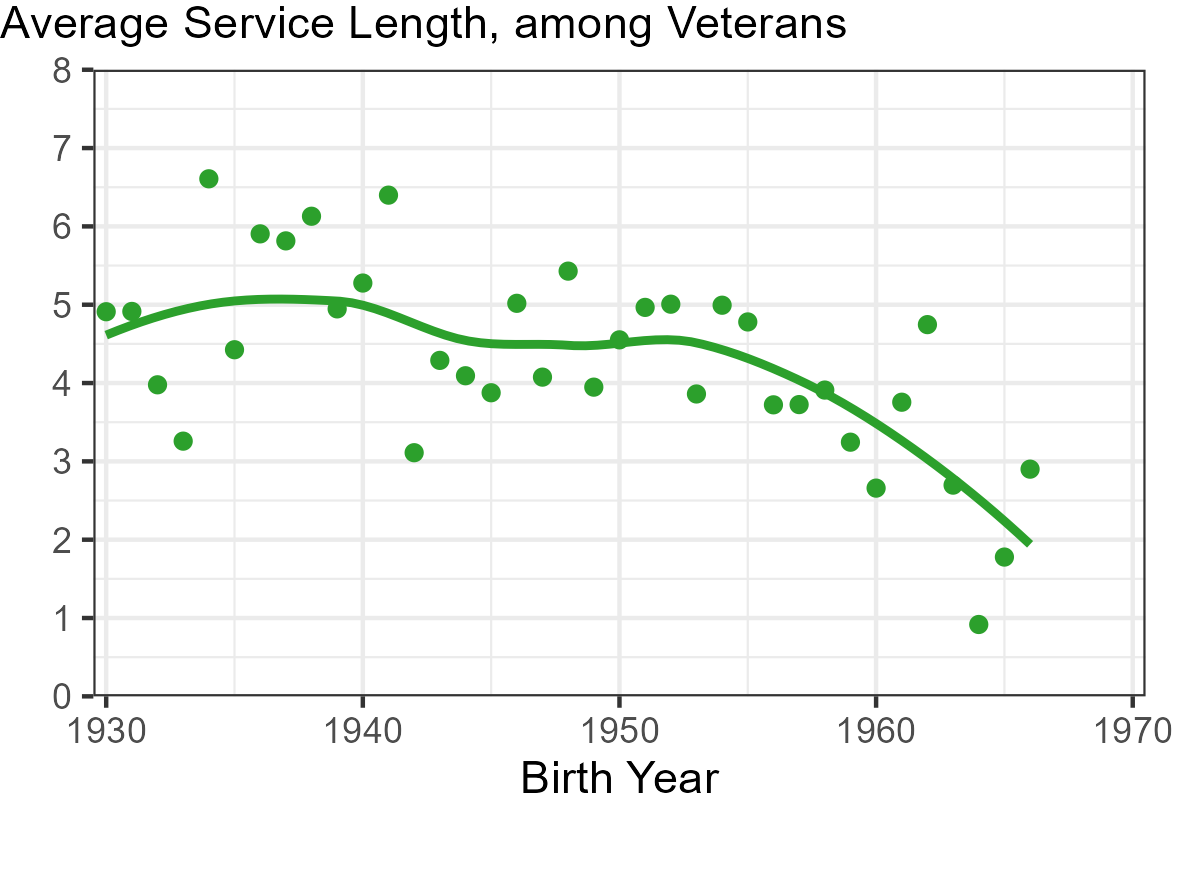}
    \label{fig:length-veterans-birth}
    \vspace{-1cm}
    \justify
    \footnotesize
    \textbf{Note:}
    This figure shows the mean number of years served in the military (among those with non-missing data on service years), separately for each birth year --- referring to veteran men household heads.
\end{figure}

\begin{figure}[h!]
    \centering
    \singlespacing
    \caption{Veteran Gain in Log Individual Income at Different Ages, Relative to Non-veterans.}
    \begin{subfigure}[b]{0.495\textwidth}
        \centering
        %\caption{OLS.}
        \includegraphics[width=\textwidth]{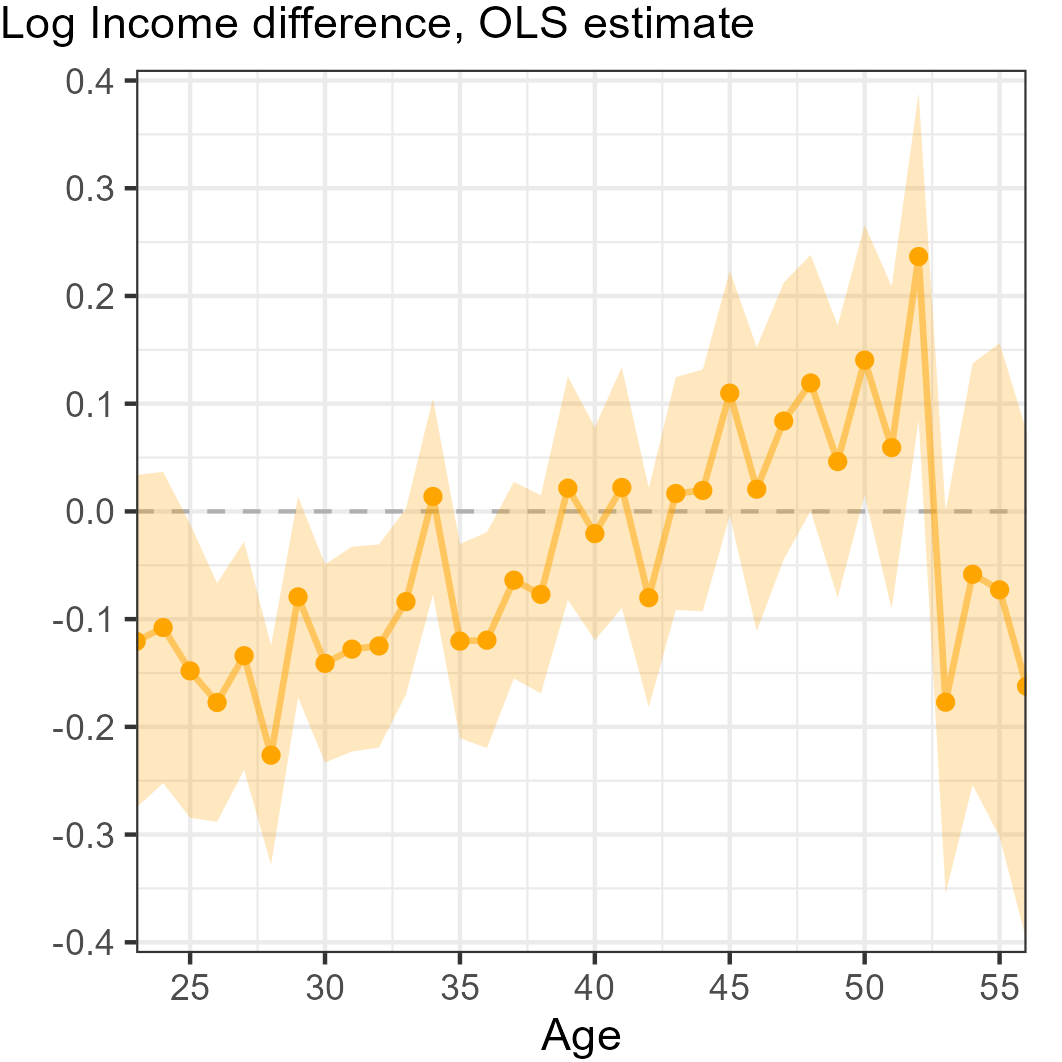}
        \label{fig:income-age-ols}
    \end{subfigure}
    \begin{subfigure}[b]{0.495\textwidth}
        \centering
        %\caption{IV.}
        \includegraphics[width=\textwidth]{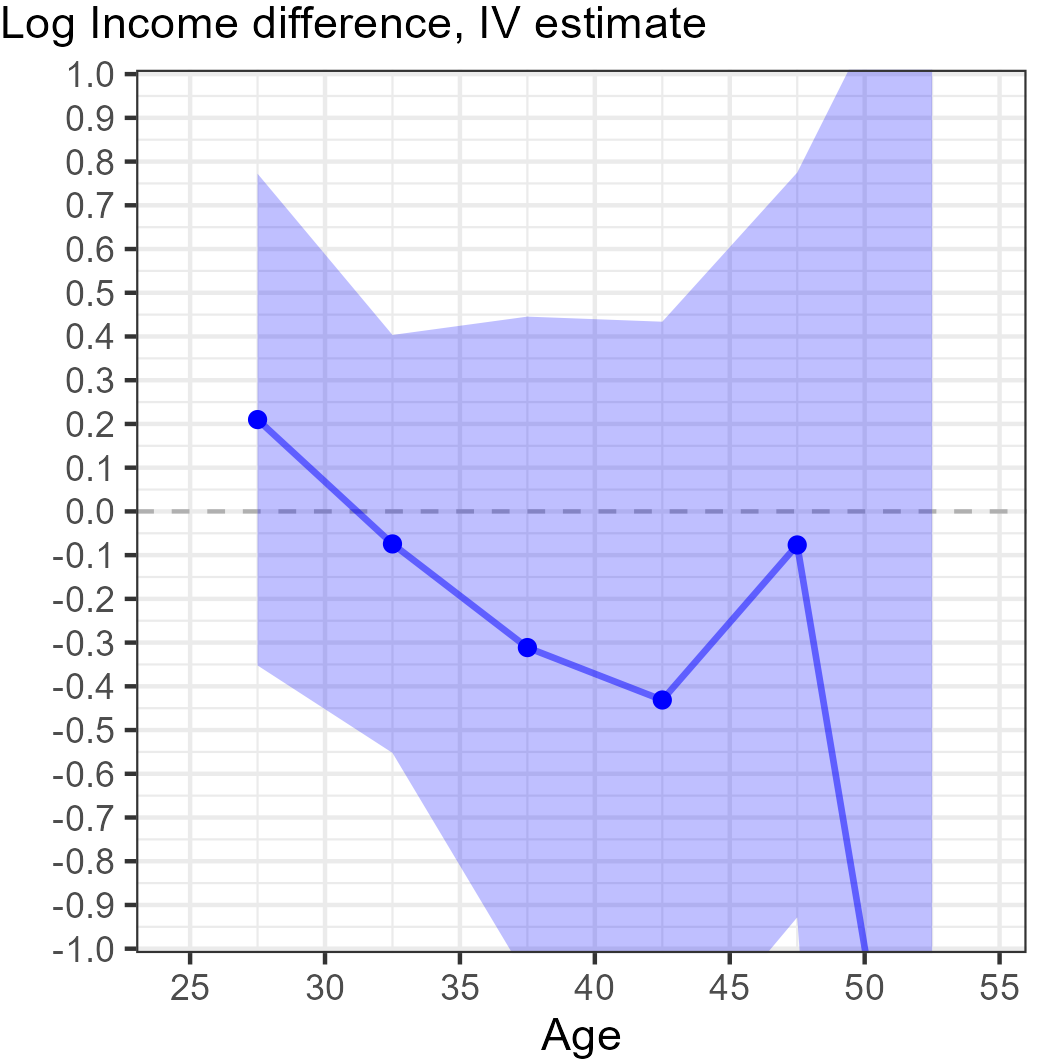}
        \label{fig:income-age-iv}
    \end{subfigure}
    \label{fig:income-age}
    \vspace{-1cm}
    \justify
    \footnotesize
    \textbf{Note:}
    These figures show the regression coefficient of veteran status on 
    log individual income, for PSID men at different ages.
    The OLS figure uses all PSID men born 1930--1970.
    The IV figures uses only men born 1950--1952, with the Vietnam-era draft lottery instrumenting for veteran status (using the binary IV specification).
    Each dot in the IV figure is a regression for men born in 5 year bins; the IV estimate for age $=$ 27.5 refers to the estimate among households where the head is aged 25--30.
\end{figure}

\autoref{fig:income-age} reports OLS and IV estimates of the effect of military service on log individual income at different ages.
The IV estimates use the binary conscription instrument employed throughout the main outcome analysis.
Both sets of estimates are noisy, particularly the IV estimates calculated within five-year age bins.
The OLS estimates show a negative association between veteran status and log individual income at younger ages, followed by associations that are close to zero or positive at older ages.
The IV estimates suggest a similar age profile, but are considerably less precise.
The income results are therefore suggestive rather than conclusive, but they are consistent with the possibility that any adverse effects of military service on individual income are concentrated at younger ages and fade later in adulthood.

\subsection{Two-Sample IV Estimates}
\label{sec:iv-binary}

As a sensitivity analysis, we also estimate the effect of military service using a two-sample IV estimator.
This approach follows the two-sample IV literature, which combines a reduced form estimated in one sample with a first-stage estimated in another \citep{arellano1992empirical,angrist1992effect,angrist1995split,inoue2010two}.
The approach is also close to the original Vietnam draft-lottery design in \citet{angrist1990lifetime}, and to recent applications using draft-lottery reduced forms with imported first-stage estimates \citep{bleemer2026vietnam}.

Let \(Z_i\) denote draft eligibility, \(D_i\) denote veteran status, and \(Y_{i,t}\) denote the outcome.
In the PSID sample, we estimate the reduced-form effect of draft eligibility on the outcome:
\begin{equation*}
    \label{eqn:twosample-reduced-form}
    Y_{i,t} = \theta Z_i + \vec\gamma' \vec X_{i,t} + u_{i,t},
\end{equation*}
where \(\vec X_{i,t}\) includes the same birth-year, and outcome-year controls used in the main IV specification.
Let \(\hat\theta\) denote the estimated reduced-form effect, with standard error \(\hat\sigma_{\theta}\).

We then combine this reduced form with an externally estimated first-stage from \citet{angrist2011schooling}, using 2000 Census data.
The 2000 Census data sampled roughly 1 in 6 of the US population, giving a population of roughly 796,000 men born in the years 1950--1952 to precisely estimate the effect of being conscripted to veteran service by the Vietnam-era draft lottery.

The corresponding first-stage equation is
\begin{equation*}
    \label{eqn:twosample-first-stage}
    D_i = \pi Z_i + \vec \phi' \vec X_i + \eta_i,
\end{equation*}
where \(\pi\) is the effect of draft conscription on veteran status.
Let \(\hat\pi_{\text{Census}}\) denote the first-stage estimate from 2000 Census data, with standard error \(\hat\sigma_{\text{Census}}\) --- both reported in Table 2, \cite[p.~101]{angrist2011schooling}.
The two-sample IV estimate of the effect of military service is then the Wald ratio
\begin{equation*}
    \label{eqn:twosample-wald}
    \hat\beta_{TSIV}
    =
    \frac{\hat\theta}{\hat\pi_{\text{Census}}}.
\end{equation*}
This is a two-sample two-stage least squares implementation: the first-stage relationship between draft eligibility and veteran status is estimated in the external Census sample, and the second-stage relationship is estimated in the PSID outcome sample.
In the just-identified Wald case, the estimator is algebraically the reduced form divided by the imported first-stage.
\citet{inoue2010two} show that the two-sample two-stage least squares implementation is generally more efficient than the original two-sample IV estimator.

\begin{table}[htbp!]
    \singlespacing
    \centering
    \vskip-0.75cm
    \caption{Effects of Military Service Among PSID Men, Conscription IV Estimates.}
    \footnotesize
    \makebox[\textwidth][c]{
        \begin{tabular}{l c c c c c c c c}
            \\[-1.8ex] \hline \hline \\[-1.8ex]
            & \multicolumn{6}{c}{IV, Men Born 1950--1952} \\
            \cmidrule(lr){2-7}
            & Non-veteran & First- & Reduced & Binary & Two-Sample & AR \\
            & Mean        & stage  & form    & IV     & IV         & 95\% CI\\
            & [Obs no.]   \\
            \\[-1.8ex]\hline \\[-1.8ex]
            \textbf{Panel A. Food Insecurity} \\
            \input{sections/tables/psid-twosample-crossA.tex}
            \\[-1.8ex]\hline \\[-1.8ex]
            \textbf{Panel B. Food Insecurity Persistence} \\
            \input{sections/tables/psid-twosample-crossB.tex}
            \\[-1.8ex]\hline \\[-1.8ex]
            %\textbf{Panel C. Welfare Participation} \\
            %\input{sections/tables/psid-twosample-crossC.tex}
            %\\[-1.8ex]\hline \\[-1.8ex]
            \textbf{Panel C. Demographic and Labor} \\
            \input{sections/tables/psid-twosample-crossD.tex}
            \\[-1.8ex]\hline \\[-1.8ex]
            Birth year controls?
                & Yes & & & Yes \\
            \hline
        \end{tabular}
    }
    \label{tab:psid-twosample-cross}
    \justify
    \footnotesize
    \vskip-0.25cm
    \textbf{Note}: This table shows the regression estimates of military veteran status on various outcomes using binary conscription as an instrument for military service.
    Observation counts are reported in square brackets, and standard errors are reported in parentheses and are clustered by birth year.
    The first-stage column reports estimates of the effect of draft conscription on veteran status.
    The reduced-form and binary IV estimates are calculated using the PSID.
    The two-sample IV column imports first-stage estimates calculated using the 2000 Census by \cite{angrist2011schooling}.
    The Anderson--Rubin column reports 95\% confidence intervals for the binary IV estimates that are robust to weak identification.
    Household food spending refers to monthly spending in 2023 US dollars, and the income variables are log transformed in the regressions.
\end{table}

Because the first-stage F statistic for the preferred within-PSID specification is 10, the final column of \autoref{tab:psid-twosample-cross} reports Anderson--Rubin 95\% confidence intervals that are robust to weak identification.
For mean PFI, the Anderson--Rubin confidence interval is $[-0.27,-0.02]$.
For the share of years food secure, the confidence interval is $[-0.04,0.77]$, while the corresponding interval for the share of years food insecure is $[-0.76,0.04]$.
These weak-IV-robust confidence intervals provide no evidence that military service adversely affected food insecurity, although the intervals for the binary food security classifications include effects in either direction.
The Anderson--Rubin confidence intervals for household food spending and household food spending per capita are also wide and include zero, consistent with the imprecision of the conventional IV estimates for these outcomes.

To account for sampling uncertainty in both the PSID reduced form and the imported first-stage, we use the delta method.
Assuming zero covariance between the estimated reduced form and the imported first-stage, the variance of \(\hat\beta_{TSIV}\) is
\begin{equation*}
    \label{eqn:twosample-var}
    \widehat{\operatorname{Var}}\left[\hat\beta_{TSIV}\right]
    =
    \frac{\hat\sigma_{\theta}^{2}}{\hat\pi^{2}_{\text{Census}}}
    +
    \frac{\hat\theta^{2}\hat\sigma_{\text{Census}}^{2}}{\hat\pi^{4}_{\text{Census}}}.
\end{equation*}
The reported SE is therefore
\begin{equation*}
    \label{eqn:twosample-se}
    \hat\sigma_{\beta}
    =
    \sqrt{
        \left[
            \frac{\hat\sigma_{\theta}}{\hat\pi_{\text{Census}}}
        \right]^2
        +
        \left[
            \frac{\hat\theta\hat\sigma_{\text{Census}}}{\hat\pi^{2}_{\text{Census}}}
        \right]^2
    }.
\end{equation*}

The imported first-stage estimates are not constant by race.
\citet{angrist2011schooling} estimate that draft eligibility increased Vietnam-era veteran status by 14.5 percentage points among white men and 9.4 percentage points among non-white men in the 1950--1952 birth cohorts among data from the 2000 Census.
We therefore do not impose a common first-stage across white and non-white men. Instead, we estimate race-specific reduced forms in the PSID by interacting draft eligibility with an indicator for white and non-white household heads, and then divide each race-specific reduced form by the corresponding race-specific first-stage estimate. This yields race-specific two-sample IV estimates, \(\hat\beta_W = \hat\theta_W / \hat\pi_W\) and \(\hat\beta_{NW} = \hat\theta_{NW} / \hat\pi_{NW}\).

Because the IV estimand is local to individuals whose veteran status is shifted by draft eligibility, the pooled estimate is constructed as a complier-weighted average of the race-specific estimates.
Specifically, each race-specific estimate is weighted by the group's contribution to the total draft-induced change in veteran status: the group's weighted share in the PSID analytic sample multiplied by the corresponding imported first-stage. This adjustment matters because white men make up most of the PSID sample and also have a larger first-stage. A simple population-weighted average would average race-specific treatment effects by sample composition, whereas the IV estimate averages effects in proportion to the composition of draft-lottery compliers.
Standard errors for the pooled estimate are computed by applying the same zero-covariance approximation to the weighted sum of the race-specific estimates.

The two-sample IV estimates for the principal food insecurity outcomes are directionally consistent with the corresponding within-PSID IV estimates, although their magnitudes differ because the estimator replaces the PSID first stage with race-specific first-stage estimates from the 2000 Census.
Precision differs considerably across outcomes, with the greatest uncertainty appearing for household food spending.
The two-sample IV estimates for food spending are extremely imprecise because the PSID reduced-form estimates differ in sign between white and non-white men, while the estimator combines race-specific reduced forms with heterogeneous external first stages.
We nevertheless report these estimates for completeness and to maintain a consistent set of corroborating outcomes across the within-PSID and two-sample IV specifications.

\subsection{Partially Identified Population Average Effects, with Unobserved Veterans}
\label{sec:unobserved}
The main IV estimates identify the effect of military service among men who are observed as PSID household heads. This is the relevant estimand for the main analysis, because the PSID measures veteran status for household heads and then links that status to the household's food insecurity outcomes. However, this estimand may differ from the average effect of military service for the broader veteran population if military service affects whether men later become household heads, or if the effects of service are concentrated among veterans outside traditional household structures. This appendix formalizes the sensitivity exercise used in the main text. The exercise does not assume a model of selection into household headship. Instead, it decomposes the population average effect into the effect among observed household heads and the unknown effect among veterans who are not observed as household heads, and then asks how large the latter would have to be to overturn the near-zero estimates in the PSID.

Write $\beta_i$ as the individual causal effect for working-age veteran $i$.
The veteran population average effect, $\E{\beta_i}$, is unknown, because the PSID analysis estimates the causal effect only among household heads, $\Egiven{\beta_i}{\text{Head}}$.
However, we can compare how large the unobserved causal effects among veteran men who are not household heads, $\Egiven{\beta_i}{\text{Non-head}}$, would have to be to rule in significantly impactful effects of military service on measures of food insecurity.
\[ \E{\beta_i}
    = \Prob{\text{Head}} \Egiven{\beta_i}{\text{Head}}
    + \Prob{\text{Non-head}}
        \underbrace{\Egiven{\beta_i}{\text{Non-head}}}_{
            \text{Unknown}}. \]

This decomposition provides an approach to sensitivity analysis by imputing alternative values for the unknown non-head effect.
We use data from the March CPS to calculate the percent of veteran men born 1950--1952 who are household heads, $\Prob{\text{Head}} = 77\%$, and percent who are not $\Prob{\text{Non-head}} = 23\%$.
Lastly, we use the IV estimates of causal effects as our estimate of $\Egiven{\beta_i}{\text{Head}}$, and the corresponding confidence intervals.
The sensitivity analysis additionally assumes that the draft-complier effect is informative about the average effect among veteran household heads.
We impose this assumption only to calibrate how adverse the unobserved non-head effect would need to be to alter the population calculation.

\begin{figure}[H]
    \centering
    \singlespacing
    \caption{Sensitivity Analysis for Veterans who are Not Household Heads.}
    \begin{subfigure}[b]{0.495\textwidth}
        \centering
        \caption{Static PFS Estimate.}
        \includegraphics[width=\textwidth]{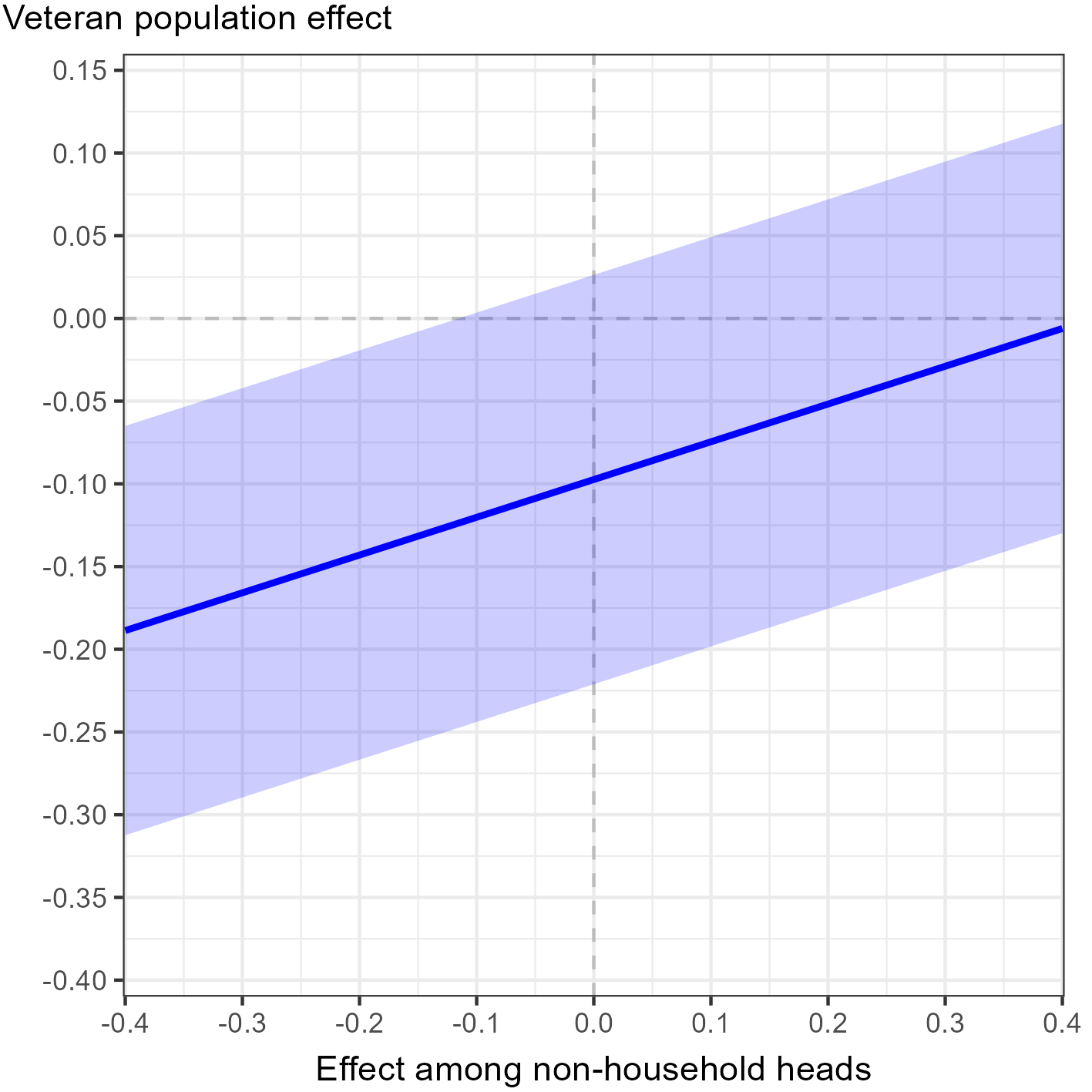}
        \label{fig:sens-pfs-static}
    \end{subfigure}
    \begin{subfigure}[b]{0.495\textwidth}
        \centering
        \caption{Dynamic PFS Estimate.}
        \includegraphics[width=\textwidth]{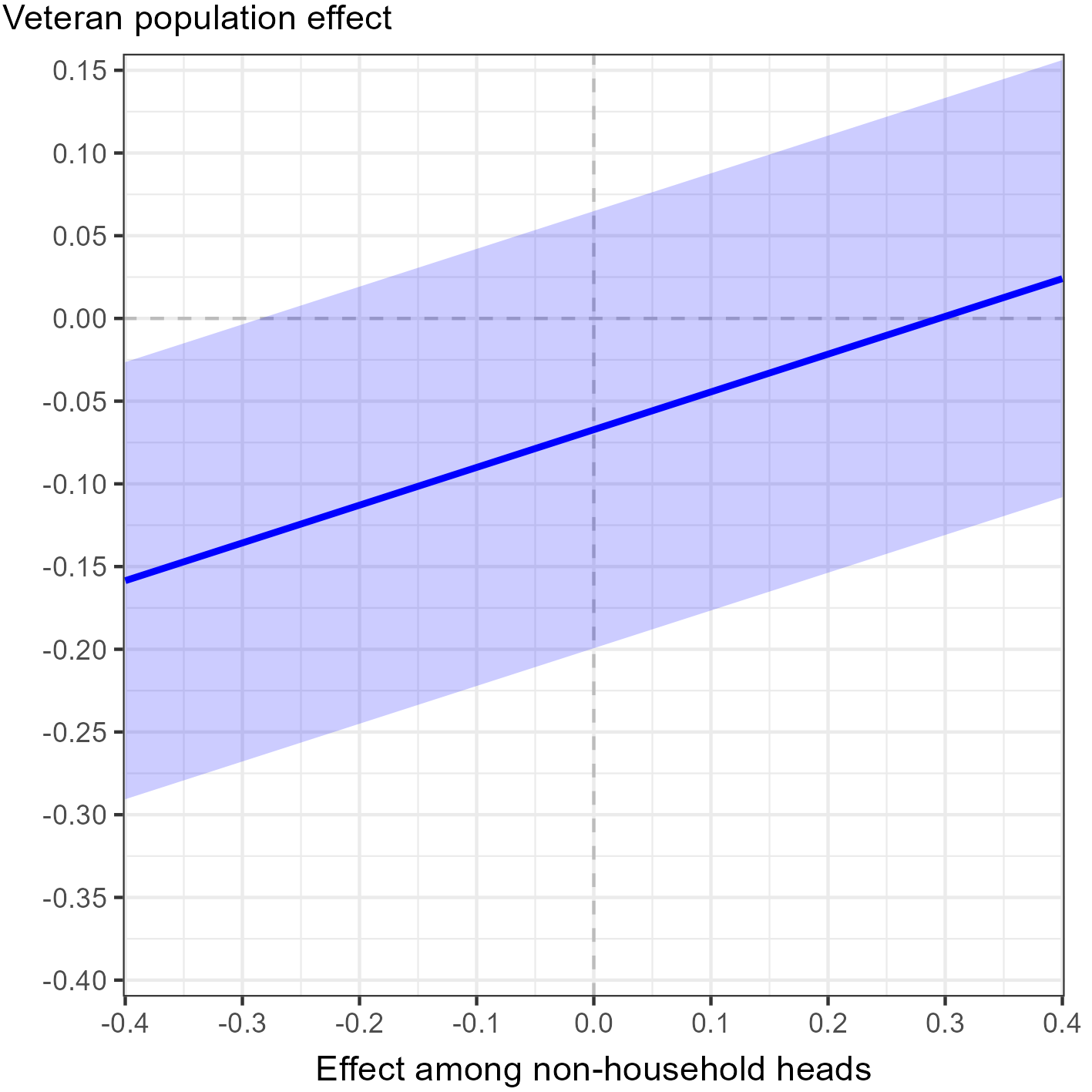}
        \label{fig:sens-pfs-dynamic}
    \end{subfigure}
    \label{fig:sens-pfs}
    \vspace{-1cm}
    \justify
    \footnotesize
    \textbf{Note:}
    These plots show the veteran population effect, $\E{\beta_i}$, relative to March CPS estimates of the percent of 1950--1952 born veterans who are household heads.
    Panel (a) uses the static estimated effect of military service on PFS in the static specification (\autoref{tab:psid-iv-cross}), and panel (b) using the dynamic specification (\autoref{tab:psid-iv-panel}).
    The orange region represents a 95\% confidence interval, using the standard errors on the IV estimates.
\end{figure}

The veteran population effect on PFS is not distinguishable from zero under the hypothetical scenario that non-household heads PFS is 30\% points worse as a result of military veteran-ship and up to 30\% points better.

\subsection{Population Average Effects, Accounting for Survival and Household Headship}
\label{appendix:selection-correction}
The main IV estimates in the paper identify the effect of military
service among men who are observed as household heads in the PSID.
This is the natural estimand for the PSID analysis, because veteran
status is observed only for household heads.
However, military service may also affect whether a man is observed in
this sample.
In particular, service may affect survival to working age, and,
conditional on survival, the probability that a man heads a household.
This appendix gives a model-based correction that combines the
PSID IV estimates with external information on survival and household
headship.

\begin{figure}[h!]
    \caption{Adult Food Security Survey Answers, PSID Men 1999--2021.}
    \centering
    \singlespacing
    \includegraphics[width=\textwidth]{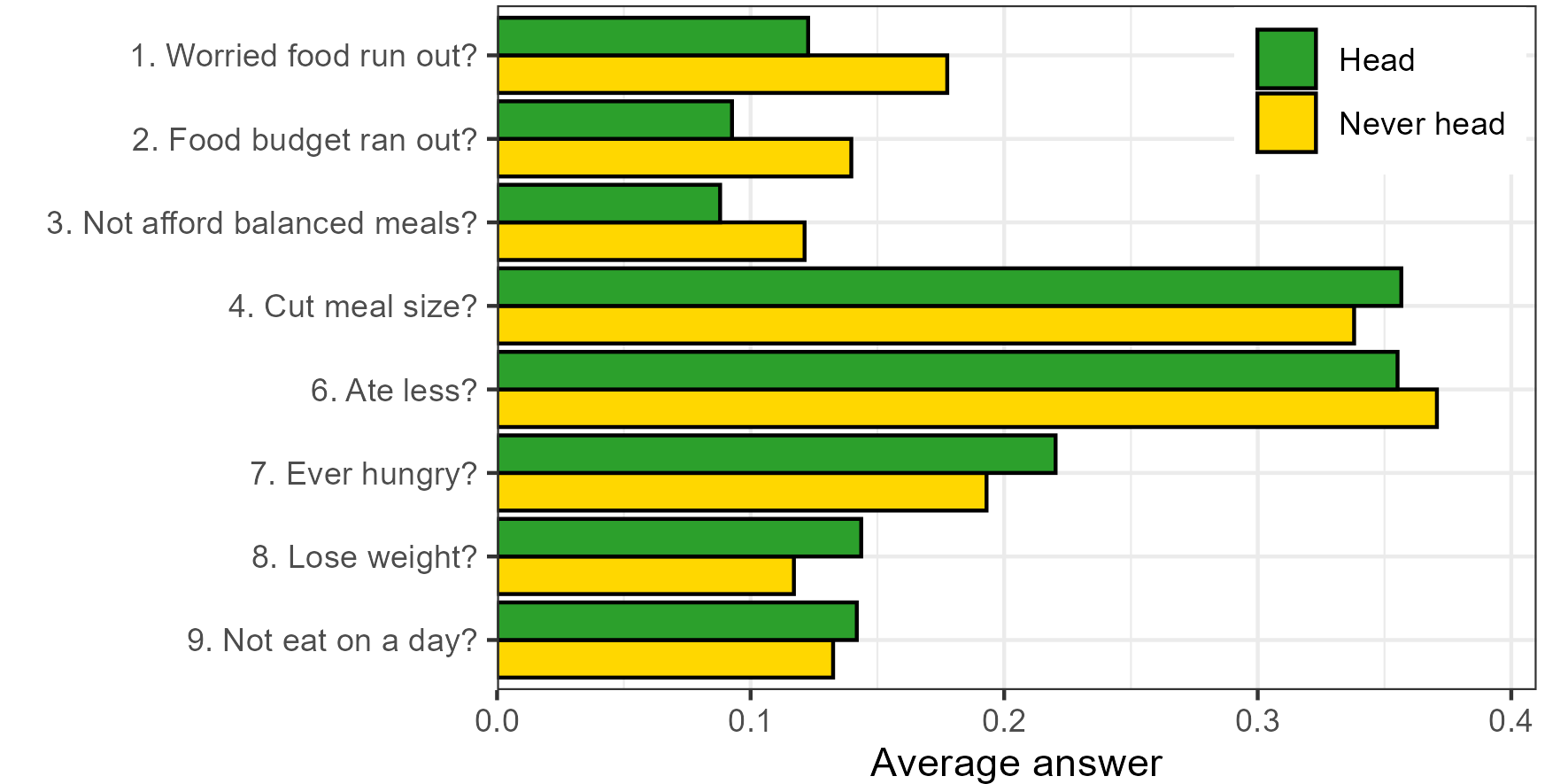}
    \label{fig:head-surveymean}
    \justify
    \footnotesize
    \textbf{Note:}
    This figure shows the average response to the eight questions of the food security survey, separately among men who ever head a PSID household (Heads) and men in the PSID never observed as a household head (Never heads).
    Veteran status is only observed among PSID men who head a household, so is missing among Never heads.
    0 in the outcome counts as answering ``never'' to the questions, 1 as ``sometimes'' or ``often.''
    The food security survey was asked of PSID households in the years 1999--2003 and 2015--2021, and we average answers for those years.
\end{figure}

The main text shows that men who are never observed as household heads have worse
food insecurity outcomes on the standard yearly measures of food insecurity, food
spending, and SNAP participation (\autoref{fig:head-food-trends}).
The same pattern appears when we classify men using the chronic food insecurity
measures developed by \cite{lee2024food}.
Never-heads are less likely to be persistently food secure, and more likely to be classified as chronically or persistently food insecure, than men who ever head a PSID household.
This reinforces the motivation for the selection sensitivity exercise: the group
omitted from the household-head veteran analysis is systematically more exposed to
long-run food insecurity.

\begin{figure}[H]
    \caption{Rates of Chronic Food Insecurity, Household Heads and Never Heads, 1980--2021.}
    \centering
    \singlespacing
    \includegraphics[width=\textwidth]{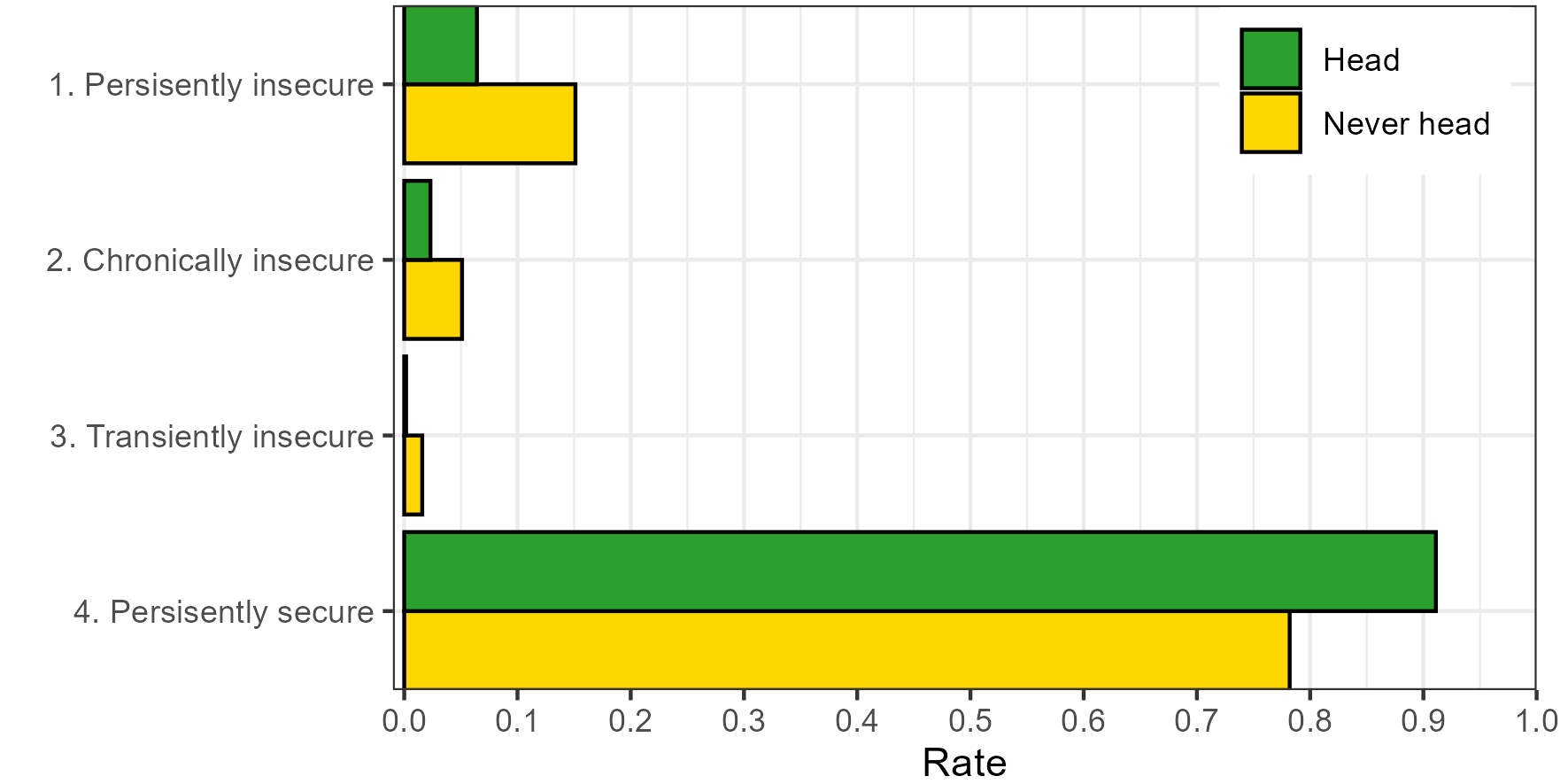}
    \label{fig:head-chronic}
    \justify
    \footnotesize
    \textbf{Note:}
    This figure shows the proportion of PSID men who ever head a household (Heads) and men who never do (Never heads) that are classified as persistently, chronically but not persistently, and transiently food insecure.
    The final category are those who are always observed as food secure; all definitions follow \cite[p.~1602]{lee2024food}.
\end{figure}

\autoref{fig:head-surveymean} and \autoref{fig:head-chronic} show that the
food insecurity disadvantage of never-heads appears both in the directly
reported food security survey questions and in the long-run food insecurity
classifications.
Never-heads report higher rates of worrying that food would run out, having
their food budget run out, and being unable to afford balanced meals; they also
appear worse off on several of the more severe survey items.
The chronic food-insecurity measures show the same pattern: never-heads are
less likely to be persistently food secure and more likely to be classified as
persistently or chronically food insecure.
These descriptive differences motivate the selection correction below, because
the group whose veteran status is not observed in the PSID is systematically
more food insecure than the household-head sample used in the main IV analysis.

\paragraph{The Model.}
Let $D_i \in \{0,1\}$ denote military service, and let
$S_i(d) \in \{0,1\}$ denote whether man $i$ would survive to the
relevant working-age observation window if treatment status were set to
$d \in \{0,1\}$ --- so that he could be observed in any data.
Let $H_i(d) \in \{0,1\}$ denote whether he would head a household,
conditional on survival --- so that he could have been observed as a
military veteran in the PSID data.

Define
\begin{equation*}
    A_i(d) \coloneq S_i(d)H_i(d),
\end{equation*}
where $A_i(d)=1$ indicates that man $i$ would both survive and be
observed as a household head under treatment status $d$.

Let $Y_i(d)$ denote the potential food-insecurity outcome under
treatment status $d$, and define the individual-level causal effect
\begin{equation*}
    \beta_i \coloneq Y_i(1)-Y_i(0).
\end{equation*}
The target parameter for the sensitivity exercise is the population
average effect,
\begin{equation*}
    \beta^{pop} \coloneq \E{\beta_i}.
\end{equation*}
Because food-insecurity outcomes are not observed for men who do not
survive to the relevant observation window, this parameter should be
interpreted as a model-informed population effect after imputing the
food insecurity consequences of selection out of the observed
household-head sample, rather than as a purely design-based effect
observed for all men.

The main PSID IV estimate identifies the effect among men who survive
and become household heads,
\begin{equation*}
    \beta_A
        \coloneq
        \Egiven{\beta_i}{A_i=1}
        =
        \Egiven{\beta_i}{S_i=1,H_i=1}.
\end{equation*}
The goal of this appendix is not to replace that design-based estimate,
but to ask how much it changes after imposing a simple model for the
missing selection margins.

Define the treatment effect of military service on survival as
\begin{equation*}
    \Delta_S
        \coloneq
        \Prob{S_i(1)=1}
        -
        \Prob{S_i(0)=1},
\end{equation*}
and define the treatment effect of military service on household
headship among survivors as
\begin{equation*}
    \Delta_{H\mid S}
        \coloneq
        \Probgiven{H_i(1)=1}{S_i(1)=1}
        -
        \Probgiven{H_i(0)=1}{S_i(0)=1}.
\end{equation*}
Let
\begin{equation*}
    \bar S
        \coloneq
        \frac{
            \Prob{S_i(1)=1}
            +
            \Prob{S_i(0)=1}
        }{2}
\end{equation*}
denote the average survival rate across treatment states, and let
\begin{equation*}
    \bar H
        \coloneq
        \frac{
            \Probgiven{H_i(1)=1}{S_i(1)=1}
            +
            \Probgiven{H_i(0)=1}{S_i(0)=1}
        }{2}
\end{equation*}
denote the average household-headship rate among survivors across
treatment states.

The use of $\bar S$ and $\bar H$ gives a symmetric decomposition of the
change in the product $S_dH_d$.
It does not impose that the relevant survival or headship rate is
literally halfway between the treated and untreated states.
Rather, it allocates the interaction between the survival and headship
margins equally across the two components.

The implied effect of military service on the probability of being
observed as a surviving household head is
\begin{align}
    \Delta_A
        &\coloneq
        \Prob{A_i(1)=1}
        -
        \Prob{A_i(0)=1}
        \nonumber \\
        &=
        \Delta_{H\mid S}\bar S
        +
        \Delta_S\bar H.
    \label{eqn:selection-margin}
\end{align}
Thus, $\Delta_A$ summarizes the combined selection margin through which
military service changes representation in the PSID household-head
sample.

To connect this selection margin to the population average effect, let
$\Delta_{Y\mid A}$ denote the food-insecurity difference associated
with being outside rather than inside the observed household-head
sample:
\begin{equation}
    \Delta_{Y\mid A}
        \coloneq
        \Egiven{Y_i}{A_i=1,\text{ marginal}}
        -
        \Egiven{Y_i}{A_i=0,\text{ marginal}}.
    \label{eqn:missing-outcome-penalty}
\end{equation}
The word ``marginal'' emphasizes that this is not the average difference
between all household heads and all non-heads.
It is the food-insecurity difference for the men whose survival or
household-headship status is changed by military service.

Under this sign convention, the model-informed population effect is
\begin{equation}
    \beta^{pop}
        =
        \beta_A
        -
        \Delta_A\Delta_{Y\mid A}.
    \label{eqn:selection-corrected-effect}
\end{equation}
Equation~\eqref{eqn:selection-corrected-effect} says that the population
effect equals the effect among observed surviving household heads,
adjusted for the degree to which military service changes selection into
that sample and the food-insecurity difference associated with being
outside the sample.

We implement this correction as a sensitivity exercise;
% The correction is point identified only after imposing three additional assumptions.
after imposing these three assumptions, we show estimates of the model-informed population effect calibrated against relevant data sources.

\paragraph{Assumption 1: The draft-complier estimate is informative about the average effect among observed household heads.}

The draft-lottery IV estimate identifies a local average treatment
effect among draft compliers who survive and become household heads:
\begin{equation}
    \beta_{A,\mathcal C}
        \coloneq
        \Egiven{\beta_i}{A_i=1,\,i\in\mathcal C},
    \label{eqn:assumption-psid-iv}
\end{equation}
where $\mathcal C$ denotes the set of draft-lottery compliers,
$D_i(0)=0$ and $D_i(1)=1$.
The selection correction, however, requires an estimate of the average
effect among all observed surviving household heads,
\begin{equation*}
    \beta_A
        \coloneq
        \Egiven{\beta_i}{A_i=1}.
\end{equation*}
We therefore use the draft-complier estimate as informative about this
household-head average effect:
\begin{equation*}
    \beta_A
        \approx
        \beta_{A,\mathcal C}.
\end{equation*}
This is an external-validity assumption, but it is weaker than assuming
homogeneous treatment effects for all men.
We do not require the effect of military service to be identical across
the population.
Rather, we assume that the IV estimate among draft compliers who become
household heads is informative about the average effect among the
broader group of surviving household heads observed in the PSID.
This assumption is appropriate for a sensitivity exercise intended to
assess whether selection into survival and household headship materially
changes the main result.

\paragraph{Assumption 2: External data identify the selection margin.}

We use CPS data to estimate household headship among living veteran and
non-veteran men born 1950--1952:
\begin{equation*}
    H_1
        \coloneq
        \Probgiven{H_i=1}{D_i=1,S_i=1},
    \qquad
    H_0
        \coloneq
        \Probgiven{H_i=1}{D_i=0,S_i=1}.
\end{equation*}
The estimated headship rates are
\begin{equation*}
    \widehat H_1=0.7774
        \quad (\text{SE }=0.0228),
    \qquad
    \widehat H_0=0.7427
        \quad (\text{SE }=0.0145).
\end{equation*}
We therefore estimate
\begin{equation*}
    \widehat{\Delta}_{H\mid S}
        =
        \widehat H_1-\widehat H_0,
    \qquad
    \widehat{\bar H}
        =
        \frac{\widehat H_1+\widehat H_0}{2}.
\end{equation*}
Because the CPS surveys living respondents, these estimates identify
headship rates conditional on survival, but do not identify the effect
of military service on survival itself.

We calibrate the average mortality rate over ages 39--49 using the 1990
male period life table reported in the SSA male 1990 period life table (Life Tables for the United States Social Security Area 1900–2100, \citealt{bell2005lifetables}).
The life table reports $l_{39}=94{,}058$ men surviving to exact age 39
and $l_{50}=89{,}937$ surviving to exact age 50 from a synthetic cohort
of 100,000 births.
We therefore calculate the probability of death before exact age 50,
conditional on survival to exact age 39, as
\begin{equation}
    \widehat m
        =
        1-\frac{l_{50}}{l_{39}}
        =
        1-\frac{89{,}937}{94{,}058}
        =
        0.0438.
    \label{eqn:mortality-calibration}
\end{equation}
This interval closely matches the ages 39--49 mortality window used by
\citet{conley2012long}.
The corresponding average survival rate is
\begin{equation*}
    \widehat{\bar S}
        =
        1-\widehat m
        =
        0.9562.
\end{equation*}
The SSA estimate is a period-life-table calibration rather than a
cohort-specific survival estimate for men born 1950--1952.
We treat $\widehat m$ and $\widehat{\bar S}$ as fixed external inputs in
the baseline variance calculations.

We estimate the effect of military service on survival using the
draft-lottery mortality estimates reported by
\citet{conley2012long}.
Let
\begin{equation*}
    r
        \coloneq
        \Probgiven{Z_i=1}{M_i=1,\text{ male}},
\end{equation*}
where $Z_i$ denotes draft eligibility and $M_i$ denotes mortality during
the observation window.
Let $q$ denote the counterfactual draft-eligible share among decedents,
and define
\begin{equation*}
    \delta_{CH}
        \coloneq
        r-q.
\end{equation*}

Our preferred calibration uses Table 2 of
\citet{conley2012long}, which compares the draft-eligible share among
male decedents with the corresponding share among female decedents:
\begin{equation*}
    r=0.3755,
    \qquad
    q=0.3736,
    \qquad
    \widehat{\delta}_{CH}=0.0019
        \quad(\text{SE }=0.0027).
\end{equation*}
Under the assumption that the female decedent distribution provides the
counterfactual eligibility distribution for men in the absence of a
draft effect, the implied reduced-form effect of draft eligibility on
mortality is
\begin{equation*}
    \widehat{\Delta}^{RF}_M
        =
        \widehat{\delta}_{CH}
        \frac{\widehat m}
             {q(1-q)}.
\end{equation*}
Because survival is one minus mortality, the corresponding reduced-form
effect on survival has the opposite sign:
\begin{equation*}
    \widehat{\Delta}^{RF}_S
        =
        -\widehat{\Delta}^{RF}_M.
\end{equation*}

Let $\widehat{\pi}_D$ denote the PSID first-stage effect of draft
eligibility on military service.
We use
\begin{equation*}
    \widehat{\pi}_D=0.1233
        \quad(\text{SE }=0.039).
\end{equation*}
The resulting effect of military service on survival is
\begin{equation}
    \widehat{\Delta}_S
        = \left( \frac{
        \widehat{\delta}_{CH} }{\widehat{\pi}_D} \right)
        \left( \frac{ -m}{q(1-q)} \right).
    \label{eqn:survival-effect}
\end{equation}
This construction treats the \cite{conley2012long} estimate as a reduced-form
draft-eligibility effect and converts it into an effect of military
service by dividing by the first-stage effect of draft eligibility on
service.

As an alternative calibration, we use Table 3 of
\citet{conley2012long}, which compares the observed draft-eligible share
among male decedents with the theoretical share implied by the
birth-year-specific draft cutoffs:
\begin{equation*}
    r=0.3755,
    \qquad
    q=0.3788,
    \qquad
    \widehat{\delta}_{CH}=-0.0033
        \quad(\text{SE }=0.0022).
\end{equation*}
The two \cite{conley2012long} specifications use different counterfactuals and
produce small survival effects on opposite sides of zero.
We use the male--female comparison as the main specification and report
the theoretical-cutoff comparison as an appendix robustness exercise.

\paragraph{Assumption 3: The headship gradient approximates the missing outcome difference.}

The food-insecurity difference for men shifted out of the observed
household-head sample is approximated by the observed food-insecurity
gradient between men who are never observed as household heads and men
who are:
\begin{equation}
    \Delta_{Y\mid A}
        =
        \Egiven{Y_i}{A_i=1}
        -
        \Egiven{Y_i}{A_i=0}.
    \label{eqn:assumption-outcome-penalty}
\end{equation}
In practice, we estimate this gradient after controlling for the same
variables as in \autoref{sec:empirics}, including birth year, and age.

This is the strongest modelling assumption in the exercise.
The observed difference between household heads and never-heads is not,
by itself, a causal effect of household headship.
Never-heads are negatively selected in ways that may be unrelated to
military service, so the observed gradient may overstate the
food-insecurity difference for men marginally displaced from household
headship.
At the same time, military service may impose specific compounding
disadvantages, such as combat-related health shocks, disrupted
labor-market attachment, or social isolation, that leave veteran
non-heads worse off than the average never-head.
The direction of the resulting approximation error is therefore
ambiguous.

\paragraph{Estimation and Uncertainty.}

Combining the external inputs gives the estimated selection margin
\begin{equation}
    \widehat{\Delta}_A
        =
        \widehat{\bar S}
        \left(
            \widehat H_1-\widehat H_0
        \right)
        +
        \widehat{\Delta}_S
        \left(
            \frac{\widehat H_1+\widehat H_0}{2}
        \right).
    \label{eqn:plugin-selection-margin}
\end{equation}
We calculate uncertainty using the delta method.
For the survival effect, we treat $\widehat m$ and $q$ as fixed and
assume that the \cite{conley2012long} estimate is independent of the PSID
first-stage estimate:
\begin{align}
    \widehat{\text{Var}}
    \left(
        \widehat{\Delta}_S
    \right)
    ={}&
    \left[
        \frac{
            \widehat m
        }{
            \widehat{\pi}_Dq(1-q)
        }
    \right]^2
    \widehat{\text{Var}}
    \left(
        \widehat{\delta}_{CH}
    \right)
    \nonumber \\
    &+
    \left[
        \frac{
            \widehat{\Delta}_S
        }{
            \widehat{\pi}_D
        }
    \right]^2
    \widehat{\text{Var}}
    \left(
        \widehat{\pi}_D
    \right).
    \label{eqn:survival-effect-variance}
\end{align}

For the combined selection margin, we treat $\widehat{\bar S}$ as fixed
and assume that $\widehat H_1$, $\widehat H_0$, and
$\widehat{\Delta}_S$ are mutually independent:
\begin{align}
    \widehat{\text{Var}}
    \left(
        \widehat{\Delta}_A
    \right)
    ={}&
    \left(
        \widehat{\bar S}
        +
        \frac{\widehat{\Delta}_S}{2}
    \right)^2
    \widehat{\text{Var}}
    \left(
        \widehat H_1
    \right)
    \nonumber \\
    &+
    \left(
        -\widehat{\bar S}
        +
        \frac{\widehat{\Delta}_S}{2}
    \right)^2
    \widehat{\text{Var}}
    \left(
        \widehat H_0
    \right)
    \nonumber \\
    &+
    \widehat{\bar H}^{\,2}
    \widehat{\text{Var}}
    \left(
        \widehat{\Delta}_S
    \right).
    \label{eqn:selection-margin-variance}
\end{align}
These assumptions are natural here because the headship rates are
estimated from CPS data, the mortality composition estimate comes from
national death records, and the first-stage and food insecurity outcomes
are estimated using the PSID.

Define the estimated selection correction as
\begin{equation}
    \widehat C
        \coloneq
        -
        \widehat{\Delta}_A
        \widehat{\Delta}_{Y\mid A}.
    \label{eqn:selection-correction}
\end{equation}
Assuming that the estimated selection margin and food-insecurity
gradient are independent, its variance is
\begin{equation}
    \widehat{\text{Var}}
    \left(
        \widehat C
    \right)
    =
    \widehat{\Delta}_{Y\mid A}^{\,2}
    \widehat{\text{Var}}
    \left(
        \widehat{\Delta}_A
    \right)
    +
    \widehat{\Delta}_A^{\,2}
    \widehat{\text{Var}}
    \left(
        \widehat{\Delta}_{Y\mid A}
    \right).
    \label{eqn:selection-correction-variance}
\end{equation}

The model-informed population estimate is
\begin{equation}
    \widehat{\beta}^{pop}
        =
        \widehat{\beta}_{PSID}
        +
        \widehat C
        =
        \widehat{\beta}_{PSID}
        -
        \widehat{\Delta}_A
        \widehat{\Delta}_{Y\mid A}.
    \label{eqn:plugin-selection-correction}
\end{equation}
Under the additional simplifying assumption that the main PSID IV
estimate is independent of the estimated selection correction, its
variance is
\begin{equation}
    \widehat{\text{Var}}
    \left(
        \widehat{\beta}^{pop}
    \right)
    =
    \widehat{\text{Var}}
    \left(
        \widehat{\beta}_{PSID}
    \right)
    +
    \widehat{\text{Var}}
    \left(
        \widehat C
    \right).
    \label{eqn:population-effect-variance}
\end{equation}
The PSID IV estimate and the estimated food-insecurity headship gradient
are constructed from overlapping PSID data, so this final independence
assumption is an approximation.
For this model-informed sensitivity exercise, we use the
zero-covariance calculation as a transparent approximation.

Identification of this model therefore comes from combining three objects: the
draft-lottery estimate among observed household-head compliers, external
estimates of the effect of military service on survival and household
headship, and the estimated food-insecurity gradient between men inside
and outside the observed household-head sample.
The resulting model's estimates should be interpreted as sensitivity analyses, rather than as a replacement for the main IV results.
Their purpose is to discipline the possible magnitude of selection bias
using the best available external information, not to claim that the
food insecurity outcomes of the missing population are directly observed.

\paragraph{Empirical Results.}
The preferred specification produces only small corrections to the main
IV estimates.
Using the male--female mortality comparison from Table 2 of
\cite{conley2012long}, the estimated combined selection margin is 0.03
(SE 0.03).
For mean PFI, the estimated headship gradient is $-0.07$ (SE 0.01),
so the implied correction rounds to zero (SE 0.00); the resulting
population estimate is $-0.12$ (SE 0.08), compared with the PSID IV
estimate of $-0.13$ (SE 0.08).
For the share of years food insecure, the corresponding gradient is -0.19 (SE 0.06), and the population estimate is -0.26 (SE 0.23), compared to the IV estimate of -0.27 (SE 0.23).
For the share of years food secure, the corresponding gradient is
$0.19$ (SE 0.06), and the population estimate is $0.26$
(SE 0.23), compared with an IV estimate of $0.27$ (SE 0.23).
The correction is also small for food spending: a headship gradient of
$-5.01$ dollars (SE 71.16) produces a correction of $0.16$ dollars
(SE 4.88), changing the total food-spending estimate from 128.83 dollars
(SE 114.14) to 128.98 dollars (SE 114.24), while the per-capita
correction is $0.74$ dollars (SE 2.91), leaving the population
estimate at 89.49 dollars (SE 84.07).
Thus, the headship gradients can be economically meaningful, but they
have little influence on the final estimates because they are multiplied
by a small estimated effect of military service on selection into the
surviving household-head sample.

\begin{table}[h!]
    \singlespacing
    \centering
    \caption{Model-Informed Population Effects, Accounting for Survival and Household Headship.}
    \small
    \makebox[\textwidth][c]{
        \begin{tabular}{l c c c c c}
            \\[-1.8ex]\hline \hline \\[-1.8ex]
            & PSID IV  & Headship & Selection & Selection  & Population \\
            & Estimate & Gradient & Margin    & Correction & Estimate \\
            & $\hat{\beta}_{PSID}$ 
            & $\widehat{\Delta}_{Y \mid A}$ 
            & $\widehat{\Delta}_{A}$ 
            & $-\widehat{\Delta}_{A}\widehat{\Delta}_{Y \mid A}$ 
            & $\hat{\beta}^{pop}$ \\
            \\[-1.8ex]\hline \\[-1.8ex] 
            \input{sections/tables/sensitivity-table-alt.tex}
            \\[-1.8ex]\hline \\[-1.8ex] 
        \end{tabular}
    }
    \label{tab:sensitivity-table-alt}
    \justify
    \footnotesize
    \textbf{Note}: This table reports model-informed population effects of military service after accounting for selection into survival and household headship.
    These estimates use the mortality estimates from Table 3 \cite{conley2012long}.
\end{table}

The alternative mortality calibration produces only a small adjustment
to the design-based IV estimates.
\autoref{tab:sensitivity-table-alt} reports each component of the
calculation separately: the PSID IV estimate among observed household
heads, the food insecurity gradient between heads and never-heads, the
combined survival and headship selection margin, the resulting selection
correction, and the model-informed population estimate.
This decomposition makes clear whether differences between the IV and
population estimates arise from a large food insecurity gradient, a large
selection margin, or both.

The alternative calibration implies a combined selection margin of
approximately 0.04 (SE 0.03).
Although several headship gradients are economically meaningful, they
are multiplied by this relatively small selection margin.
For mean PFI, the headship gradient is $-0.07$ (SE 0.01), generating
a correction that rounds to zero (SE 0.00); the resulting population
estimate is therefore $-0.12$ (SE 0.08), compared with a PSID IV
estimate of $-0.13$ (SE 0.08).
The same pattern holds for the share of years food secure and food
insecure.
The respective population estimates are 0.26 (SE 0.23) and $-0.26$
(SE 0.23), compared with IV estimates of 0.27 (SE 0.23) and
$-0.27$ (SE 0.23).

The largest corrections arise for food spending because these outcomes
have larger headship gradients in their original dollar units.
For total household food spending, a headship gradient of $-5.01$ dollars
(SE 71.16) produces a correction of $0.19$ dollars (SE 6.94), changing
the estimate from 128.83 dollars (SE 114.14) to 129.01 dollars
(SE 114.35).
For per-capita food spending, the corresponding correction is
$0.88$ dollars (SE 3.97), leaving the population estimate at 89.63
dollars (SE 84.12), compared with an IV estimate of 88.75 dollars
(SE 84.02).

Even where the headship gradient is sizable, the correction
remains small because the mortality literature and CPS headship
estimates imply that military service changes selection into the
observed household-head sample only modestly.
The alternative mortality calibration therefore leads to the same
substantive conclusion as the preferred calibration: accounting for
survival and household headship does not materially change the
design-based IV results.

%% file: sections/tables/firststage-binary-panelA.tex
% latex table generated in R 4.4.1 by xtable 1.8-8 package
% Wed Jul 22 16:57:44 2026
 Draft eligibility effect & 0.011 & 0.127 & 0.139 & 0.136 & 0.118 \\ 
    & (0.039) & (0.051) & (0.093) & (0.091) & (0.095) \\ 
  F statistic & 0.1 & 6.3 & 2.2 & 2.2 & 1.6 \\ 
  Observations & 735 & 470 & 139 & 160 & 171 \\ 
  

%% file: sections/tables/firststage-categorical-panelA.tex
% latex table generated in R 4.4.1 by xtable 1.8-8 package
% Wed Jul 22 16:57:45 2026
 RSN 1--95 & 0.581 & 0.058 & -0.018 & 0.02 &  \\ 
    & (0.039) & (0.067) & (0.149) & (0.113) &  \\ 
  RSN 96--195 & 0.532 &  & 0.027 & -0.076 & -0.142 \\ 
    & (0.036) &  & (0.14) & (0.123) & (0.116) \\ 
  RSN 196--366 & 0.551 & -0.087 & -0.196 & -0.22 & -0.085 \\ 
    & (0.031) & (0.059) & (0.134) & (0.106) & (0.103) \\ 
  

%% file: sections/tables/firststage-binary-panelB.tex
% latex table generated in R 4.4.1 by xtable 1.8-8 package
% Wed Jul 22 16:57:45 2026
 Draft eligibility effect & 0.041 & 0.11 & 0.141 & 0.167 & 0.061 \\ 
    & (0.055) & (0.062) & (0.119) & (0.112) & (0.115) \\ 
  F statistic & 0.6 & 3.2 & 1.4 & 2.2 & 0.3 \\ 
  Observations & 363 & 346 & 104 & 126 & 116 \\ 
  

%% file: sections/tables/firststage-categorical-panelB.tex
% latex table generated in R 4.4.1 by xtable 1.8-8 package
% Wed Jul 22 16:57:45 2026
 RSN 1--95 &  & 0.082 & 0.199 &  &  \\ 
    & & (0.085) & (0.15) &  &  \\ 
  RSN 96--195 & 0.024 &  & 0.297 & -0.225 & -0.117 \\ 
    & (0.083) &  & (0.189) & (0.137) & (0.155) \\ 
  RSN 196--366 & 0.004 & -0.013 &  & -0.189 & 0.009 \\ 
    & (0.077) & (0.08) &  & (0.126) & (0.133) \\ 
  

%% file: sections/tables/psid-twosample-crossA.tex
% latex table generated in R 4.4.1 by xtable 1.8-8 package
% Wed Jul 22 17:02:03 2026
 PFI                                 & 0.17   & 0.12   & -0.02   & -0.13    & -0.3       & -0.27, -0.02   \\ 
                                      & [816]  & (0.02) & (0.01)  & (0.08)   & (0.03)     &                \\ 
  Food secure, percent years          & 0.91   & 0.12   & 0.03    & 0.28     & 0.44       & -0.04, 0.77    \\ 
                                      & [816]  & (0.02) & (0.02)  & (0.23)   & (0.08)     &                \\ 
  Food insecure, percent years        & 0.09   & 0.12   & -0.03   & -0.27    & -0.44      & -0.76, 0.04    \\ 
                                      & [816]  & (0.02) & (0.02)  & (0.23)   & (0.08)     &                \\ 
  Household food spending             & 375.93 & 0.12   & 15.87   & 128.66   & 188.73     & -47.63, 324.61 \\ 
                                      & [816]  & (0.02) & (13.52) & (114.18) & (19011.04) &                \\ 
  Household food spending, per capita & 183.9  & 0.12   & 10.87   & 88.11    & 105.11     & -58.04, 242.18 \\ 
                                      & [816]  & (0.02) & (10.22) & (84.44)  & (11252.16) &                \\ 
  SNAP, ever participated             & 0.2    & 0.12   & -0.02   & -0.16    & -0.47      & -0.42, 0.21    \\ 
                                      & [816]  & (0.02) & (0.02)  & (0.18)   & (0.28)     &                \\ 
  SNAP, percent years participated     & 0.07   & 0.12   & -0.01   & -0.11    & -0.19      & -0.44, 0.13    \\ 
                                      & [816]  & (0.02) & (0.02)  & (0.14)   & (0.1)      &                \\ 
  

%% file: sections/tables/psid-twosample-crossB.tex
% latex table generated in R 4.4.1 by xtable 1.8-8 package
% Wed Jul 22 17:02:22 2026
 Chronic and persistent     & 0.06  & 0.12   & -0.01  & -0.11  & -0.21  & -0.69, 0.23  \\ 
                             & [816] & (0.02) & (0.03) & (0.27) & (0.14) &              \\ 
  Chronic and not persistent & 0.02  & 0.12   & -0.02  & -0.15  & -0.16  & -0.5, 0.09   \\ 
                             & [816] & (0.02) & (0.02) & (0.15) & (0.07) &              \\ 
  Transiently insecure       & 0     & 0.12   & -0.01  & -0.06  & -0.05  & -0.08, -0.03 \\ 
                             & [816] & (0.02) & (0)    & (0.02) & (0.01) &              \\ 
  Persistently secure         & 0.92  & 0.12   & 0.04   & 0.31   & 0.42   & -0.28, 1.27  \\ 
                             & [816] & (0.02) & (0.05) & (0.42) & (0.22) &              \\ 
  

%% file: sections/tables/psid-twosample-crossD.tex
% latex table generated in R 4.4.1 by xtable 1.8-8 package
% Wed Jul 22 17:03:10 2026
 Education years              & 13.74  & 0.13   & 0.4    & 3.17   & 4.14   & 0.45, 6.43  \\ 
                               & [786]  & (0.01) & (0.16) & (1.34) & (8.07) &             \\ 
  Employed, percent years      & 0.84   & 0.13   & 0.01   & 0.1    & 0.25   & -0.38, 0.73 \\ 
                               & [806]  & (0.02) & (0.03) & (0.27) & (0.13) &             \\ 
  Individual income            & 82.78  & 0.14   & -0.05  & -0.38  & 0.41   & -2.2, 1.43  \\ 
                               & [734]  & (0.01) & (0.13) & (0.97) & (1.01) &             \\ 
  Household income             & 122.92 & 0.12   & 0.06   & 0.51   & 0.99   & -0.47, 2.11 \\ 
                               & [816]  & (0.02) & (0.08) & (0.73) & (0.54) &             \\ 
  Household income, per capita & 48.26  & 0.12   & 0.11   & 0.87   & 1.37   & -0.31, 2.67 \\ 
                               & [816]  & (0.02) & (0.1)  & (0.88) & (0.59) &             \\ 
  Household size               & 4.05   & 0.12   & 0.04   & 0.3    & -0.47  & -0.54, 2.46 \\ 
                               & [816]  & (0.02) & (0.08) & (0.67) & (1.66) &             \\ 
  

%% file: sections/tables/sensitivity-table-alt.tex
% latex table generated in R 4.4.1 by xtable 1.8-8 package
% Wed Jul 22 17:04:06 2026
 Mean PFI & -0.13 & -0.07 & 0.04 & 0 & -0.12 \\ 
    & (0.08) & (0.01) & (0.03) & (0) & (0.08) \\ 
  Food secure, percent years & 0.27 & 0.19 & 0.04 & -0.01 & 0.26 \\ 
    & (0.23) & (0.06) & (0.03) & (0) & (0.23) \\ 
  Food insecure, percent years & -0.27 & -0.19 & 0.04 & 0.01 & -0.26 \\ 
    & (0.23) & (0.06) & (0.03) & (0) & (0.23) \\ 
  Household food spending & 128.83 & -5.01 & 0.04 & 0.19 & 129.01 \\ 
    & (114.14) & (71.16) & (0.03) & (6.94) & (114.35) \\ 
  Household food spending, per capita & 88.75 & -23.89 & 0.04 & 0.88 & 89.63 \\ 
    & (84.02) & (51.21) & (0.03) & (3.97) & (84.12) \\ 
  SNAP, ever participated & -0.16 & 0.12 & 0.04 & 0 & -0.16 \\ 
    & (0.18) & (0.02) & (0.03) & (0) & (0.18) \\ 
  SNAP, percent years participated & -0.11 & -0.06 & 0.04 & 0 & -0.11 \\
    & (0.14) & (0.01) & (0.03) & (0) & (0.14) \\
  